\documentclass[acmsmall,screen]{acmart}

\setcopyright{none}
\renewcommand\footnotetextcopyrightpermission[1]{}
\usepackage{array}
\usepackage{enumerate}
\usepackage{graphicx}
\usepackage{xcolor}
\usepackage{multirow}
\usepackage{colortbl}
\usepackage{longtable}
\usepackage{tabularx}

\newcommand{\nff}[1]{{\textcolor{black}{#1}}}

\newcommand{\nfff}[1]{\textcolor{black}{#1}}

\AtBeginDocument{%
  \providecommand\BibTeX{{%
    \normalfont B\kern-0.5em{\scshape i\kern-0.25em b}\kern-0.8em\TeX}}}

\setcopyright{acmcopyright}
\copyrightyear{2025}
\acmYear{2025}
\acmDOI{XXXXXXX.XXXXXXX}

\begin{document}

\title{When Deep Learning Meets Information Retrieval-based Bug Localization: A Survey}

\author{Feifei Niu}
\email{feifeiniu96@gmail.com}
\orcid{0000-0002-4123-4554}

\author{Chuanyi Li}
\email{lcy@nju.edu.cn}

\affiliation{%
  \institution{State Key Laboratory for Novel Software Technology, Software Institute, Nanjing University}
  \city{Nanjing}
  \country{China}
}

\author{Kui Liu}
\email{brucekuiliu@gmail.com}
\affiliation{%
  \institution{Software Engineering Application Technology Lab, Huawei}
  \city{Hangzhou}
  \country{China}}

\author{Xin Xia}
\email{xin.xia@acm.org}
\affiliation{%
  \institution{College of Computer Science and Technology, Zhejiang University}
  \country{China}}

\author{David Lo}
\email{davidlo@smu.edu.sg}
\affiliation{
\institution{School of Computing and Information Systems, Singapore Management University}
\country{Singapore}
}

\renewcommand{\shortauthors}{Niu et al.}

\begin{abstract}
Bug localization is a crucial aspect of software maintenance, running through the entire software lifecycle. Information retrieval-based bug localization (IRBL) identifies buggy code based on bug reports, expediting the bug resolution process for developers. Recent years have witnessed significant achievements in IRBL, propelled by the widespread adoption of deep learning (DL). To provide a comprehensive overview of the current state of the art and delve into key issues, we conduct a survey encompassing 61 IRBL studies leveraging DL. We summarize best practices in each phase of the IRBL workflow, undertake a meta-analysis of prior studies, and suggest future research directions. This exploration aims to guide further advancements in the field, fostering a deeper understanding and refining practices for effective bug localization. Our study suggests that the integration of DL in IRBL enhances the model's capacity to extract semantic and syntactic information from both bug reports and source code, addressing issues such as lexical gaps, neglect of code structure information, and cold-start problems. Future research avenues for IRBL encompass exploring diversity in programming languages, adopting fine-grained granularity, and focusing on real-world applications. Most importantly, although some studies have started using large language models for IRBL, there is still a need for more in-depth exploration and thorough investigation in this area.

\end{abstract}



\received{14 May 2024}
\received[revised]{16 December 2024}
\received[accepted]{29 April 2025}

\maketitle

\section{Introduction}
The scale and complexity of software systems have expanded progressively, resulting in a tremendous increase in software bugs. Software bugs can trigger software failures, resulting in significant losses and widespread inconvenience. For example, \nff{on July 19, 2024, a faulty update distributed by CrowdStrike to its Falcon Sensor security software caused approximately 8.5 million Microsoft Windows systems to crash and fail to restart. This incident, described as ``historic in scale,'' resulted in widespread disruption across industries such as airlines, banks, hospitals, and governmental services, with an estimated financial loss of at least US\$10 billion.\footnote{https://en.wikipedia.org/wiki/2024\_CrowdStrike\_incident}}
\nff{Given the significant losses caused by software bugs, software maintenance, particularly bug fixing, plays a critical role throughout the software lifecycle, consuming approximately 70\% of the time and cost associated with software development~\cite{4526688}. As an vital process of bug fixing, bug localization aims to identify code entities (e.g., classes, methods or changesets) related to software bugs~\cite{zou2019empirical}.
Over the years, researchers have proposed multiple approaches for bug localization, leveraging development history~\cite{kim2007predicting, rahman2011bugcache}, bug report similarity~\cite{zhou2012should}, and others. Zou et al.~\cite{zou2019empirical} categorized bug localization techniques into seven main families based on their inputs: spectrum-based bug localization (SBBL)~\cite{xie2013theoretical, abreu2007accuracy, harrold2000empirical}, information retrieval-based bug localization (IRBL)~\cite{zhou2012should}, mutation-based bug localization~\cite{papadakis2015metallaxis, moon2014ask}, dynamic program slicing~\cite{agrawal1995fault, renieres2003fault}, history-based bug localization~\cite{kim2007predicting,rahman2011bugcache}, stack trace analysis~\cite{wong2014boosting, wu2014crashlocator}, predicate switching~\cite{zhang2006locating}.
Among these, SBBL and IRBL are among the most studied and widely recognized techniques~\cite{zamfirov2022literature, zou2019empirical}. SBBL relies on run-time behavior, such as test coverage information, to compare program spectra on passed and failed test cases, enabling the ranking of program elements based on their likelihood of being faulty.
In contrast, IRBL does not require run-time information. Instead, it takes a bug report-typically a natural language document describing the symptoms of a software bug-as input. The bug report serves as a query, while the source code entities of the project are treated as a document corpus. IRBL techniques identify textual overlap between the query (bug report) and the documents (source code entities) to locate the root cause of a bug. The source code entities with the highest overlap are returned in a ranked list, starting with the most relevant ones~\cite{zamfirov2022literature}. This study focuses primarily on IRBL techniques, aiming to explore their effectiveness and challenges in bug localization tasks.}


It has been two decades since the introduction of the first IRBL technique~\cite{marcus2004information}, and during the two decades, many new technologies have emerged in response to the development of machine learning and deep learning (DL)~\cite{akbar2020large}. Akbar et al.~\cite{akbar2020large} carried out a large-scale comparative evaluation of IRBL approaches, and they divided the approaches into three generations: (1) The first generation approaches, developed between the years 2004 and 2011, leveraged Bag-of-Words (BoW)~\cite{kuhn2007semantic, lukins2008source, marcus2004information, rao2011retrieval} and laid foundations for such approaches. Marcus et al.~\cite{marcus2004information} demonstrated that Latent Semantic Indexing (LSI) could be used for concept location. (2) The second generation approaches, scattered from 2010 to 2016~\cite{davies2013bug, davies2012using, moreno2014use, nichols2010augmented, saha2013improving, sisman2012incorporating, wang2014version, wen2016locus, wong2014boosting, ye2014learning, youm2015bug, zhou2012should}, augment BoW-based approaches with additional software information, including version history~\cite{zhou2012should, wang2014version, rath2018analyzing, youm2015bug, sisman2012incorporating, wang2016amalgam+, rahman2011bugcache, wen2016locus, kim2007predicting, lewis2013does}, code structure~\cite{saha2013improving, wang2016amalgam+, wang2014version}, and stack trace~\cite{wong2014boosting, wang2016amalgam+}. These studies revealed that similar bug reports, code structure, version history, and stack trace, play an important role in localizing buggy files. (3) Recently, the emergency of DL technology has ignited the development of the third generation of IRBL techniques. Lam et al.~\cite{lam2015combining} introduced the HyLoc approach in 2015, pioneering the utilization of DL for bug localization. Different from the first two generations, the third generation emphasises more on leveraging semantic information within both bug reports and source codes. Additionally, code syntactics structures, such as Abstract Syntax Trees (AST) and Control Flow Graphs (CFG), have also been incorporated~\cite{ma2022flowingcflow, xiao2023bugradar, liang2019deepcast}. The integration of abundant code information has significantly enhanced the accuracy of third-generation IRBL.

\nff{As of now, we are aware of four surveys~\cite{agarwal2014fault, wong2016survey, xia2023information, xia2023information} related to IRBL. Agarwal et al.~\cite{agarwal2014fault} reviewed 30 key papers on fault localization approaches in 2014. Wong et al.~\cite{wong2016survey, wong2023software} provided a comprehensive classification of fault localization approaches into eight categories, including spectrum-based, machine learning-based, and model-based approaches. Xia and Lo~\cite{xia2023information} summarized techniques utilized in IRBL. In addition, six empirical studies~\cite{saha2014effectiveness, wang2015evaluating, lee2018bench4bl, li2022empirical, akbar2020large, tsumita2023large} have extensively evaluated the effectiveness of IRBL techniques, primarily focusing on non-DL-based approaches. Zou et al.~\cite{zou2018practitioners} and Kochhar et al.~\cite{kochhar2016practitioners} explored practitioners' perspectives on IRBL approaches. Despite these efforts, there is no in-depth survey dedicated to the application of DL in the IRBL field. Our research indicates that most DL-based approaches have emerged since 2020 (as shown in Fig.~\ref{fig:year}). Given the rapid growth and increasing popularity of DL-based approaches, a comprehensive and systematic investigation into this third generation of IRBL techniques is imperative., which would provide researchers and practitioners with an overview of state of the art and future trends.}

To this end, we systematically select a pool of 61 IRBL studies that leverage DL techniques, and then synthesize quantitative and qualitative analysis using the data extracted from these studies. The objectives of this study are (1) to help researchers to gain a comprehensive understanding about typical concerns and current techniques at different stages of DL-based IRBL approaches, enabling researchers entering this field swiftly familiarize themselves; (2) to help researchers pinpoint gaps and opportunities for future studies on this topic; and (3) to benefit practitioners in selecting gaps and tailoring IRBL models to their quality assurance requirements. 

This work differs from the related studies from the following aspects: (1) focusing particularly on the use of DL for IRBL, (2) with substantial level of depth on several aspects of DL-based IRBL, (3) achieving comprehensive coverage of the literature including 61 primary studies published until \nff{November 2024}, and (4) following a systematic literature survey research method. To the best of our knowledge, we are the first to perform a comprehensive survey on DL-based IRBL techniques. \nff{This survey encompasses the latest deep representations of text and code, including both semantic and syntactic information, along with various deep models and language models. Additionally, we highlight the research challenges and propose future directions for the field.} 







The remaining of the paper is structured as follows: 
Section~\ref{sec:methodology} demonstrates the research methodology. Section~\ref{sec:results} presents an overview of the results. Section~\ref{sec:approach} presents the overview on the various DL-based approaches for IRBL. Section~\ref{sec:evaluation} shows evaluation aspects of this problem. Section~\ref{sec:challenge} outlines the challenges faced by IRBL approaches. 
Section~\ref{sec:threats} reports the threats to validity. 
Section~\ref{sec:conclusion} concludes the study and offers our vision on the future developments on the field.

\section{Methodology} \label{sec:methodology}
In this paper, we carry out a systematic literature survey under the guidelines of Kitchenham~\cite{kitchenham2004procedures} and Zhang et al.~\cite{zhang2011identifying}, to ensure an unbiased and repeatable procedure. It was carried out in three steps: planning, execution, and analysis. At the planning stage, we firstly identified the research questions and the survey protocols serving for our research aim and objectives. Then, at the execution stage, we carried out the study according to the protocols, study selection and snowballing to get a diverse set of literature on the subject of DL-based IRBL. Finally, we had four of our co-authors to analyze the selected literature and address the research questions at the analysis stage. Our research process can be visualized in Fig.~\ref{fig:process}.

\subsection{Research Questions}
The overall objective of this survey is to gain insights into the application of DL in IRBL. In order to have a detailed view of this topic, the survey addresses three research questions, which would allow us to systematically categorize and understand the current research and identify limitations and future research directions in deep bug localization.

\begin{enumerate}[RQ1.]
    \item What IRBL approaches leveraging DL have been proposed so far?
    
    This RQ identifies and analyzes the \textbf{techniques} that have been proposed in the literature, including the model structure, the adopted text and code representation, as well as other features utilized for bug localization. We also provided an overview of the performance of all models.
    
    \item How are the DL-based approaches evaluated?
    
    This RQ spots how the literature \textbf{evaluates} their approaches, including the dataset employed for evaluation, evaluation metrics and validation approaches, as well as the granularity of different approaches, to facilitate future usage. 

    \item What are the challenges faced when applying DL in IRBL?
    
    This RQ figures out the \textbf{challenges and problems} faced in deep bug localization, as well as points out the road head.
\end{enumerate}

\subsection{Search Strategy}

We began with a search strategy to adequately and effectively search for relevant studies from academic digital libraries. We designed our search string based on the PICO (population, intervention, comparison and outcome) framework~\cite{keele2007guidelines}, which has been widely used by surveys and systematic mapping studies~\cite{zakari2019software, zakari2020multiple, croft2022data}. The relevant terms for population, intervention and outcomes are as follows:

\begin{itemize}
    \item Population: Software, Program
    \item Intervention: Deep Learning, Information Retrieval
    \item Outcome: Bug Localization
\end{itemize}

Based on the PICO structure, the following search string was utilized for searching relevant articles:

(``software'' OR ``program'') AND (``deep learning'') AND (``information retrieval'') AND (``bug'' OR ``fault'' OR ``defect'') AND (``localize'' OR ``localization'')

We applied this search string to the eight electronic databases listed in Table~\ref{tab:digitallibrary} to search for relevant articles. We conducted our search on November 30, 2024, identifying studies published up to that date.

\begin{figure}[htpb]
\centering
\includegraphics[width=0.7\linewidth]{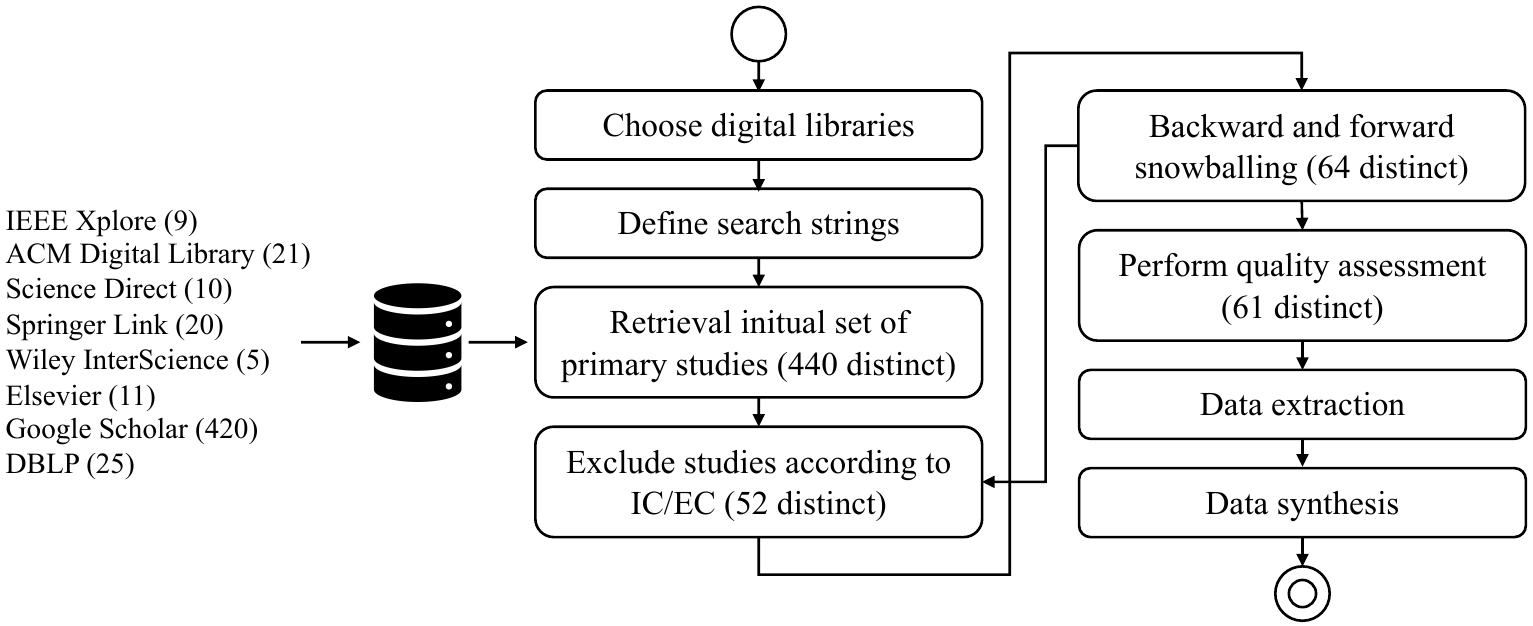}
\caption{Primary Study Selection Process.}
\Description[]{}
\label{fig:process}
\end{figure}

As shown in Fig.~\ref{fig:process}, we initially retrieved 440 distinct studies: 9 studies from IEEE Xplore, 21 studies from ACM Digital Library, 10 studies from Science Direct, 20 studies from Springer Link, 5 studies from Wiley InterScience, 11 studies from Elsevier, 420 studies from Google Scholar, and 25 studies from DBLP.

\subsection{Study Selection}

\subsubsection{Inclusion/Exclusion Criteria}
To identify the most relevant articles to address the research questions in our survey, we specified our inclusion criterias (ICs) and exclusion criterias (ECs), inspired by similar studies~\cite{giray2023use, zakari2020multiple, croft2022data}. The ICs and ECs are outlined in Table~\ref{tab:criteria}. By applying the ICs and ECs on title, abstract and keywords, \nff{we first ensured that the selected studies focus on information retrieval-based approaches to bug localization (IC1) and employ DL algorithms in their methodologies (IC2). We adopt the definition of DL from the seminal work Deep Learning by Yann et al.~\cite{lecun2015deep}, which defines it as the use of multilayer neural networks, e.g., Convolutional Neural Networks (CNNs)~\cite{lecun1998gradient}, Long Short-Term Memory Networks (LSTMs)~\cite{hochreiter1997long}, Recurrent Neural Networks (RNNs)~\cite{rumelhart1986learning}, and transformers~\cite{vaswani2017attention}.}

\nff{Additionally, we excluded literature that was not written in English (EC1), technical reports and theses (EC2), and duplicate or redundant studies reporting the same findings (EC3). Further filtering involved verifying that the research was published before November 2024 (IC3) in peer-reviewed venues or arXiv (IC4). To confirm the publication sources, we extracted information such as the ``journal,''``URL,'' ``DOI,'' and ``series.'' For papers from arXiv, particularly those from recent years, we opted to include them due to the emerging nature of this field, where many works are still in the submission process. Although these papers had not undergone peer review, we implemented a quality assessment process to exclude low-quality papers. We excluded papers that did not employ information retrieval techniques (EC4) or DL techniques (EC5). Additionally, we omitted research focused on fault localization methods, such as spectrum-based fault localization, that did not use bug reports as a basis for bug localization (EC6). Furthermore, we did not include studies that targeted bug localization within DL models themselves, as our emphasis was on the broader application of DL techniques for bug localization in software systems rather than within the architecture of DL models.}

\nff{For empirical studies, we included those that introduced a new IRBL technique. However, if the paper was a secondary study, and the primary study was already included in our pool, we excluded it. Overall, 388 studies were removed based on our criteria, and as a result of this first screening step, we retained 52 studies in our pool.}

\subsubsection{Snowballing}
In order to obtain a thorough literature, we performed snowballing on our initial set of literature, under the guide line of~\cite{wohlin2014guidelines}. We carried out both forward and backward snowballing, and iteratively look through the references and citations of the literature. The snowballing process finishes when there are no more new studies adding in. For each round of snowballing, we screened the articles with the IC and EC. 
Finally, we included additional 12 papers in our final set. So far, we obtained 64 studies in our article pool.

\subsubsection{Quality Assessment}
Quality assessment is a vital process of survey to ensure that we form a proper and fair representation of the research works~\cite{kitchenham2004procedures}. We used a quality checklist to assess the quality of studies and excluded studies that could not pass the checklist. We derived our quality checklist from Hall et al.~\cite{hall2011systematic} and Hosseini et al.~\cite{hosseini2017systematic}. We mainly assessed the primary studies from three aspects: the data, the model, and the evaluation criteria, as listed in Table~\ref{tab:qualityassessment}. We assessed all three criteria to ensure high-quality paper.

The quality assessment checklist was independently applied to all 64 primary studies by two authors. Discussions were conducted in the event of disagreement to achieve consensus. Finally, three studies were excluded from our primary study pool for not meeting the quality assessment criteria. Finally, 61 primary studies were included for the data extraction phase. 


\subsection{Data Analysis}

\subsubsection{Data Extraction}
After primary study selection, we formed a data extraction form (Table~\ref{tab:extractionform}) to extract data from primary studies, to answer the research questions. As indicated in the table, there are 17 fields in total. The initial five rows constitute the metadata of the studies, with six fields specifically pertained RQ1, another four fields collectively related to RQ2, and the remaining two fields associated with RQ3. The first author firstly formed an initial extraction form according to previous surveys~\cite{catal2022applications}. Then two authors conducted a pilot study on five randomly selected primary studies to assess the completeness and usability of the form. The two authors continuously discussed and refined the structure of the form until they reached a consensus. All the primary study was divided between the two authors to extract data from each set of studies independently. The data extraction form was collaboratively collected by two authors using an online sheet. Finally, the third author inspected the extraction form to ensure the correctness of the results.


\subsubsection{Data Synthesis}
The ultimate goal of a survey study is the aggregation of information to provide an overview of the state of the art. We extracted quantitative data of our data extraction form to identify and report the results for RQ1 and RQ2. For RQ3, we carried out qualitative analysis to synthesize the outputs of our data extraction form. Specifically, this is to identify the reported challenges and solutions, as well as gaps of current research. Any discussion in a paper that explicitly mentioning a challenge or future work was extracted to the data extraction form during data extraction. We extract the main themes and revised manually to categorize the challenged towards application of DL in IRBL.

\section{Results}\label{sec:results}

\begin{figure}[htbp]
\centering
\begin{minipage}{0.35\textwidth}
    \raggedright 
    \includegraphics[width=\textwidth]{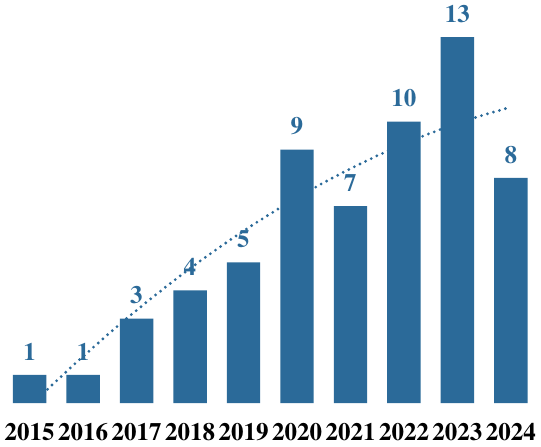}
    \caption{Number of primary studies over the years.}
    \Description[<Number of primary studies over the years.>]{<Number of primary studies over the years.>}
    \label{fig:year}
\end{minipage}%
\hspace{0.05\textwidth} 
\begin{minipage}{0.55\textwidth}
    \raggedleft 
    \includegraphics[width=\textwidth]{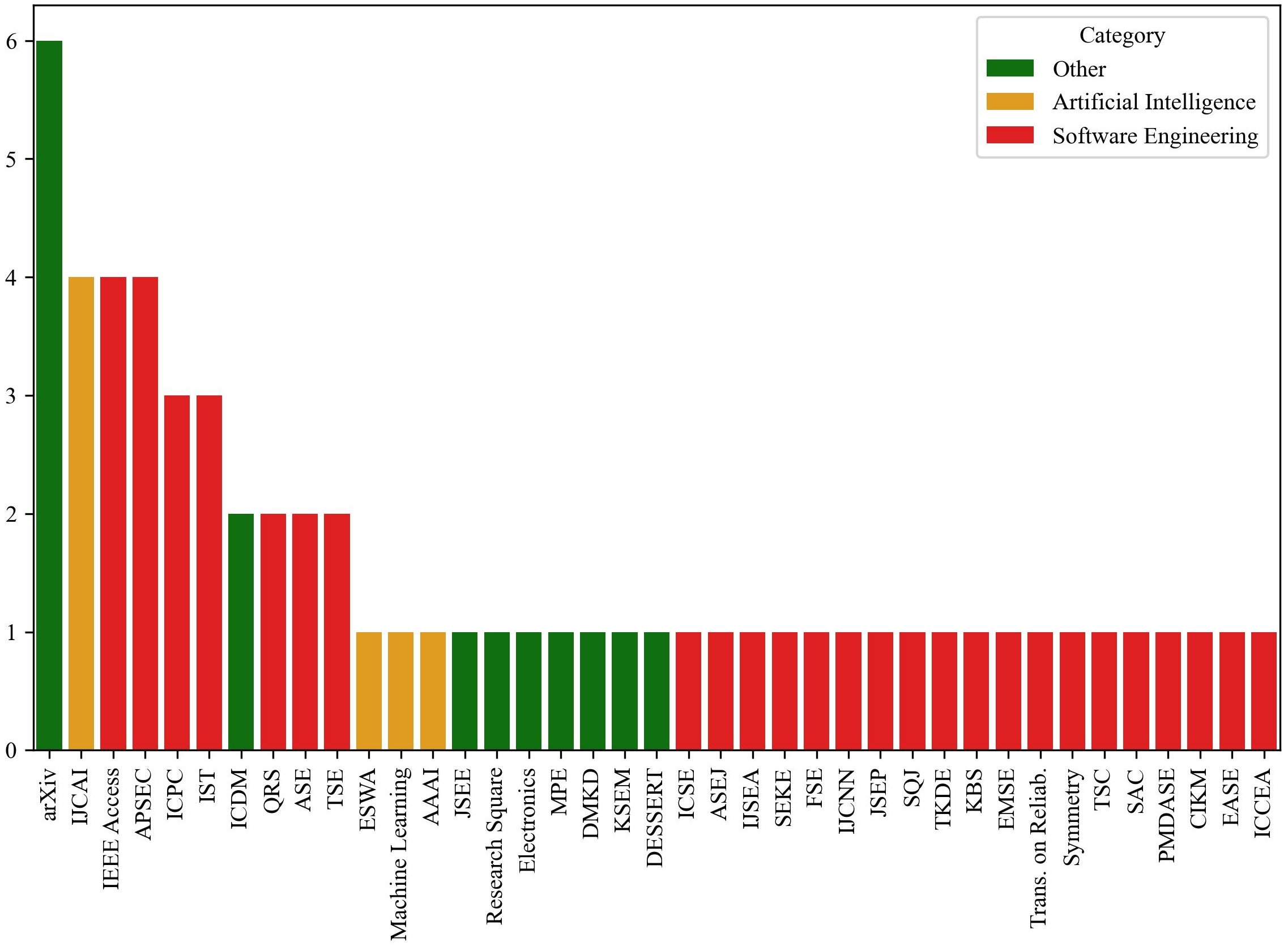}
    \caption{Distribution of venues.}
    \Description[<Distribution of venues.>]{<Distribution of venues.>}
    \label{fig:venue}
\end{minipage}
\end{figure}

There are 61 articles in the final study pool.
Fig.~\ref{fig:year} presents the distribution of selected primary studies over the years. The first approach that employs DL for IRBL appears in 2015. Then there is a significant increase in the number of papers published each year since 2020, which aligns with the rapid development and popularization of DL techniques. This indicates that DL has attracted the attention of an increasing number of researchers and practitioners, and is still undergoing huge growth at the time of this study. \nff{Figure~\ref{fig:venue} shows the distribution of venues for the 61 studies (classified according to the category provided on the SJR\footnote{https://www.scimagojr.com} website). Among them, 55 papers were published in peer-reviewed venues, including 39 in software engineering venues (such as ICSE, TSE, ASE), 7 in artificial intelligence venues (such as IJCAI), and 9 in other venues (such as Electronics)}.



The goal of bug localization is to identify and pinpoint the possible location or locations in the source code where a software bug is present. The formulation of bug localization problem is:
Let $\mathcal{B}$ = \{$b_1$, $b_2$, ..., $b_{N_b}$\} denote the set of bug reports, and $\mathcal{C}$ = \{$c_1$, $c_2$, ..., $c_{N_c}$\} denote the set of source code entities (e.g., files, methods, statements, or changesets) a software project, where $N_b$, and $N_c$ denote the number of bug reports and source code entities, respectively. The target of bug localization is to learn a function $f$: $\mathcal{B}$ $\times$ $\mathcal{C}$ $->$ $\mathcal{Y}$, where $\mathcal{Y}$ $\in$ $\{+1, -1\}$ indicates whether a source code entity $c_i$ $\in$ $\mathcal{C}$ is related to a bug report $b_j$ $\in$ $\mathcal{B}$.

\begin{figure}[htpb]
\centering
\includegraphics[width=0.6\textwidth]{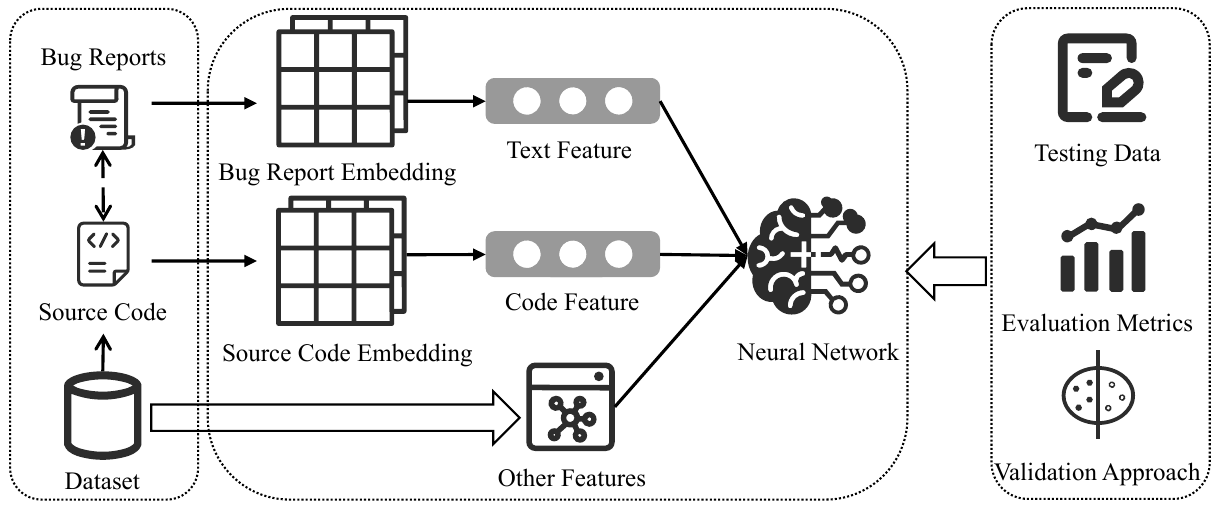}
\caption{Framework of IRBL Approaches.}
\Description[]{}
\label{fig:framework}
\end{figure}

The overall framework of DL-based IRBL approaches is depicted in Fig. \ref{fig:framework}. Initially, the link between bug reports and source code entities is constructed using heuristic rules proposed by Bachmann and Bernstein~\cite{bachmann2009software}. IRBL approaches take bug reports and source code entities as text and utilize embedding to transform them into vectors. Subsequently, text and code features are extracted employing DL techniques. Finally, matching approaches are applied to correlate bug reports with source code entities.

Bug reports can be linked to different kinds of code entities, e.g., files, classes, methods, statements, or changesets. Therefore, IRBL can be categorized into different granularity, such as file-level, method-level, statement-level, and changeset-level.

For evaluation purposes, the dataset is divided into training and test set based on the validation approach. Validation approaches in DL include holdout validation, \textit{k}-fold cross-validation, leave-one-out validation, and time series validation. Model performance is assessed by the evaluation metrics, including MAP, MRR, and Top \textit{k}.

Since the second generation of IRBL approaches relies on code historical information, such as version history and similar bug reports, cross-project bug localization (CPBL) is challenging to implement. However, the third generation of IRBL approaches primarily focuses on comparing the similarity between bug reports and source code, with a few approaches also incorporating code historical information. To address the cold-start problem in newly established projects, DL-based CPBL approaches have been introduced. Additionally, leveraging the robust learning capabilities of language models, cross-language bug localization (CLBL) approaches have also been proposed.
Among the 61 studies, three studies (4.9\%)~\cite{huo2019deeptranp, zhu2021trobo, zhu2020cooba} propose solutions for CPBL, while four studies (6.6\%)~\cite{loyola2018bug, yang2021locatingmram, liang2022modelingflim, zhu2022bl-gan} are evaluated on both within-projects and cross-projects. Besides, two studies (3.3\%)~\cite{chandramohan2024supporting, chakraborty2024blaze} propose both CPBL and CLBL approaches. The remaining 52 studies (85.2\%) exclusively put forth within-project bug localization (WPBL) approaches.


\section{RQ1: What IRBL approaches leveraging DL have been proposed so far?}\label{sec:approach}
\begin{figure}[htbp]
\centering
\includegraphics[width=\textwidth]{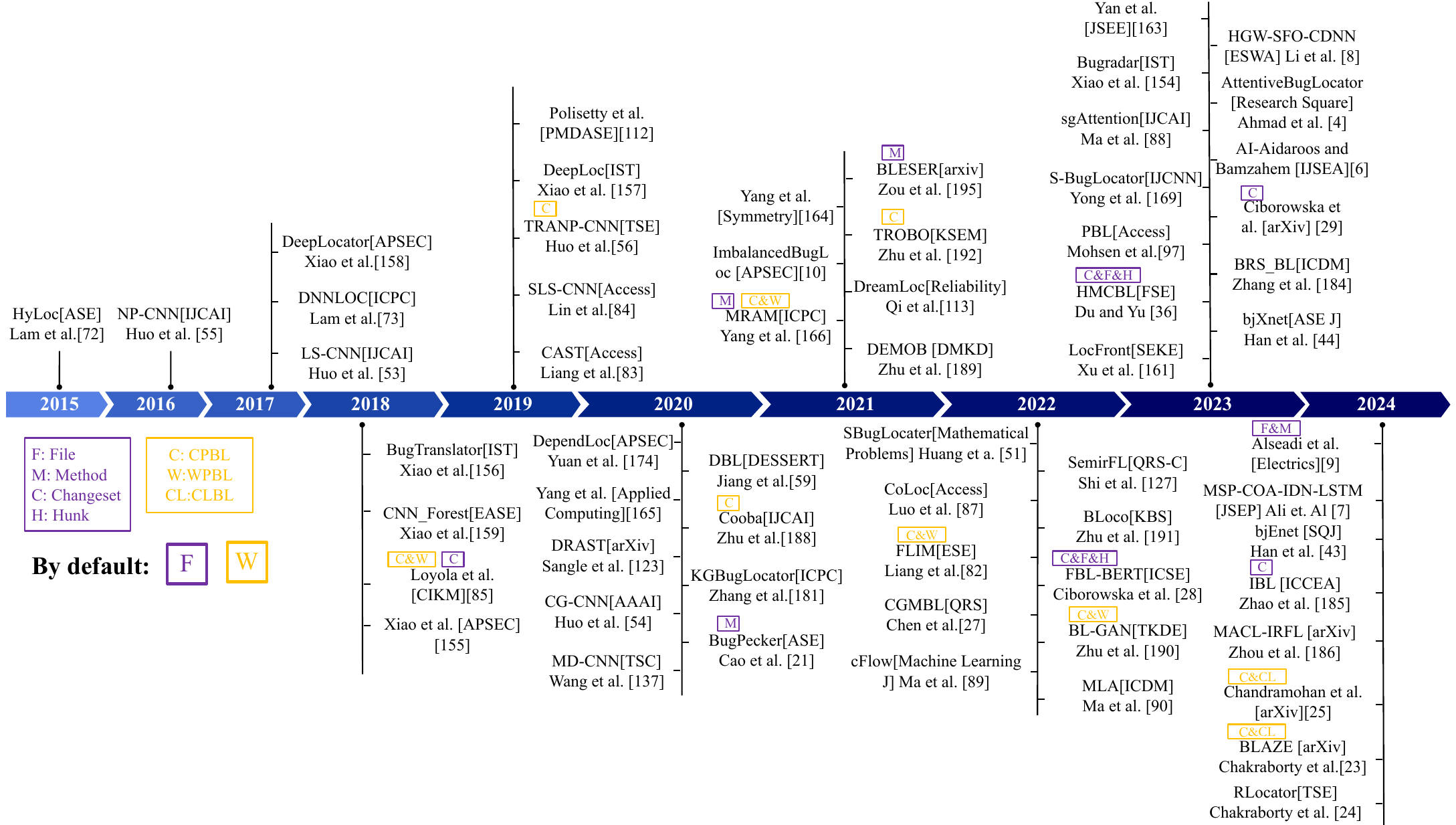}
\caption{Primary studies over the years.}
\label{fig:approaches}
\Description[]{}
\vspace{-0.5cm}
\end{figure}

Figure~\ref{fig:approaches} illustrates a timeline of IRBL approaches that utilize DL, encompassing 61 papers published from 2015 to October 2024. This timeline highlights the evolution of the third generation of IRBL approaches to date. In the following sections, we categorize these approaches based on their model structures, representation for bug reports and code, as well as other features employed for IRBL.

\subsection{Model Taxonomy}
\begin{figure}[htbp]
\centering
\includegraphics[width=0.9\textwidth]{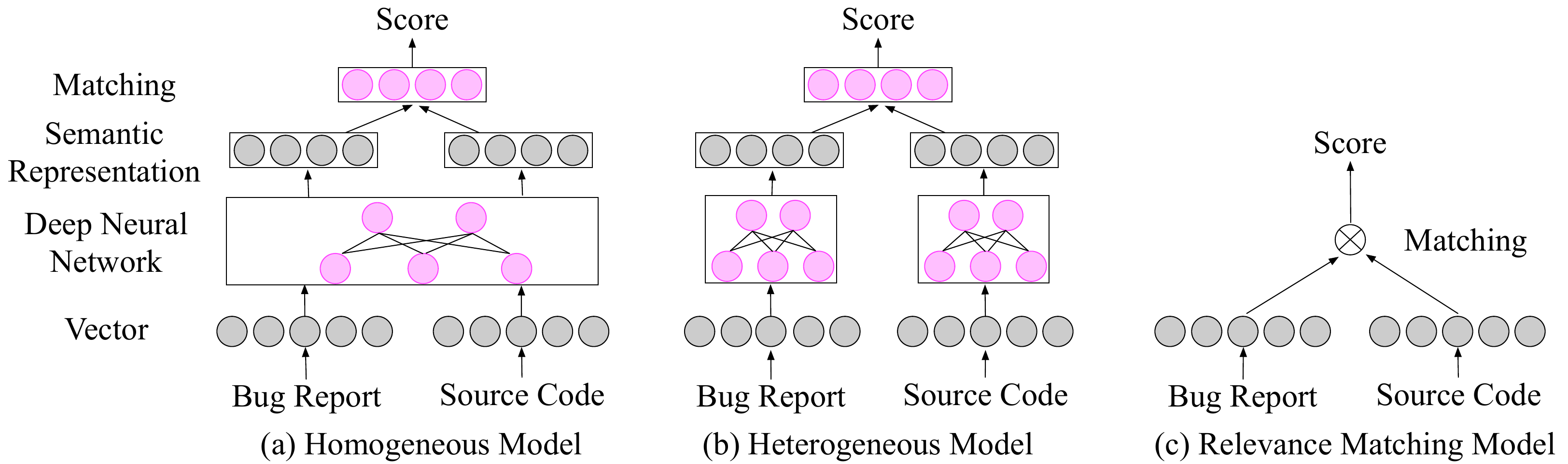}
\caption{Three types of model structures.}
\Description[]{}
\label{fig:modelstructure}
\vspace{-0.5cm}
\end{figure}

The essence of IRBL lies in the matching between bug reports and source code elements. The text matching includes two types: semantic matching and relevance matching~\cite{huang2022sbuglocater}. Semantic matching evaluates the similarity between two pieces of text based on semantic information. Relevance matching necessitates the identification of documents related to the given query, which is typically keyword-based. The first two bug localization generations are primarily relevance matching-based. With the rapid development of Natural Language Processing (NLP) technology, the third generation approaches extract semantic features from bug reports and source code with deep neural networks and then compare the semantic features. There can be three kinds of model structures: (1) Some approaches employ the same neural network structure to coherently learn semantic features from bug reports and source code, which is \textbf{homogeneous network} (as shown in Figure~\ref{fig:modelstructure} (\nfff{a})). (2) On the other hand, considering that bug reports and source code may have different syntactic structures, some approaches use different neural networks to extract semantic features individually, which is \textbf{heterogeneous network} (as shown in Figure~\ref{fig:modelstructure} (\nfff{b})). \nfff{(3) The third category of models, referred to as relevance matching models, convert bug reports and source code into vector representations—using techniques such as TF-IDF, word embeddings, or pre-trained language models (e.g., BERT)—and compute their semantic or lexical similarity through vector similarity measures (e.g., cosine similarity or dot product), as illustrated in Figure~\ref{fig:modelstructure} (c). These models focus purely on content relevance between bug reports and code entities and do not incorporate any structural or relational information.}
Besides, some approaches also integrate other features (r.f. Section~\ref{sec:otherfeather}), such as code fixing history, code comments, and more. In table~\ref{tab:overview} (column ``Model'', ``HE'' stands for heterogeneous, ``HO'' for homogeneous, ``R'' for relevance matching, and ``O'' for others ), we list the category of each approach using different model structures. A total of 33 studies utilized heterogeneous models, 13 studies employed homogeneous models, six papers adopted relevance matching models, and the remaining nine papers utilized other models, including encoder-decoder, and adversarial model. Overall, the majority of papers still favor the use of heterogeneous models, extracting distinct features from bug reports and source code, respectively.

\begin{center}
\begin{table}[htbp]%
\caption{Overview of Approaches.}
\label{tab:overview}
\resizebox{\textwidth}{!}{
\begin{tabular}{llcccp{4cm}cp{4cm}ccc}
\toprule
Year & Approach  & W\&C & Chronological & Consecutive & Split Ratio & Before-fix version & Sample & Granularity & Model \\
\midrule
2015    & HyLoc~\cite{lam2015combining} & W    & \checkmark & \checkmark   &      &  -   & Top 300 similar files & F & HO \\ \midrule
2016    & NP-CNN~\cite{huo2016learningnp-cnn}& W    &  & & 10-fold cross validation &  -   &     & F & HE \\ 
\midrule
\multirow{3}{*}{2017}  
& DeepLocator~\cite{xiao2017improvingdeeplocator}& W    &  & &      & \checkmark   & Files that were ever buggy as negative samples      & F & O \\
& DNNLOC~\cite{lam2017bugdnnloc}    & W    & \checkmark & \checkmark   &      &  -   & Top 300 similar files  & F & HO \\
& LS-CNN~\cite{huo2017enhancingls-cnn}    & W    &  & & 10-fold cross valiation  & \checkmark   &       & F &HE\\ 
\midrule
\multirow{4}{*}{2018}  
& BugTranslator~\cite{xiao2018machinebugtranslator}  & W    & \checkmark & & 75:25 & \checkmark   &       & F &O \\
& CNN\_Forest~\cite{xiao2018bugcnn_forest}& W    & \checkmark & &      & \checkmark   &       & F &HE \\
& Loyola et al.~\cite{loyola2018bug}  & W\&C & \checkmark & & 80:20 \& held out one  &  O   &       & C &HE \\
& Xiao et al.~\cite{xiao2018improving}& W    &  & &      &  \checkmark   &       & F & O \\
\midrule
\multirow{5}{*}{2019}  
& Polisetty et al.~\cite{polisetty2019usefulness} & W    &  & & 10-fold cross validation & \checkmark   &       & F & HO \\
& DeepLoc~\cite{xiao2019improvingdeeploc}   & W    & \checkmark & & 60:20:20 & \checkmark   & Top 300 similar files    & F &HE \\
& TRANP-CNN~\cite{huo2019deeptranp}  & C    & \texttimes & & 80:20, repeat 10 times&  -   &       & F &HE \\
& SLS-CNN~\cite{liu2019convolutionalsls-cnn}   & W    & \checkmark & & 60:20:20 &  -   & Top 300 similar files & F & HE \\
& CAST~\cite{liang2019deepcast}      & W    & \checkmark & \checkmark   &      & \checkmark   &       & F &HE \\ 
\midrule
\multirow{9}{*}{2020}  
& DependLoc~\cite{yuan2020dependloc}  & W    & \texttimes & & 80:10:10,10-fold cross validation     & \checkmark   & Top 200 similar files & F &HE \\
& Yang et al.~\cite{yang2020applying}& W    &  & & 70:30,10-fold cross validation       &  -   &       & F &HE \\
& DRAST~\cite{sangle2020drast}     & W    & \checkmark & \checkmark   &      &   -  & Over-sampling, under-sampling & F(C) & HO \\
& CG-CNN~\cite{huo2020controlcg-cnn}    & W    & \texttimes & & 10-fold cross validation  &-& Randomly drop negative samples & F & HE \\
& MD-CNN~\cite{wang2020multimd-cnn}    & W    & \checkmark & & (90:10):20   & -    & Loss function & F & HO \\
& DBL~\cite{jiang2020deepdbl}       & W    &  & &      &  -   &       & F &HE\\
& Cooba~\cite{zhu2020cooba}    & C    &  & & one project+20 : 80, 10-fold cross validation &-& & F &HE  \\
& KGBugLocator~\cite{zhang2020exploitingkgbuglocator}   & W    & \texttimes & & 10-fold cross validation &  -   &       & F &HE \\
& BugPecker~\cite{cao2020bugpecker}  & W    & \checkmark & & 80:20  & \checkmark   &       & M &HE \\
\midrule
\multirow{7}{*}{2021}  
& Yang et al.~\cite{yang2021utilizing}& W    &  & & 10-fold cross validation &  -   &       & F &HO \\
& ImbalancedBugLoc~\cite{anh2021imbalanced} & W    & \checkmark & \checkmark   &      & \checkmark   & Loss function with bootstrapping & F & HE \\
& MRAM~\cite{yang2021locatingmram}     & W\&C & \checkmark & \checkmark   &      & \checkmark   & randomly select 300 negative samples & M &HE \\
& TROBO~\cite{zhu2021trobo}    & C    &  & & one project+20 : 80 & \checkmark   &       & F &O\\
& DreamLoc~\cite{qi2021dreamloc}  & W    & \checkmark & & 80:10:10 & \checkmark   & Randomly sample 800  & F &HE \\
& DEMOB~\cite{zhu2021deepdemob}    & W    &  & & 10-fold cross validation & -& Select part of irrelevant files     & F &HE \\
& BLESER~\cite{zou2021bleser} & W & \checkmark & & 80:10:10 & \checkmark & Random over-sampling & M &HE \\
\midrule
\multirow{10}{*}{2022} 
& SBugLocater~\cite{huang2022sbuglocater} & W    & \checkmark & & 60:20:20 & \checkmark   & Top 300 unrelated files & F & HO\\
& CoLoc~\cite{luo2022improvingcoloc}    & W    &  & &      & -   &       & F&R \\
& FLIM~\cite{liang2022modelingflim}     & W\&C & \checkmark & \checkmark   &      &  -   &       & F &R\\
& CGMBL~\cite{chen2022cgmbl}    & W    & \checkmark & & 70:30  & \checkmark   & Files that were ever buggy files & F &O \\
& cFlow~\cite{ma2022flowingcflow}     & W    &  & & 80:10:10 &-     & Randomly select the same number of positive instances   & F &HE \\
& SemirFL~\cite{shi2022semirfl}   & W    &  & & 80:20  &  -   &       & F &HE \\
& BLoco~\cite{zhu2022enhancingbloco}     & W    &  & & 10-fold cross validation, 10 times & -  & Simple sampling strategy& F &HE \\
& FBL-BERT~\cite{ciborowska2022fastfbl-bert}  & W    & \checkmark & & 50:50  &  O   & Randomly choose a code change  & F\&C\&H &HO \\
& BL-GAN~\cite{zhu2022bl-gan}   & W\&C & \texttimes & & 80:20:20 \& one project+20 : 80, 10-fold cross validation &- & & F &O \\
& MLA~\cite{ma2022learningmla}       & W    &  & & 80:10:10, repeat 3 times&  -   & Randomly select a subset  & F &HE\\ \midrule
\multirow{13}{*}{2023} 
& Bugradar~\cite{xiao2023bugradar}  & W    & \checkmark & & 90:10  & \checkmark   & Top 300 irrelevant source files  & F & HE \\
& sgAttention~\cite{ma2023capturingsgattention} & W    & \texttimes & & 80:10:10 & -    &       & F& HE \\
& S-BugLocator~\cite{yong2023decomposings-buglocator}   & W    &  & & 80:10:10(10-fold cross validation)&  -   & Randomly drop some negative pairs   & F&HE \\
& PBL~\cite{mohsen2023enhancingpbl}       & W    & \texttimes & & 10-fold cross validation  & \checkmark   &       & F&R \\
& HMCBL~\cite{du2023prehmcbl}    & W    & \checkmark & & 50:50  &  O   & Memory bank\cite{wu2018unsupervised}   & F\&C\&H &HE \\
& LocFront~\cite{xu2023buglocfront}  & W    & \checkmark & & 80:10:10& \checkmark   & Randomly selected 200 unrelated samples & F &O \\
& HGW-SFO-CDNN~\cite{ali2023automatedhgwsfocdnn}   & W    &  & &      &  -   &       & F &HE \\
& AttentiveBugLocator~\cite{ahmad2023attentivebuglocator}      & W    &  & & 70:30  &    - &       & F&HO \\
& AI-Aidaroos et al.~\cite{Al-Aidaroosimpact} & W    &  & &      &   -  &       & F &HE \\
& Ciborowska et al.~\cite{ciborowska2023too}& W    & \checkmark & & 50:50  & O   & Randomly select a hunk      & C & HO\\
& bjXnet~\cite{han2023bjxnet}   & W    & \checkmark & & 60:40(10 folds on test set)&  -   & Randomly select s- for each s+ & F &HE \\
& Yan et al.~\cite{yan2023bug} & W    & \checkmark & & 60:20:20 &  -   & Top 300 similar files & F &HE \\
& BRS\_BL~\cite{zhang2023enhancing} & W & \checkmark & & 80:10:10 & \checkmark & Randomly sample 200 files & F &O \\\midrule
\multirow{8}{*}{2024}
& RLocator~\cite{chakraborty2024rlocator}  & W    & \checkmark & & 60:40  & \checkmark   & Top \textit{k} relevant files ElasticSearch (ES)  & F &HO \\
& Alsaedi et al.~\cite{alsaedi2024two} & W & &&&-&&F\&M&R \\
& MSP-COA-IDN-LSTM~\cite{ali2024softwaremsp-coa-idn-lstm} & W &  & && - & & F &HO \\
& bjEnet~\cite{han2024bjenet} & W & \checkmark & & 60:40 &- & Randomly select \textit{n} files& F &O\\
& IBL~\cite{zhao2024fineibl} & W & & & & O & & C &HE\\
& MACL-IRFL~\cite{zhou2024multimacl-irfl} &W&\checkmark&&80:10:10&\checkmark& & F &HO\\
& Chandramohan et al.~\cite{chandramohan2024supporting} & C\&CL & && -&- &Top 10 most similar files& F &R\\ 
& BLAZE~\cite{chakraborty2024blaze} & C\&CL && & - & - & & F &R\\
\bottomrule
\end{tabular}}
\end{table}
\end{center}

HyLoc~\cite{lam2015combining}, DNNLOC~\cite{lam2017bugdnnloc}, DRAST~\cite{sangle2020drast}, and Bugradar~\cite{xiao2023bugradar} utilized rSVM to extract feature vectors from bug reports and source code, subsequently employing DNN to match them against each other. Anh and Luyen~\cite{anh2021imbalanced} adopted a pre-trained GloVe model to build word embedding vectors for bug reports and source files, and then used DNN to matching the bug reports and source code. SBugLocater~\cite{huang2022sbuglocater} leveraged ALBERT to extract semantic feature from bug reports and source code, and then used DNN for semantic matching. Coloc~\cite{luo2022improvingcoloc} and FLIM~\cite{liang2022modelingflim} approach used pre-trained CodeBERT to convert bug reports and source code into vectors, and them calculated the cosine similarity between vectors. FBL-BERT~\cite{ciborowska2022fastfbl-bert} used ColBERT to encode bug reports and changesets, and then matched each other with DNN models. SgAttention~\cite{ma2023capturingsgattention} used Transformer encoder to encode bug reports, and used CodeBERT model to encode source code, then predicted the relevance of bug reports and source code with DNN~. Ciborowska et al.~\cite{ciborowska2023too} used BERT to convert bug reports and source code into vectors, and then matched them with DNN models. Instead of using DNN models, Chandramohan et al.~\cite{chandramohan2024supporting}, Alsaedi et al.~\cite{alsaedi2024two}, the PBL~\cite{mohsen2023enhancingpbl} and the Blaze~\cite{chakraborty2024blaze} approach utilized different embedding techniques — UnixCoder~\cite{guo2022unixcoder}, GloVe, and SentenceTransformer~\cite{sentencetransformer}, BERT, as well as CodeSage~\cite{zhang2024code} — to vectorize bug reports and source code, subsequently calculating their cosine similarity.

Apart from the above three model structures, some other model structures have also been proposed for IRBL. Xiao et al.~\cite{xiao2017improvingdeeplocator, xiao2018improving} used LSTM-based encoder-decoder architecture,  which took the feature vector of bug reports as input of the encoder, and the feature vector of source code as input of the decoder. Similarly, BugTranslator~\cite{xiao2018machinebugtranslator} used an attention-based RNN encoder-decoder model to translate natural languages (i.e., bug reports) into code tokens. TROBO~\cite{zhu2021trobo}, CGMBL~\cite{chen2022cgmbl} and BL-GAN~\cite{zhu2022bl-gan} used adversarial learning to bridge the semantic gap between code and bug reports, which includes a generator and a discriminator. The generator is designed to generate new synthetic instances from the task domain that can fool the discriminator, whereas the discriminator is desired to classify instances as either real (i.e., from the real task domain) or fake (i.e., generated by the generator)~\cite{zhu2022bl-gan}. LocFront~\cite{xu2023buglocfront} embedded bug reports and source code with Word2Vec, and then computed the scaled dot-product attention score for each project information with respect to bug report. The BRS\_BL~\cite{zhang2023enhancing} approach generates summaries of bug reports, while the bjEnet~\cite{han2024bjenet} approach creates summaries of code methods, followed by embedding processes.

\nff{Different models adopt varying strategies for linking bug reports with source code. The heterogeneous model offers flexibility and specialization by employing separate networks for bug reports and source code, allowing for better representation of each input type but increasing both complexity and resource demands~\cite{zeng2021deep}. In contrast, the homogeneous model simplifies the architecture by using a single network for both inputs, resulting in greater efficiency and ease of training; however, it lacks the ability to generate specialized representations for bug reports and source code~\cite{ma2023capturingsgattention}. The relevance matching model, on the other hand, focuses on computing a direct matching score between inputs, making it computationally efficient but potentially less capable of capturing complex semantic relationships~\cite{ciborowska2022fastfbl-bert}. While the heterogeneous model excels at handling diverse input types, the homogeneous model is more streamlined but may underperform when the inputs vary significantly in structure and semantics. Moving forward, challenges for IRBL models include improving generalization across different programming languages and domains, managing ambiguous or incomplete bug reports, and ensuring scalability for large systems. Additionally, models must enhance explainability to foster trust and adoption, enable real-time detection, and incorporate multimodal data like runtime logs and version histories. Human-AI collaboration and model interpretability will also be critical for future advancements.}

\subsection{Text and Code Representation}

\begin{figure}[!htbp]
\centering
\includegraphics[width=\textwidth]{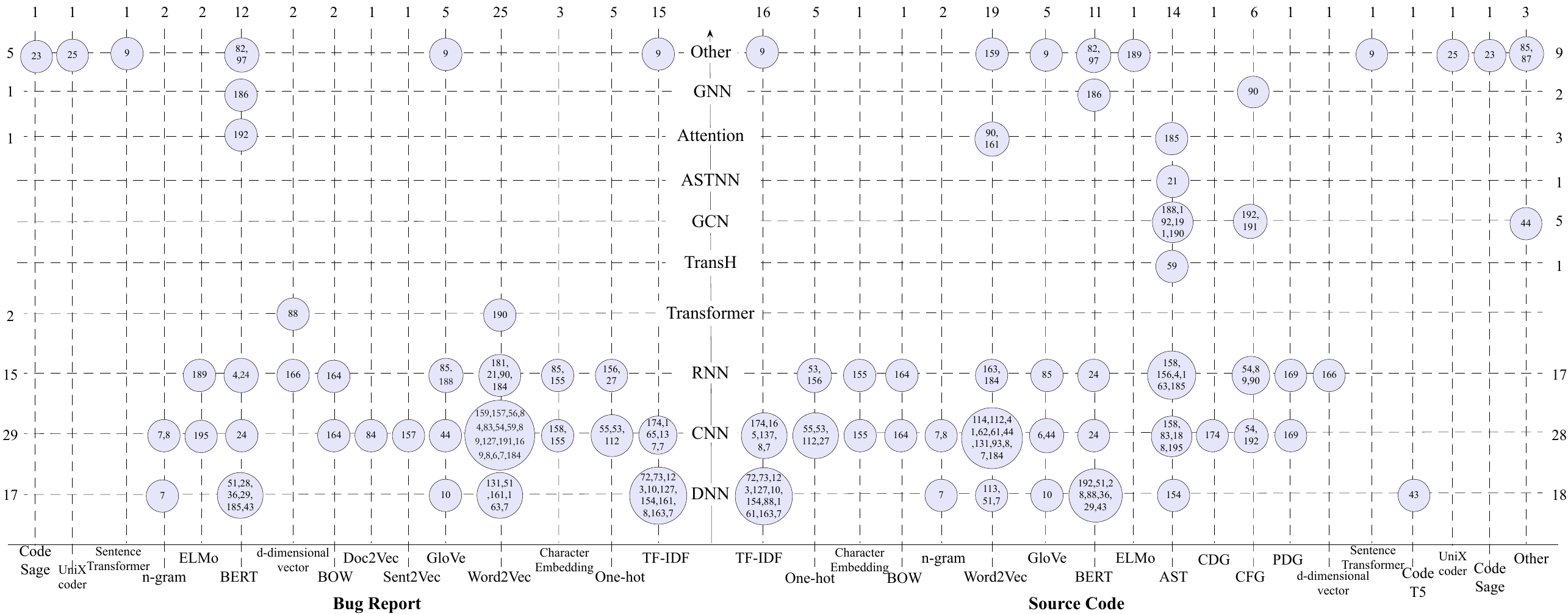}
\caption{Deep techniques used for text representation and feature extraction.}
\Description[]{}
\label{fig:techniques}
\vspace{-0.5cm}
\end{figure}

Given that bug reports and source code are textual data, the process of vectorization and feature extraction are essential for DL algorithms to evaluate their relevance.
Figure~\ref{fig:techniques} shows representation and feature extraction approaches used in the primary studies in our study pool.

\subsubsection{Bug Report Representation}
Various approaches have been adopted to convert the bug reports into easy-to-process representations, such as lists of features or embedding-based vector representations~\cite{sonbol2022use}. 
The basic representation model is Vector Space Model (VSM), which represents text as a term-by-document matrix~\cite{salton1989automatic}. Such techniques include one-hot, Bag-of-Words (BOW), and TF-IDF. However, since these representation approaches mainly rely on word frequencies and such, they overlook semantic context. To address this issue, some advanced embedding techniques have been proposed. Word embedding excels in capturing the contextual essence of a word within a document, enabling words with similar meanings to possess analogous vector representations. Famous pre-trained word embeddings include Word2Vec~\cite{mikolov2013efficientword2vec}, GloVe~\cite{pennington2014glove}, BERT~\cite{devlin2018bert}, and ELMo~\cite{matthew1802deep}.

\textbf{Word2Vec.} According to Figure~\ref{fig:techniques}, the most popular bug report representation technique is Word2Vec, which has been adopted in 25 studies. There are two main architectures used in Word2Vec: Skip-Gram and Continuous Bag of Words (CBOW). Skip-Gram predicts the context words given a target word, while CBOW predicts the target word given its context words. Both architectures have been used to learn semantic feature from bug reports. 

After represent the text with Word2Vec, DL techniques including DNN, CNN, RNN and Transformer are used for extracting semantic features from Word2Vec. Among these four DL techniques, CNN is the most popular, which has been adopted in fourteen studies. Six studies~\cite{qi2021dreamloc, huang2022sbuglocater, xu2023buglocfront, yan2023bug, zhang2023enhancing, ali2024softwaremsp-coa-idn-lstm} use DNN to extract features. Zhang et al.~\cite{zhang2020exploitingkgbuglocator} used LSTM to extract semantic information from Word2Vec embedding of bug reports. Cao et al.~\cite{cao2020bugpecker} and Ma et al.~\cite{ma2022learningmla} leverage GRU to extract textual feature from Word2Vec embedding. Besides, the self-attention layer~\cite{vaswani2017attention} is also employed to enhance the word vector in MLA~\cite{ma2022learningmla}. BL-GAN~\cite{zhu2022bl-gan} approach use Word2Vec as the embedding and attention-based Transformer network as the encoder, to achieve a better trade-off between the ability to model long-range dependency and computational efficiency. 

\textbf{TF-IDF.} TF-IDF is the second most commonly used technique for bug reports representation. This weighting scheme helps identify key terms in a document and is widely used in various NLP tasks. Moreover, Zhou et al.~\cite{zhou2012should} proposed rVSM, which takes the document length into consideration, and could optimize the classic VSM model for bug localization. 
Experimental results show that rVSM performs better than classical VSM. As a result, ten out of fourteen studies use rVSM and four studies~\cite{lam2017bugdnnloc, yuan2020dependloc, anh2021imbalanced, ali2024softwaremsp-coa-idn-lstm} use the classical TF-IDF. 

CNN and DNN are the most popular DL network for extracting features from TF-IDF vectors of bug reports. Ten studies used DNN to compute the similarity of TF-IDF vector with source code vectors. Four studies~\cite{yuan2020dependloc, yang2020applying, wang2020multimd-cnn, ali2024softwaremsp-coa-idn-lstm} used CNN for extracting features from TF-IDF vector of bug reports. Alsaedi et al.~\cite{alsaedi2024two} used cosine similarity. 




\textbf{BERT.} The third popular bug report representation is BERT, which is pre-trained on a large corpus of text data using unsupervised learning~\cite{devlin2018bert}. It learns to predict missing words in a sentence, considering the context of the surrounding words. Du et al.~\cite{du2023prehmcbl} proposed HMCBL approach which extracts bug report feature vectors from bug reports and then use multi-layer perception neural network as a projector to compress the vector. Ahmad et al.~\cite{ahmad2023attentivebuglocator} initialized each word using pre-training BERT to produce a dynamic context-dependent representation for each sentence based on the overall context. Then BiLSTM produces hidden for bug report embedding vectors. Ciborowska et al.~\cite{ciborowska2023too} used BERTOverflow~\cite{tabassum2020code} to embed bug reports, which is pre-trained on the StackOverflow corpus. Huang et al.~\cite{huang2022sbuglocater} used ALBERT (A Lite BERT)~\cite{lan2019albert} as encoder and the \textit{k}-max pooling layer to extract the feature information and obtain the final semantic matching score through the dense layer. FBL-BERT~\cite{ciborowska2022fastfbl-bert} approach used a massive corpora of relevant text to pre-train the BERT model and then fine-tune on bug reports and bug-inducing changesets. CodeBERT is a bimodal BERT pre-trained on both natural language and programming language~\cite{feng2020codebert}. Chakraborty et al.~\cite{chakraborty2024rlocator} utilized CodeBERT to encode bug reports and a DL model composed of CNN and LSTM to identify the most potential buggy files. Zhu et al.~\cite{zhu2021trobo} mapped bug reports into corresponding embedding sequences with CodeBERT and employ soft attention~\cite{bahdanau2014neural} to automatically highlight the key information. IBL~\cite{zhao2024fineibl} and bjEnet~\cite{han2024bjenet} approach used BERT to extract semantic information from bug reports and employ DNNs to assess the relevance with the corresponding code information. MACL-IRFL~\cite{zhou2024multimacl-irfl} approach leveraged CodeBERT to convert bug reports into vector representations.

\textbf{GloVe.} GloVe aims to generate dense vector representations (embeddings) for words by leveraging statistical information on the co-occurrence of words in a large corpus~\cite{pennington2014glove}. Unlike sparse one-hot encoding, GloVe embeddings capture semantic relationships between words in a continuous vector space. GloVe have proven effective in various NLP tasks, facilitating tasks such as text classification~\cite{moreo2021word, stein2019analysis, selva2021review}, sentiment analysis~\cite{yu2017refining}, and machine translation~\cite{baliyan2021multilingual, font2019equalizing}. Loyola et al.~\cite{loyola2018bug} used GloVe pre-trained word embeddings and character-level embeddings, splitting tokens into characters, to learn vector representations of bug reports, which were processed through an LSTM module. Zhu et al.~\cite{zhu2020cooba} exploited the pre-trained GloVe to map each word into a \textit{k}-dimensional embedding and then exploit a BiLSTM to encode the input sequence. Anh et al.~\cite{anh2021imbalanced} adopted a pre-trained GloVe model for bug reports and source code to address the lexical mismatch between them. Han et al.~\cite{han2023bjxnet} chose the GloVe as the pre-trained word vectors and then TextCNN network takes the text vector as input and outputs the report feature. Alsaedi et al.~\cite{alsaedi2024two} utilized GloVe to extract vector representations of bug reports and source code for similarity calculations.

\textbf{Others.} In addition to the commonly used word embedding approaches mentioned above, there are also others.
Zhu et al.~\cite{zhu2021deepdemob} used pre-trained ELMo~\cite{matthew1802deep} to generate dynamic context-dependent word representations in bug reports, followed by a BiLSTM to learn sequential features efficiently. Yang et al.~\cite{yang2021utilizing} clustered topics using the frequency of words in the bug report, and each created topic consisted of topic words. Xiao et al.~\cite{xiao2019improvingdeeploc} noted that summaries in bug reports are concise, while descriptions are more detailed. To differentiate between them, they used Word2Vec with the Skip-gram model for summaries and Sent2Vec~\cite{pagliardini2017unsupervised} for descriptions. These representations were then input into an enhanced CNN model to localize buggy files. Similarly, Liu et al.~\cite{liu2019convolutionalsls-cnn} used Word2Vec to convert summaries into word vectors and Doc2Vec~\cite{mikolov2013distributed} to create sentence vectors from descriptions, extracting features from bug reports using CNN. Chakraborty et al.~\cite{chakraborty2024blaze} adopted CodeSage~\cite{zhang2024code}, a GPT-based multi-modal embedding model, to align bug reports and source code files. Chandramohan et al.~\cite{chandramohan2024supporting} fine-tuned a UniXcoder~\cite{guo2022unixcoder} model for bug reports and source code.

Apart from the above word embeddings, character-level embeddings are also adopted, where each character in a word is transformed into a \textit{k}-dimensional character embeddings and then processed by CNN~\cite{xiao2017improvingdeeplocator} or LSTM~\cite{loyola2018bug}, or both~\cite{xiao2018improving}. Ali et al~\cite{ali2023automatedhgwsfocdnn, ali2024softwaremsp-coa-idn-lstm} employed three techniques such as Word2Vec, bags of n-grams model, and TF-IDF for acquiring the essential features from the ``source files and bug reports.'' Then it used PCA~\cite{iqbal2020determining} for reducing the feature dimension, before using CNN for feature extraction. Besides, there are also two studies~\cite{yang2021locatingmram, ma2023capturingsgattention} that only mention embedding the bug reports into a \textit{d}-dimensional vector without pointing out the specific techniques.

\subsubsection{Code Representation}
On the one hand, source code can be considered as plain text. In this way, source code can be embedded in the same way as bug reports, such as TF-IDF, Word2Vec, BERT, etc. On the other hand, as a kind of machine language, source code follows specific code syntax rules, and structure dictated by the programming language. Therefore, embedding source code may involve capturing the hierarchical structure, relationships between code elements, and even semantic meanings. Du et al.~\cite{du2023prehmcbl} concluded that there are majorly three categories of code representation ways within the literature: token-based ways, syntactic-based ways and semantic-based ways. \textbf{Token-based ways} represent text or code by breaking it down into individual tokens. Tokens can be words, subwords, characters, or any other unit that the approach uses to divide the input. Although the simplicity facilities learning of token-based representation, it ignores the structural nature of code and thus captures limited semantics. \textbf{Syntactic-based ways} focus on capturing the structural relationships and grammar of the input text or code. They often involve parsing the input based on the language's syntax rules to extract syntactic information. Since the tree structure typically has an unusually deep hierarchy, significant refinement efforts of the raw tree representation are often required to enable successful learning in practice. As a result, the learning performance is constrained. \textbf{Semantic-based ways} aim to capture the meaning or semantics of the input text or code. These approaches often leverage pre-trained language models or embeddings to understand the context and meaning of the words or code elements. 

As depicted in Figure~\ref{fig:sunburst}, 44 studies employ token-based representation for source code, 22 studies utilize syntactic-based code representation, 12 studies adopt semantic-based approaches, with two exceptions. 


\textbf{Token-based Code Representation.}
According to Figure~\ref{fig:techniques}, most studies primarily adopt the token-based representation for source code. Specifically, these representation include TF-IDF, One-hot encoding, Character embedding, BOW, Word2Vec, GloVe, and n-gram. Among them, the most widely adopted is Word2Vec, used in 19 studies, followed by TF-IDF, which is employed in 16 studies. Next are One-hot (five studies) and GloVe (five studies), n-gram (two studies), character embedding (one study), and BOW (one study).
Similar to bug reports, token-based code representation mainly use DNN, CNN and RNN to extract textual features from the code, except that Ma et al.~\cite{ma2022learningmla} embeded code tokens with Word2Vec and enhanced the token vectors with self-attention layer, followed by the max-pooling layer. Xu et al.~\cite{xu2023buglocfront} also obtained the dot-product attention score between program information and bug reports. Xiao et al.~\cite{xiao2018bugcnn_forest} converted the words in source code into Word2Vec word vectors and then use the ensemble of random forests with multi-grained scanning to extract code feature.

\textbf{Syntactic-based Code Representation.} Different from human language, code follows the syntax structure of machine language. When it comes to mining syntax feature from source code, the most straightforward approach is the AST structure, which is a hierarchical tree-like data structure that represents the syntactic structure of source code in a programming language. According to Figure~\ref{fig:sunburst}, AST is the most dominant syntax-based code representation, which has been used in 14 primary studies. Besides, CFG, class dependency graph (CDG) and Program Dependency Graph (PDG) have also been employed.

Recent studies have revealed that neural models based on ASTs more accurately represent source codes, and programming languages can benefit from syntax and structured representations. In addition to extracting structural information from ASTs using DNN~\cite{xiao2023bugradar}, CNN~\cite{xiao2017improvingdeeplocator, liang2019deepcast, zhu2020cooba}, and RNN~\cite{xiao2017improvingdeeplocator, xiao2018machinebugtranslator, ahmad2023attentivebuglocator, yan2023bug}, TransH~\cite{jiang2020deepdbl}, GCN~\cite{zhu2020cooba, zhu2021trobo, zhu2022enhancingbloco, zhu2022bl-gan}, ASTNN~\cite{cao2020bugpecker}, and GRU~\cite{zhao2024fineibl} have also been employed.

The second widely adopted syntactic-based code representation is CFG, which has been adopted in five primary studies. CFG is a graphical representation of the flow of control or the execution flow within a program. It is a directed graph that models the possible paths that a program can take during its execution. In a CFG, nodes represent basic blocks of code, and edges represent the flow of control between these basic blocks. CG-CNN~\cite{huo2020controlcg-cnn} leverages DeepWalk to learn the semantic representation by considering the neighboring statements, after processed by CNN. TROBO~\cite{zhu2021trobo} combines a CNN layer and multi-layer GCN to process code file based on the CFG. Ma et al.~\cite{ma2022flowingcflow} designed a flow-based GRU for feature learning from the CFG, which transmits the semantics of statements along the execution path. Zhu et al.~\cite{zhu2022enhancingbloco} proposed Code-NoN, a hierarchical network that integrates CFG and AST properties to represent source code files. They then employed DGP (Dense Graph Propagation), a variant of Graph Convolutional Networks (GCN)~\cite{kipf2016semi, zhang2021kcrec}, to effectively capture directional information between nodes by treating information propagation from parent and child classes differently. Ma et al.~\cite{ma2022learningmla} generated multi-level abstraction of CFG and then design a GNN model for feature learning from the multi-level abstraction of the CFG, where the block feature is alternately propagated within and between abstraction levels. Yuan et al.~\cite{yuan2020dependloc} built a CDG for all source files in a project to illustrate class reference relationships. They then applied a customized Ant Colony algorithm on the CDG to simulate possible reference paths and quantify the intrinsic dependency relationships. Yong et al.~\cite{yong2023decomposings-buglocator} constructed PDG of each source code file by analyzing the control flow and data flow and then use a CNN to extract semantic representation of each statement, followed by a BiLSTM for further representation.

\textbf{Semantic-based Code Representation.} Semantic-based code representation aim to represent the inherent meanings and relationships between different code elements rather than just focusing on syntax. It is crucial for tasks that require a deeper understanding of the functionality and intent of the code. In the 61 primary studies, BERT and ELMo are two typical semantic-based code representations. There are different variant of BERT, including basic BERT~\cite{ciborowska2023too, han2024bjenet}, CodeBERT~\cite{zhu2021trobo, ma2023capturingsgattention, chakraborty2024rlocator, zhou2024multimacl-irfl}, ALBERT~\cite{huang2022sbuglocater} and ColBERT~\cite{ciborowska2022fastfbl-bert}. Besides, Du et al.~\cite{du2023prehmcbl} introduced the Semantic Flow Graph (SFG) for compactly representing deep code semantics. Building on SFG, they proposed SemanticCodeBERT, a BERT-like model for learning code representations that consider deep code structure. Experimental results show that SemanticCodeBERT outperforms FBL-BERT~\cite{ciborowska2022fastfbl-bert}, GraphCodeBERT~\cite{guo2020graphcodebert}, and UniXcoder~\cite{guo2022unixcoder}. The BERT-based code representations are then input to DNN, CNN, or RNN networks to extract semantic code information. Liang et al.~\cite{liang2022modelingflim} constructed a training dataset to fine-tune the CodeBERT model and then calculate the distance between bug reports and source code with Cosine Similarity or Manhattan Distance. Mohsen et al.~\cite{mohsen2023enhancingpbl} fine-tuned BERT model with bug reports and source code and then calculate cosine similarity between the semantic representation of bug reports and source code. Zhu et al.~\cite{zhu2021deepdemob} embeded source code with ELMo-based embedding layer and then leverage MDCL encoder to extract the multi-grained features of the source file, which contains multiple DCNNs~\cite{kalchbrenner2014convolutional} and a BiLSTM layer. Chandramohan et al.~\cite{chandramohan2024supporting} fine-tuned a UniXcoder~\cite{guo2022unixcoder} model for their source code. Blaze~\cite{chakraborty2024blaze} relied on fine-tuned CodeSage for code representation. Alsaedi et al.~\cite{alsaedi2024two} used SentenceTransformers~\cite{sentencetransformer} for method names embedding. bjEnet~\cite{han2024bjenet} approach firstly leveraged CodeT5+~\cite{wang2023codet5+} for code summarization, and then fine-tune the BERT model based on the code summarization.

\textbf{Others.} Except for the above most common code representations, Loyola et al.~\cite{loyola2018bug} structured code changes into a code change genealogy~\cite{10.1145/1454247.1454257}, which considers additions or modifications of method calls. They then compute random walks over the directed graph starting from each node. Luo et al.~\cite{luo2022improvingcoloc} proposed CoLoc, initially pre-trained on a large bug report corpus in an unsupervised manner and then further refined using a contrastive learning objective to capture semantic differences between bug reports and buggy files. Yang et al.~\cite{yang2021locatingmram} converted the source code into \textit{d}-dimensional vector without pointing out the specific embedding approach. Apart from extracting code semantic information with GloVe, Han et al.~\cite{han2023bjxnet} used code property graph (CPG)~\cite{yamaguchi2014modeling} and encoded both the node and edge information with Gated Graph Convolution Networks (GatedGCN)~\cite{li2015gated}. CPG consists of AST, CFG, and DDG, and can well reflect the information on the source code structure, the statement execution process, the control dependence, and the data dependence~\cite{han2023bjxnet}.

\subsection{Feature Extraction Models}
From Figure~\ref{fig:techniques}, we observe that currently CNN, DNN, and RNN are the most commonly used deep models for extracting features from textual representations of bug reports and source code. Additionally, graph-based structure, such as AST, is predominantly processed using GCN to extract features from the graph structure.





\subsection{Other Features} \label{sec:otherfeather}
\nff{Despite the advancements in DL technology that have enhanced the models' understanding of bug reports and source code, treating bug reports solely as natural language texts for NLP processing fails to capture their unique characteristics. Unlike formal texts, such as press releases, which adhere to structured grammar and style guidelines, bug reports are shaped by their practical purpose and context. They tend to be more informal, often incorporating technical jargon, abbreviations, and incomplete sentences. Moreover, bug reports frequently include domain-specific terminology and contextual clues (such as class names) that demand a deep understanding of the software and its environment. They often combine structured data, such as error codes and stack traces, with unstructured narrative descriptions, which can aid in bug localization. To fully leverage these features, existing research has also explored extracting additional features from bug reports,}
such as stack trace, bug fixing history, collaborative filtering score, class name, code comments. We listed such features and the corresponding papers in Table~\ref{tab:otherfeature}.


\textbf{Stack Trace.} Stack trace in bug reports can be a valuable source of localization hints, which is information regarding the suspicious file and the specific suspicious buggy line in the file, provided by the compiler when a program error occurs. 
Schroter et al.~\cite{schroter2010stack} discovered that approximately 60\% of the bugs within bug reports containing stack trace information can be resolved by modifying the functions identified in the stack trace. Additionally, they found that around 40\% of the bug-related source files can be located in the first frame of the stack trace, and an impressive 90\% of the bug-related source files can be identified within the first 10 frames. These findings underscore the significance of stack traces in bug localization.
In this bug report example,
\footnote{\url{https://bz.apache.org/bugzilla/show_bug.cgi?id=531}} it also includes the error stack trace, which would facilitate the fixing of the reported bug.
Yang et al.~\cite{yang2020applying} applied BRTracer~\cite{wong2014boosting} to analyze the stack trace and then combine the extracted information from bug reports, source code and stack traces into an autoencoder~\cite{lam2017bugdnnloc}. Xiao et al.~\cite{xiao2023bugradar} extracted stack traces with regular expression, and then construct a knowledge graph called TriGraph. Xu et al.~\cite{xu2023buglocfront} used regular expressions to extract effective stack frame information and then rank each file in the stack trace to score the source file. Ahmad et al.~\cite{ahmad2023attentivebuglocator} leveraged BRTracer~\cite{wong2014boosting} to calculate stack trace score and then combine with other features to improve bug localization performance. Yan et al.~\cite{yan2023bug} and Alsaedi et al.~\cite{alsaedi2024two} extracted stack trace with regular expressions and then calculate the score as Schroter et al.~\cite{schroter2010stack}. Ciborowska and Damevski~\cite{ciborowska2023too} used infozilla~\cite{bettenburg2008extracting} to distinguish stack traces and code snippets from natural language in bug reports.

\textbf{Bug Fixing History.} Several researchers have shown that recently fixed files in the recent past are most likes to be fixed in the near future~\cite{kim2007predicting, lam2015combining, ye2014learning}. To this end, bug fixing history has been used to improve performance of IRBL approaches~\cite{lam2015combining, lam2017bugdnnloc, xiao2019improvingdeeploc, sangle2020drast, wang2020multimd-cnn, cao2020bugpecker, anh2021imbalanced, yang2021locatingmram, qi2021dreamloc,shi2022semirfl, xiao2023bugradar, xu2023buglocfront, Al-Aidaroosimpact, alsaedi2024two, zhang2023enhancing}, which includes bug fixing recency and frequency. Ye et al.~\cite{ye2014learning} calculated the bug fixing recency as the inverse of the time interval between the creation of the bug report and the last time the source file was fixed, as shown in Equotion~\eqref{eq:combined}. In the formula, \textit{b.month} represents the month in which bug report \textit{b} was reported and \textit{f.month} denotes the month in which source file \textit{f} was last fixed before bug report \textit{b} was created.
The bug-fixing frequency is expressed by the number of times source file \textit{f} was fixed before bug report \textit{b} was submitted~\cite{ye2014learning}. Google's developers proposed BugCache algorithm, which maintains the commit history of each file in the system by the day of a new bug fix~\cite{lewis2013does}. They calculated the cache score based on Equotion~\eqref{eq:combined}, where $f$ is one of the buggy files in commit $c \in C$, $t_i$ is the elapsed time in days since the file's creation. This formula tends to select the most recently and frequently changed files.




\begin{align} \label{eq:combined}
    \text{Recency}(i) &= \frac{1}{b.\text{month} - f.\text{month} + 1}
    \hspace{2cm}
    \text{BugCache}(f) &= \sum_{c \in C \wedge f \in c} \frac{1}{1 + e^{-12t_i + 12}}
\end{align}

\textbf{Collaborative Filtering Score.} Researchers assume that similar bug reports are more likely to be corresponded to the same files~\cite{zhou2012should}. To this end, they proposed collaborative filtering score to measure the similarity of a bug report and previously fixed bug reports by the same file~\cite{ye2014learning}. Many DL-based bug localization approaches also take this score as an additional feature to enhance the model performance~\cite{lam2017bugdnnloc, xiao2019improvingdeeploc, sangle2020drast, cao2020bugpecker, yang2021locatingmram, qi2021dreamloc, shi2022semirfl, xu2023buglocfront, alsaedi2024two, zhang2023enhancing}. It is calculated by Formula~\ref{eq:cfs}, where \textit{br(b, f)} indicates the set of fixed bug reports related with the source file \textit{f} before \textit{b} was reported, and the collaborative  filtering score \textit{cfScore(b, f)} is the VSM similarity between \textit{b} and the summaries of all the bug reports in \textit{br(b, f)}.

\begin{equation}
    cfScore(b, f) = simi(n,br(b, f))
    \label{eq:cfs}
\end{equation}

\textbf{Class Name.} At times, bug reports may include error messages that reference specific elements such as class names, functions, APIs, etc.~\cite{sangle2020drast}. These terms serve as vital indicators for bug localization, demanding careful consideration. Lam et al.~\cite{lam2017bugdnnloc} extracted name of identifiers, name of API classes and interfaces comments and string literals in the source code files and calculate the similarity with bug reports. Xiao et al.~\cite{xiao2019improvingdeeploc} calculated class name similarity as Ye et al.~\cite{ye2014learning} did. DRAST~\cite{sangle2020drast} approach extracted the function names, identifiers, macros, unions, typedef, struct, cpp from the C file and function names, class names, identifiers from Java files. The SemirFL~\cite{shi2022semirfl} model combined both CNN and rVSM, which is fed with four metadata features (bug-fixing recency, bug-fixing frequency, collaborative filtering score, and class name similarity.

\textbf{Code Comments.} Lam et al.~\cite{lam2015combining} used an DNN to learn the relations between bug reports’ and code comments, and another DNN to learn between bug reports’ and code tokens (i.e., identifiers, APIs). Huang et al.~\cite{huang2022sbuglocater} took code comments as natural language data for source code files.

\textbf{Others.} There are also other features employed to improve the performance of IRBL approaches. BugTranslator~\cite{xiao2018machinebugtranslator} also used API documents, project-specific documents, and older bug reports with corresponding buggy files for training the model. Liu et al.~\cite{liu2019convolutionalsls-cnn} fused surface lexical correlation matching between bug reports and source code with semantic correlation matching. BL-GAN~\cite{zhu2022bl-gan} approach trained a generator to generate file paths for each given bug report. DreamLoc~\cite{qi2021dreamloc} and BRS\_BL~\cite{zhang2023enhancing} approach also integrated cyclomatic complexity~\cite{qi2021dreamloc} as one feature, which is a measure of alternative execution paths in code segments caused by control flow statements and is used to measure the program quality. Chandramohan et al.~\cite{chandramohan2024supporting} also involved commit message-level analysis to enhance their model's performance.

\nff{In summary, due to the informality of bug reports, the variability in how users describe bugs can lead to ambiguity and inconsistency, presenting a significant challenge for models designed to process them. For instance, users might describe the same issue in different ways, using subjective terms that may not always align with the technical language used in the source code. Consequently, models aimed at understanding and linking bug reports to source code need to handle this informality, technical specificity, and occasional incompleteness or vagueness. Overcoming these challenges is crucial for accurate bug localization, as these tasks require deep contextual understanding and flexible language processing capabilities.}
\section{RQ2: How are the DL-based approaches evaluated?}\label{sec:evaluation}
As shown in Figure~\ref{fig:framework}, in the evaluation process, there mainly involve three parts: dataset, evaluation metrics, and validation approaches. Moreover, approaches can also be categorized based on varying levels of granularity.

\subsection{Datasets}

\begin{table}[htbp]
\caption{Projects used in no less than ten primary studies.}
\label{tab:projects}
\centering
\resizebox{\textwidth}{!}{%
\begin{tabular}{lcccc|lcccc}
\toprule
Project   & \# of studies & \# of bug reports & Granularity & \# of code & Project   & \# of studies & \# of bug reports & Granularity & \# of code \\ \midrule
\multirow{3}{*}{JDT}     & \multirow{3}{*}{49}   & \multirow{3}{*}{6274} & File & 8184 & \multirow{3}{*}{AspectJ} & \multirow{3}{*}{44}   & \multirow{3}{*}{593}  & File & 4439 \\
                        &&& Method & 49152 & &&& Method & 32816  \\
                        &&& Changeset & 13860 & &&& Changeset & 2939  \\ \cline{1-5} \cline{6-10}
\multirow{3}{*}{SWT}     & \multirow{3}{*}{44}   & \multirow{3}{*}{4151} & File & 2056 & \multirow{3}{*}{Tomcat}  & \multirow{3}{*}{37}   & \multirow{3}{*}{1056} & File & 1552 \\
                        &&& Method & 13456 & &&& Method & 36569  \\
                        &&& Changeset & 10206 & &&& Changeset & 10034  \\ \cline{1-5} \cline{6-10}
Eclipse   & 37            & 6495              & File & 3454 & \multirow{2}{*}{Birt}      & \multirow{2}{*}{14}           & \multirow{2}{*}{4178}              & File & 6841 \\\cline{1-5}
 \multirow{2}{*}{PDE} &     \multirow{2}{*}{11}            &   \multirow{2}{*}{4034}              &  File   &  2970    &               &        &        & Method            & 100625  \\
      &           &              & Changeset & 834 &               &                &                   &          \\
\bottomrule
\end{tabular}
}
\end{table}

\nff{\textbf{Distribution.}} In order to understand which dataset has been used for evaluation, we extracted the datasets used in the primary studies, which usually consist of a few projects. Table~\ref{tab:projects} presents information of projects that have been used in no less than ten primary studies. All these projects are developed using Java, and they are open source. In addition to the frequency of project data usage, we also recorded the size of each project, which includes the number of bug reports, the granularity of the dataset for bug localization, and the number of code entities corresponding to different granularities. Since different studies may choose different time duration, in Table~\ref{tab:projects} we only present the mode of dataset sizes across all the studies. JDT, AspectJ, and SWT are the top three projects that have been used in over 40 primary studies. Meanwhile, they are also widely used in the first two generations of IRBL studies~\cite{zhou2012should, wang2014version, wang2016amalgam+, wong2014boosting}. According to the table, JDT has 8,184 files with 6,274 bug reports. The substantial number of bug reports and code contributes to its popularity.


In addition to the typical projects mentioned above, there are also studies that contribute unique datasets. Table~\ref{tab:dataset} presents the open sourced dataset published in the 61 primary studies. Most datasets are in Java, but the authors of DRAST\cite{sangle2020drast} provided a dataset in C, and BLAZE~\cite{chakraborty2024blaze} contributed a dataset including C++, Go, Java, JavaScript and Python.

Xiao et al.~\cite{xiao2018bugcnn_forest} used AspectsJ, Eclipse UI, JDT, SWT and Tomcat to evaluate the performance of CNN-Forest. The before-fixed version~\cite{ye2014learning} of the source code in each project are available in the dataset, which includes 10,754 bug reports in total. The BugC~\cite{sangle2020drast} dataset consists of 2,462 bug reports from 21 open-source C projects. Sangle et al.~\cite{sangle2020drast} evaluated DRAST on seven projects from BugC dataset, as well as Tomcat and AspectsJ projects from the benchmark dataset. Zhu et al.~\cite{zhu2020cooba} proposed a cross-project bug localization approach COOBA. It is evaluated on four projects of the dataset by Ye et al.~\cite{ye2014learning}: AspectJ, SWT, JDT Eclipse Platform UI. DreamLoc~\cite{qi2021dreamloc} is evaluated on Birt, Eclipse, JDT, SWT, Tomcat from dataset by Ye et al.~\cite{ye2014learning}. BLESER~\cite{zou2021bleser} was evaluated on the Defects4J dataset~\cite{just2014defects4j}, which is a benchmark for bug localization and program repair~\cite{b2016learning, li2019deepfl, martinez2016astor}. FLIM~\cite{liang2022modelingflim} used a fine-tuned language model to extract code semantics at the function level, and was evaluated on Eclipse UI, JDT, BIRT, SWT, Tomcat, AspectJ from benchmark dataset by Ye et al.~\cite{ye2015mapping}. HGW-SFO was evaluated on AspectJ, JDK, SWT, Tomcat and ZXing, which are collected by the authors~\cite{ali2023automatedhgwsfocdnn}. Zhang et al.~\cite{zhang2023enhancing} created a new dataset based on Ye et al.~\cite{ye2014learning}, which includes Tomcat, SWT, JDT, Eclipse. Chakraborty et al.~\cite{chakraborty2024blaze} cured the BeetleBox dataset, which comprises 26,321 bugs sourced from 29 projects across C++, Go, Java, JavaScript, and Python.

We note that mainstream research continues to concentrate on the Java language, while only a limited number explore other languages~\cite{sangle2020drast, chakraborty2024blaze}. In the future, it is crucial to shift more attention towards alternative languages like Python and C++. Furthermore, more explanation on cross-language research avenues holds promising potential.


\nff{\textbf{Construction.}} The predominant evaluation dataset, as contributed by Ye et al.~\cite{ye2014learning} \footnote{http://dx.doi.org/10.6084/m9.figshare.951967}, is widely utilized and encompasses six open-source projects: AspectJ, Birt, Eclipse Platform UI, JDT, SWT, and Tomcat. In this dataset, bug reports are linked to their fixed files based on heuristics proposed by Dallmeier and Zimmermann~\cite{dallmeier2007extraction}. To prevent the incorporation of future bug-fixing information in the dataset, a before-fix version of the project corresponding to each bug report was checked out.
It is noteworthy that the dataset for changeset-level bug localization differs from that of method-level or file-level. Four studies at the changeset-level~\cite{loyola2018bug, ciborowska2022fastfbl-bert, ciborowska2023too, du2023prehmcbl} utilized the dataset contributed by Wen et al.~\cite{wen2016locus}. In this dataset, each bug report is linked to the respective bug-inducing changesets using the SZZ algorithm~\cite{sliwerski2005changes}, and has been validated manually by the authors. Zhao et al.~\cite{zhao2024fineibl} constructed their own dataset using the SZZ algorithm to evaluate IBL.

\nff{\textbf{Data Quality and Bias.} In DL-based approaches, data quality plays a critical role. Kochhar et al.~\cite{kochhar2014potential} highlights that the quality of the bug localization dataset can impact the validity of the results reported in the studies. Potential biases in bug localization include: 1) \textit{Misclassified reports}; 2) \textit{Already localized reports}; 3) \textit{Incorrect ground truth}. Widyasari et al.~\cite{widyasari2022influence} revealed that bias 1 and bias 3 have no significant impact on bug localization results, while bias 2 has a statistically significant effect.
For bias 1 (misclassified reports), bug reports in the datasets are typically collected from issue tracking systems (IST) such as JIRA and Bugzilla. Research shows that issue classification within these systems is sometimes erroneous~\cite{herzig2013s, kochhar2014s}. In our study, most of the studied research employed datasets from Ye et al.~\cite{ye2014learning}, where the verification of bug reports classification was not mentioned, which may introduce bias to the reported results. In addition, the changeset-level dataset~\cite{wen2016locus}, and the Defects4J~\cite{just2014defects4j} dataset has been manually validated, reducing the impact of bias 1 to some extent.
For bias 2 (already localized reports), studies~\cite{bettenburg2008makes, bettenburg2007quality} found that developers emphasize stack traces, program entity names, and test cases as useful information, which may contain clues regarding bug locations. Their investigation reveals that around 50\% (10\% includes stack traces, 30\% contains test cases, and 45\% mentions program entities) of bug reports contain such information. 
Yang et al.~\cite{yang2021locatingmram} further demonstrated that the presence of such information in bug reports significantly boosts model performance. Notably, in our study on IRBL leveraging DL technologies, six~\cite{ma2022learningmla, ma2022flowingcflow, huo2020controlcg-cnn, cao2020bugpecker, ma2023capturingsgattention, huo2019deeptranp} studies explicitly mentioned excluding fully localized reports (bug reports already describe which files and methods are faulty) to mitigate bias 2. Furthermore, FBL-BERT~\cite{ciborowska2022fastfbl-bert} excluded log messages from bug reports to avoid boost the performance of models. While some studies~\cite{liang2019deepcast, han2023bjxnet, huang2022sbuglocater,shi2022semirfl, chen2022cgmbl, yang2021utilizing, anh2021imbalanced, zhu2021deepdemob, qi2021dreamloc, wang2020multimd-cnn, jiang2020deepdbl, liu2019convolutionalsls-cnn, yan2023bug, xu2023buglocfront, ali2023automatedhgwsfocdnn, alsaedi2024two, han2024bjenet, zhao2024fineibl, zhou2024multimacl-irfl}, clearly specified that their data comes from bug reports within a specific time frame, they do not provide details about additional filtering processes. Other studies among the 61 studies did not discuss any data filtering or validation.
For bias 3 (incorrect ground truth), previous research acknowledges that code commits often combine different objectives, such as bug-fixing, refactoring, and feature-implementation changes, leading to code tangling~\cite{herzig2013impact}. However, current automatically constructed datasets have not addressed the noisy data caused by code tangling. Additionally, Niu et al.~\cite{niu2023rat} found that file renaming caused file paths from previous versions to no longer exist in the new version, which can impact the accuracy of similarity-based retrieval approaches. Duplicate bug reports are also quite common~\cite{wang2008approach}. Although Lee et al.~\cite{lee2018bench4bl} argued that duplicated reports are often not attached to fixed files, if such duplicates appear separately in the training and testing data, it can lead to data leakage. Despite this risk, there has been limited discussion on handling this issue in DL-based bug localization approaches.}



\subsection{Granularity}
\nff{From the perspective of localization granularity, IRBL approaches can be categorized into file-level, class-level, method-level, statement-level, and change-set-level approaches. As their names suggest, file-level, class-level, method-level, and statement-level localization aim to identify the buggy file, class, method, or statement, respectively. Change-set-level localization, on the other hand, focuses on the set of changes made by developers in a specific commit. While statement-level localization provides a micro-level perspective by honing in on individual buggy statements, change-set-level localization offers a macro-level perspective by examining code modifications as a whole. These varying levels of granularity enable diverse strategies for bug localization, catering to both detailed analysis and broader contextual understanding.}

Out of all the studies, 52 specifically concentrate on file-level bug localization, four~\cite{cao2020bugpecker, yang2021locatingmram, zou2021bleser, alsaedi2024two} delve into method-level bug localization, while the remaining five~\cite{ciborowska2022fastfbl-bert, ciborowska2023too, loyola2018bug, du2023prehmcbl, zhao2024fineibl} explore changeset-level bug localization. Ciborowska et al.~\cite{ciborowska2022fastfbl-bert, ciborowska2023too} investigated both file-level, changeset-level and hunk-level bug localization. Within the scope of this study, there is no line-level or statement-level approaches identified yet, while file-level bug localization has been widely researched. 

\nff{As illustrated in Figure~\ref{fig:framework}, IRBL models that leverage DL technologies essentially follow the same workflow across different granularities—specifically, file-level, method-level, and changeset-level (with no existing research at the statement-level). These models utilize bug reports and source code as input, employing various embedding techniques to extract semantic features from both text and code for matching purposes. The primary differences among the granularities can be categorized into three aspects:}

\noindent \textbf{(1) Construction of Ground Truth: }The ground truth for file-level and method-level bug localization is established by linking bug reports to the corresponding bug-fixing commits, which are then associated with the files or methods involved in those commits. In contrast, changeset-level datasets are more closely aligned with just-in-time defect prediction~\cite{ni2022best}, where bug reports are linked to bug-inducing commits (typically preceding the bug-fixing commits). Using \textit{git diff}, the code changes from these commits are then associated with the bug reports to establish the ground truth.

\noindent \textbf{(2) Search Space:} Based on the construction of the ground truth, the search space for file-level and method-level bug localization consists of the files or methods in a specific version of the code repository. Conversely, the search space for changeset-level bug localization encompasses the entire commit history, treating the code changes in each commit as a distinct document.

\noindent\textbf{(3) Code Processing:} For embedding source code, file-level and method-level approaches typically embed each file or method separately. In the case of changeset-level code embedding, Ciborowska et al.~\cite{ciborowska2022fastfbl-bert} proposed three encoding strategies for changeset-level code embedding: treating the changeset as a single document, splitting it into grouped lines by modification type, and splitting it into ordered lines with special tokens for each modification type. Their results showed that the third strategy, which preserves line order, outperforms the others.

File-level bug localization benefits from the richness of code text for precise localization, leading to higher accuracy in current research. However, it still requires developers' significant effort to locate the relevant code within large files~\cite{murali2021industry, wen2016locus, zou2018practitioners}. In contrast, finer-grained approaches, such as method-level~\cite{cao2020bugpecker, yang2021locatingmram} and changeset-level~\cite{loyola2018bug, du2023prehmcbl, ciborowska2022fastfbl-bert, ciborowska2023too} localization, have emerged to alleviate some of this burden. Method-level localization, while reducing the effort required to locate buggy code~\cite{polisetty2019usefulness}, introduces complexities due to the intricate interactions between methods and the presence of multiple candidates within a single file, which can hinder accuracy. Changeset-level localization focuses on analyzing code changes within commits but demands a comprehensive understanding of the entire commit history, making it computationally intensive and complicating the isolation of relevant changes, particularly when multiple modifications are present.

\nff{In future research, the challenges of file-level localization lie in enhancing contextual analysis to better capture inter-file dependencies, as these relationships often influence bug occurrences; developing DL models that effectively represent these interactions to significantly improve localization accuracy. Furthermore, noise reduction techniques are needed to filter out irrelevant changes within large files, potentially utilizing advanced neural networks to prioritize significant modifications. Scalability poses another challenge, necessitating the development of efficient DL algorithms capable of managing vast codebases without compromising performance. In the realm of method-level localization, improving dependency tracking is crucial for analyzing the intricate interactions among methods, as this understanding is vital for accurately pinpointing bugs. Researchers should also explore granularity optimization, aiming for hybrid models that balance detailed method-level analysis with higher-level insights, alongside the creation of automated context extraction tools that summarize relevant information around methods to enhance alignment with bug reports. For changeset-level localization, advancing techniques for commit history analysis is critical, enabling the efficient isolation of relevant changes from extensive commit histories through DL-driven summarization methods. Additionally, enhancing the quality of commit messages remains a significant challenge; research could focus on employing natural language processing to generate clearer documentation, which is vital for tracing bug origins. Finally, improving contextual understanding of code changes will require developing models that analyze not only the code modifications but also the rationale behind them, facilitating better correlation with reported bugs. Across all granularities, the integration of DL techniques is paramount for enhancing pattern recognition capabilities, while the development of user-centric tools can simplify the localization process for developers. Emphasizing interdisciplinary approaches that merge software engineering insights with advanced DL methodologies could drive innovation in IRBL techniques, ultimately leading to more effective debugging tools and enhanced software quality.}



\subsection{Evaluation Metrics}
Evaluation metrics are necessary for evaluating the performance of the proposed approach. Different evaluation metrics have been adopted for assessing the effectiveness of IRBL techniques. Figure~\ref{fig:evaluation} shows the metrics that have been adopted. As most of the IRBL techniques generate ranking lists, which is basically ranking list of code entities (e.g., files, classes, methods, changesets, or statements), studies often use MAP, MRR and Top \textit{k} as the evaluation metrics. In all 61 primary studies, almost all studies used MAP (59 studies) for evaluation, closely followed by MRR (54 studies) and Top \textit{k} (54 studies). Then, a few studies also adopt Area Under Curve (AUC)~\cite{huo2016learningnp-cnn, liu2019convolutionalsls-cnn, polisetty2019usefulness, ma2022flowingcflow}, precision~\cite{xiao2017improvingdeeplocator, mohsen2023enhancingpbl}, recall~\cite{xiao2017improvingdeeplocator}, F-measure~\cite{xiao2017improvingdeeplocator}, and accuracy~\cite{ali2024softwaremsp-coa-idn-lstm} for evaluation, though these metrics are primarily associated with classification tasks. The definition of MAP, MRR and Top \textit{k} are as follows:


\textbf{Mean Average Precision (MAP)}~\cite{manning2008introduction} is a standard metric widely used in information retrieval to evaluate ranking approaches. It considers all the ranks of all buggy files into consideration. It is calculated as the mean of the Average Precision over all queries. Average Precision of a given bug report aggregates precision of positively recommended files as:

\begin{equation}
AP = \sum_{i=1}^{N}\frac{P(i) * pos(i)}{\# \, of \, positive \, instances}, \quad
P(i) = \frac{\# \, of \, buggy \, files \, in \, top \, i}{i}
\end{equation}

where \textit{i} is a rank of the ranked files, \textit{N} is the number of ranked files and \textit{pos}(\textit{i}) $\in$ \{0,1\} indicates whether the \textit{i}th  file is a buggy file or not. \textit{P}(\textit{i}) is the precision at a given top \textit{i} files.


\textbf{Mean Reciprocal Rank (MRR})~\cite{voorhees1999trec} computes the average of the reciprocal of the positions of the first correctly located buggy files in the ranked files, following this equation:

\begin{equation}
MRR = \frac{1}{Q}\sum_{i=1}^{|Q|}\frac{1}{rank_i} 
\label{eq:mrr}
\end{equation}

\textbf{Top \textit{k}}~\cite{schutze2008introduction} measures the percentage of bug reports in which at least one of the buggy files is a top \textit{k} ranked file. The most common values for \textit{k} are 1, 3, 5, 10, 20, etc. The higher the Top \textit{k} value, the better the performance of the model/approach.

\nff{MAP, MRR, and Top-\textit{k} accuracy are essential metrics for evaluating IRBL tasks, each with a distinct focus. MAP assesses the precision across all ranks of buggy files by averaging the precision at each position, providing a comprehensive measure of retrieval quality. In contrast, MRR zeroes in on the rank of the first relevant buggy file, making it particularly valuable for scenarios where quickly finding any relevant item is crucial. Top-\textit{k} accuracy is more practical, measuring how often the correct bug appears within the top \textit{k} retrieved results, thus highlighting the system's effectiveness in presenting relevant results at the highest ranks. While MAP offers a broad perspective on precision across multiple buggy files, allowing us to assess the overall quality of the retrieval process, MRR emphasizes the importance of the first relevant hit, and Top-\textit{k} is highly practical for real-world applications, where developers typically engage only with the top few candidates for bug fixes without having to sift through extensive lists of results~\cite{kochhar2016practitioners}.}


\subsection{Validation Approaches}
Validation approaches are frequently employed in software bug localization to guarantee the precision and reliability of the techniques. Researchers use different data partition for training and testing. The validation process of bug localization contains data \textbf{partitioning}, \textbf{sampling} and \textbf{validation}. Since DL is essentially based on training and testing, so it is important to construct the training data and testing data, that is data partitioning. Moreover, in bug localization, since one bug report is typically relevant to a small number of code entities while irrelevant to much larger number of entities. The data is significantly imbalanced. Some studies propose to sample the data~\cite{sangle2020drast, huo2020controlcg-cnn, zou2021bleser, ciborowska2022fastfbl-bert}. The validation strategy of each approach has been listed in Table~\ref{tab:overview}.
Apart from several studies~\cite{xiao2018improving, luo2022improvingcoloc, ali2023automatedhgwsfocdnn, Al-Aidaroosimpact, jiang2020deepdbl} that did not elaborate on specific validation details, how the studies validate their approaches are as follows:

\textbf{Partitioning and Validation.} For WPBL, the validation approach can be classified according to that if all the bug reports are sorted chronologically: (1) Anachronistically Validation (Cross-Validation), the most common validation approach is \textit{k}-fold cross-validation, typically with 10 folds. Specifically, for each project in the dataset, the bug reports are randomly divided into 10 equal size folds (as shown in Figure~\ref{fig:validation} (a)). Among the 10 folds, the bug reports of a single fold are retained as the unfixed bug reports for testing models, and the remaining nine folds are used as fixed bug reports for training models. This process is repeated 10 times to ensure each fold has been evaluated. Sometimes, researchers may also leave one fold as validation set~\cite{yong2023decomposings-buglocator}. Yang et al.~\cite{yang2020applying} divided bug reports in each project into 10 folds and chose seven folds as training set and the rest three folds as test set. 
Pendlebury et al.~\cite{pendlebury2019tesseract} criticized that it introduces bias if a model is trained from future data and tested on past data. To resolve this problem, some studies started to sort the bug reports chronologically.
(2) Chronologically Validation (Consecutive Validation), firstly all the bug reports in one project are sorted by time of creation. Then the studies either divide the dataset into two parts (the ratio can be 8:2, 7:3, or 6:4, etc), the newer bug reports are evaluated with model trained on older bug reports. Or the studies sort the bug reports chronologically by their report timestamps, and divide the bug reports into \textit{k} (typically 10) folds with equal sizes, in which $fold_1$ is the oldest and $fold_{k}$ is the newest. The model is trained on $fold_i$ and tested on $fold_{i+1}$. The final results are obtained by taking the averages of all \textit{k}-1 folds. This is called \textit{k}-fold consecutive validation (as shown in Figure~\ref{fig:validation} (b)). 

\nff{As shown in Figure~\ref{fig:validation}, \textit{k}-fold cross-validation shuffles the data randomly between folds to ensure that each fold serves equally as both a training and testing set through \textit{k} iterations. This approach helps mitigate bias in model evaluation by exposing the model to all data points as part of both the training and testing sets. Conversely, \textit{k}-fold consecutive validation utilizes historical data for training and predicts using subsequent fold, iterating \textit{k}-1 times. This strategy is particularly suitable when the order of data is significant, ensuring that future data points are not used to predict past outcomes. Given that the submission and resolution of bug reports occur in chronological order, we believe that consecutive validation provides a more accurate evaluation method.}

For CPBL, four studies~\cite{huo2019deeptranp, zhu2021trobo, zhu2020cooba, zhu2022bl-gan} trained their model on one source project with 20\% dataset from the target project, subsequently identifying buggy files for the remaining 80\% of bug reports from the target project. To mitigate the impact of randomness, this process was repeated ten times to reduce the influence of randomness~\cite{huo2019deeptranp, zhu2021trobo, zhu2020cooba}.
Two studies used one project as training data and the data from another project as the testing data~\cite{yang2021locatingmram, loyola2018bug}.
Liang et al.~\cite{liang2022modelingflim} compared all three strategies: 1) using the early 20\% data of the target project to fine-tune the model; 2) using the data of the other five projects; and 3) mixing the above two sets of data. They found out that the third strategy, i.e., mixing the data from the other five projects and the early 20\% data from the target project as the training set, performs the best.

Recently, two studies~\cite{chakraborty2024blaze, chandramohan2024supporting} on CLBL came out. Blaze~\cite{chakraborty2024blaze} fine-tuned the CodeSage~\cite{zhang2024code} model on the BeetleBox training set and tested it on the BeetleBox test set, SWE-Bench~\cite{jimenez2023swe}, and the dataset from Ye et al.~\cite{ye2014learning}. The BeetleBox dataset includes C++, Go, Java, JavaScript, and Python languages. Chandramohan et al.~\cite{chandramohan2024supporting} fine-tuned UniXcoder on Java-only dataset and evaluated the performance on C/C++ and Golang projects. Although CLBL approaches are not as effective as non-cross-language approaches (as shown in Table~\ref{tab:performance}), these two studies still demonstrate that it is a promising direction, worthy of further research in the future.


\textbf{Before-fix Version.} As bug reports are submitted at various times and stages throughout the project, each report is associated with a specific project version. As projects undergo evolution over time, distinct versions may encompass varying source code spaces. Previous bug localization studies used just one code revision to match all the bug reports. However, software bugs exists in different versions of source code. Consequently, using only on revision of source code does not match the actual scenarios and may cause bias to the evaluation results. For example, the adopted revision of code may already have been fixed towards the specific bug report, and thus contains future bug-fixing information. Further, the buggy file might already have been deleted in the evaluated revision. To this end, Ye et al.~\cite{ye2014learning} pointed out that using \textit{before-fix} version of code file can avoid leaking future bug-fixing information for older bug reports. \nff{The ``Before-fix version'' column in Table~\ref{tab:overview} presents some studies adopted \textit{before-fix} version of source code (\checkmark), while some studies did not mention this evaluation setting in the paper (-). For studies involving changeset-level bug localization (O), the target for localization is already at the level of individual commits.
Additionally, many papers explicitly state that their datasets follow ``the oracle of bug-to-file mappings''~\cite{lam2015combining, lam2017bugdnnloc, zhu2020cooba, zhu2021deepdemob, ciborowska2022fastfbl-bert, liang2022modelingflim, ciborowska2023too, du2023prehmcbl}, but no further details were provided. LS-CNN~\cite{huo2017enhancingls-cnn} only mentioned using multiple versions, without offering additional information. MRAM~\cite{yang2021locatingmram} built code revision graphs to address single revision problem, illustrating the relationships between code entities across different revisions. Previous empirical study \cite{lee2018bench4bl} pointed out that using a single version (typically the latest) may seem simpler but can undermine accuracy due to file additions and deletions. Niu et al.~\cite{niu2023ablots, niu2024extensive} further demonstrated that selecting an improper revision can introduce significant bias in evaluation results.}
Associating bug reports with their corresponding code versions may require some effort, but it is a more accurate assessment approach that aligns better with the actual situation. Therefore, in the future, it is recommended that the researchers utilize proper versions for evaluation.

\textbf{Class Imbalance Handling.}
In the source code repository, there can be thousands of files/metho-ds/statements. However, one bug report is only relevant to several positive samples, while the number of negative samples is significantly huge. This leads to an extremely imbalanced dataset, where the model may struggle to learn patterns associated with positive classes. Nine studies~\cite{ciborowska2022fastfbl-bert, zhu2021deepdemob, yong2023decomposings-buglocator, qi2021dreamloc, yang2021locatingmram, xu2023buglocfront, ciborowska2023too, han2024bjenet, zhang2023enhancing} studies randomly chose a certain number of negative samples, the number includes 200, 300, and 800. Nine studies~\cite{lam2015combining, lam2017bugdnnloc, xiao2019improvingdeeploc, huang2022sbuglocater, yuan2020dependloc, xiao2023bugradar, chakraborty2024rlocator, yan2023bug, chandramohan2024supporting} studies chose negative samples by calculating the textual similarity between samples and bug reports. Among these nine studies, all chose the top \textit{k} similar files as negative samples, while only Huang et al.~\cite{huang2022sbuglocater} chose the 300 least similar files as negative samples. Besides, 
Sangle et al.~\cite{sangle2020drast} experimented with various sampling strategies to address class imbalance, including SMOTE~\cite{chawla2002smote}, ADASYN~\cite{he2008adasyn}, random over-sampling and under-sampling, Kmeans SMOTE~\cite{douzas2018improving}, and TOMEK links~\cite{kotsiantis2006handling}. They found that SMOTE provided the best results for their dataset and models, outperforming other methods and no over-sampling.
Zou et al.~\cite{zou2021bleser} tested the BLESER approach using two re-sampling strategies (random over-sampling and under-sampling) and two cost-sensitive strategies (weight-based binary-cross entropy and Focal Loss). They found that random over-sampling significantly outperformed the other methods on their dataset and model.
Chen et al.~\cite{chen2022cgmbl} selected files that were ever buggy files as the negative samples. Wang et al.~\cite{wang2020multimd-cnn} set the weight \textit{W} in the binary cross-entropy lost function to solve this problem. Liu et al.~\cite{liu2019convolutionalsls-cnn} and Anh et al.~\cite{anh2021imbalanced} used focal loss function to rectify samples of the minority class within iterative training batches to the proposed models. Du et al.~\cite{du2023prehmcbl} used a memory bank~\cite{wu2018unsupervised} to store rich changesets obtained from different batches for later contrast. There are still some studies did not employ resampling, or the use of resampling was not mentioned in the papers. Over all, re-sampling is currently the most commonly used and most effective strategy for handling class imbalance issue.

\section{RQ3: What are the challenges faced when applying DL in IRBL?}\label{sec:challenge}
This section begins with a comprehensive performance overview, with an in-depth analysis of several key studies. It then summarizes the issues mitigated by third-generation technology compared to the previous two generations, as well as the challenges that remain unresolved.

\subsection{Performance Overview and Key Studies}
\nff{As shown in Figure~\ref{fig:evaluation}, the most commonly used evaluation metrics are MAP, MRR, and Top \textit{k} (where \textit{k} is typically 1, 5, or 10).
We have compiled the reported experimental results from each DL-based approach. Specifically, we calculated the average performance on the datasets studied, based on the reported results in the respective papers. The performance of all approach is presented in Table~\ref{tab:performance} (with ``-'' indicating that the metric was not reported in the corresponding study). We have highlighted high performance approaches (MAP, MRR, Top 1 $\geq$ 50\%, Top 5 $\geq$ 70\%, Top 10 $\geq$ 80\%) in \textbf{bold}.
It is important to note that the datasets and projects analyzed vary across different studies, and the experimental setups (e.g., version, data splits, sampling methods) also differ. As a result, this table provides only a rough comparison. A more precise comparison would require systematic evaluation.}

\nff{Table~\ref{tab:performance} shows that 18 approaches achieve an average MAP of 50\% or higher, 29 approaches have an average MRR of 50\% or higher, 11 approaches attain a Top 1 of 50\% or higher, 19 approaches reach a Top 5 of 70\% or higher, and 15 approaches achieve a Top 10 of 80\% or higher. Among them, five approaches have all five evaluation metrics highlighted in bold: Coloc\cite{luo2022improvingcoloc}, bjXnet~\cite{han2023bjxnet}, Yan et al.~\cite{yan2023bug}, bjEnet~\cite{han2024bjenet}, MACL-IRFL~\cite{zhou2024multimacl-irfl}. At the same time, the table also indicates that although file-level bug localization has achieved commendable results, there is still room for improvement in bug localization at other levels, such as method-level and commit-level. Additionally, there is significant potential for enhancement in cross-project and cross-language bug localization.
}

\nff{In order to further look into the approaches, we selected seven key studies (published in top venues in software engineering) for a more in-depth analysis of their methodology, contributions, and limitations, as shown in Table~\ref{tab:keystudy}. Based on these in-depth analyses, we summarize the challenges of existing research in the following sections.}

\subsection{\nff{Challenges Mitigated by DL Techniques}}

\textbf{Lexical Gap.} Lexical gap refers to that the terms used in bug reports to describe a bug are different from the terms and code tokens used in source files~\cite{lam2015combining}. Yang et al.~\cite{yang2021locatingmram} verified through experiments by measuring the textual similarity with TF-IDF. The result shows that the average textual similarity between bug reports and their fixed methods is 0.0153, while that between bug reports and their irrelevant methods is 0.0149. Obviously, there is a big semantic gap between bug reports and methods, and a typical IR-based approach is not able to identify those faulty methods by matching textual similarity. The proceeding two generations primarily used VSM to calculate exact term for bug localization, the effectiveness will be compromised in the common case where there exists a significant lexical gap between the descriptions in the bug reports and naming practices adopted by developers in the software artifacts~\cite{du2023prehmcbl}. To this end, the third generation of IRBL approaches use semantic word embedding techniques, especially language models (BERT, ELMo, GloVe, Word2Vec, etc) to capture the semantic information within bug reports and source code, so that terms with similar meaning will have high similarity.

\textbf{Ignoring Structural Information within Code.}  Many previous studies took the source code as natural language, and correlated the bug report and source code by measuring similarity in the same lexical feature space. However, these approaches fail to consider the structure information of source code which carries additional semantics beyond the lexical terms~\cite{huo2016learningnp-cnn}. The program structure specifies how different statements interact with each other to accomplishing certain functionality, which provides additional semantics to the program functionality besides the lexical terms. To extract such structural information from code, syntactic-based code representations have been employed to simulate code structures, including AST~\cite{xiao2017improvingdeeplocator, xiao2018machinebugtranslator}, CFG~\cite{huo2020controlcg-cnn, ma2022flowingcflow}, and PDG~\cite{yong2023decomposings-buglocator}.

\textbf{Cold-start.} The proceeding two generations of IRBL approaches leverage previous bug fixing history. However, they may face the cold-start problem, when a bug localization approach needs to be applied to new projects having a limited bug fixing history. These approaches cannot perform well, because there is not sufficient bug fixing data (i.e., bug reports labeled with corresponding buggy code files) for training such supervised models~\cite{zhu2021trobo}. To address this issue, Huo et. al.~\cite{huo2019deeptranp} presented the deep transfer bug localization task, and proposed the TRANP-CNN as the first solution for the cold-start problem which combines cross-project transfer learning and CNN for file-level bug localization. Subsequently, other approaches have also been proposed in response to the cold-start issue~\cite{zhu2020cooba,yang2021locatingmram, zhu2021trobo, liang2022modelingflim,zhu2022bl-gan}. Recently, there has also been work on cross-programming language approaches~\cite{chakraborty2024blaze, chandramohan2024supporting}.

\nff{Although the introduction of DL has indeed mitigated above challenges, these challenges still persist in the third generation of IRBL approaches and have not been fully resolved. For instance, there still remain gaps between bug reports and source code, as the terminology used by users often differs from the terms used in code, resulting in a persistent gap in bug localization. Therefore, this continues to be a significant challenge. Moreover, while several cross-project approaches~\cite{huo2019deeptranp, zhu2020cooba,yang2021locatingmram, zhu2021trobo, liang2022modelingflim,zhu2022bl-gan} have been proposed to address the cold-start problem, research has shown that using a portion of the target project's data (typically 20\% labeled data) for training yields the best results, outperforming predictions made with entirely new projects~\cite{huo2019deeptranp}. Thus, the cold-start problem still remains an open issue.}

\subsection{Open Questions}
Although significant progress has been made in software bug localization, several issues still remain to be addressed. \nff{We highlight some open challenges for future research, with the first three challenges being specific to DL-based approaches, while the following challenges are general issues relevant to IRBL.}


\textbf{Sampling.} Given the imbalanced nature of the bug report dataset, characterized by a substantial surplus of negative files (i.e., non-buggy files) compared to positive files (i.e., buggy files), the sampling strategy plays a pivotal role in influencing the model's performance. Some researchers have chosen to sample the top \textit{k} most similar files, while others have focused on the top \textit{k} least similar files. There are also researchers randomly select a certain amount of files as negative sample. Determining an effective sampling approach to establish a balanced dataset requires further investigation. Additionally, it is crucial for researchers to be mindful of the impact of class rename refactoring, which alters file names. This precaution is necessary to avoid potential misinterpretation during the sampling process.

\textbf{The Code Size Problem.} The size of code files is not fixed; that is, some files are short, while others are long. It is difficult to handle the sparse representation when localizing short files. These short files cannot be simply filtered out for their relatively large proportion of faulty files~\cite{yang2021locatingmram}. Furthermore, language models such as BERT and CodeBERT impose token count limitations, leading to the truncation or discarding of tokens that surpass the specified threshold. As a result, some valuable or buggy lines within large files may be omitted or overlooked.

\nff{\textbf{Interpretability of DL Models.} The decisions made by DL models are often opaque and lack clarity, making it challenging for developers to comprehend how predictions are generated. This complexity can diminish trust in the model's outputs, which are essential for effective bug identification and resolution.}


\textbf{Diversity in Programming Language.} According to the statistical data in programming language, the majority of current research is primarily focused on bug localization in the Java programming language, possibly due to Java having been one of the most popular programming languages. However, in recent years, with the widespread adoption of DL technologies, Python has gradually replaced Java in its predominant position~\cite{b2016learning, benton2019defexts}. This has led to a strong demand for bug localization studies in the Python language. Understanding the bugs and faults in large software repositories built in Python is therefore important~\cite{widyasari2020bugsinpy}. Future research is needed to delve into the language diversity in bug localization, like Python, C, C++, etc, examine variations in bug localization among different languages, propose corresponding bug localization approaches for specific languages, and even suggest language-agnostic bug localization approaches across diverse programming languages. Language diversity stands as an important topic for future research.

\textbf{Generalization of Approaches.} Table.~\ref{tab:dataset} shows that most of the studies were evaluated on the few commonly used projects, such as JDT, AspectJ, SWT, Eclipse, and Tomcat. However, in practical use, it is also important to generalize the approach to other projects. Therefore, it is important to evaluate the approach on more projects, instead of always on the few projects. Also, generalization of replication study can also be practical.

\textbf{Bug Report Category.} The existing state-of-the-art practices often consider all bug reports as natural language descriptions and apply a uniform procedure to identify related buggy files. However, diverse bug reports may emphasize various aspects such as capability, security, function, performance, and more. Distinct approaches may yield varying results when applied to different types of bug reports. \nff{Zou et al.~\cite{zou2018practitioners} also revealed that there is a growing need for effective approaches to locate
performance bugs, memory leaks, and environment-related bugs, which are particularly challenging
to identify.} To the best of our knowledge, there has been no research thus far dedicated to investigating bug localization specifically for different types of bug reports.

\textbf{The Scale and Quality of Dataset.} The AI pioneer Andrew Ng said it is time for smart-sized, ``data-centric'' solutions to big issues~\cite{andrew}. Previously, researchers and developers primarily focus on identifying more effective models to improve bug localization performance while keeping the data largely unchanged. However, the potential quality issues and undesirable flaws of data needs more attention, such as missing values, incorrect labels, and anomalies. Zimmermann et al.~\cite{zimmermann2009cross} have highlighted that sufficient defect data is often unavailable for many projects and companies. A large-scale and high-quality dataset is a crucial step in ensuring the accuracy of bug localization~\cite{bhandari2023data}. \nff{To this end, future research on datasets can focus on developing high-quality training datasets, guided by practitioners' recommendations~\cite{yu2024dataset}, where key characteristics include reliability, relevance, accuracy, compliance, scale, and so on. Additionally, high-quality benchmarks are essential for unbiasedly evaluating the performance of LLMs}.

\textbf{Dealing with Additional Real-World Characteristics.} 
While current state-of-the-art DL approaches have demonstrated promising performance in bug localization, existing studies have primarily concentrated on assessing the first two generations of IRBL approaches in industry projects~\cite{jarman2021legion, murali2021industry}. Significantly, there is a gap in research specifically evaluating DL approaches within the industry context. Although numerous approaches utilizing diverse DL models have been proposed, empirical research to determine the most effective DL model, particularly in real-world scenarios, is currently scarce.

\nfff{Empirical evaluations of first- and second-generation IRBL approaches on industrial projects show that while these techniques perform well on small-scale systems, their effectiveness diminishes in large-scale industrial settings~\cite{li2022empirical}. Key challenges include software product lines, bilingual issues, and low-quality bug reports. However, leveraging product-specific information, capturing multiple perspectives from bug reports, and handling multilingual content can improve localization accuracy by enhancing lexical similarity~\cite{li2022empirical}. Moreover, using historical bug reports and applying collaborative filtering—prioritizing frequently modified files—further boosts performance in industrial contexts.
Murali~\cite{murali2021industry} found that their approach for Facebook projects relied heavily on word-level overlap between bug reports and commits, requiring both to share relevant terms. The method lacked the ability to assign different weights to features and faced a trade-off between complexity and interpretability. Jarman et al.~\cite{jarman2021legion} demonstrated comparable performance on Adobe repositories to that of previous open-source evaluations, and highlighted the benefits of incorporating additional sources, such as commit messages and bug report comments. They also noted that optimal configurations and relevant data can vary significantly across different industrial projects.}

When it comes to the application of existing approaches to real-world industry projects, the existing research is performed under idealized assumption, whereas real-world situations can be significantly more complex~\cite{zou2018practitioners}. For example, current studies often assume a one-to-one or one-to-many relationship between commits and bug reports, meaning that one or several commits fix a single bug. In such cases, all code modifications within these commits are considered as buggy code to that bug report. However, in practice, developers may address not only one but multiple bugs in a single commit. Furthermore, the code changes submitted in a commit may involve code refactoring or the addition of new features. In this way, not all code changes in the commit are necessarily buggy.

Another scenario is that current research assumes independence between different bugs, with no mutual influence. In reality, multiple bugs might emerge simultaneously and impact each other. These real-world situations are more intricate than the assumptions made in existing research. When applying existing solutions to practical scenarios, these issues are inevitable. 

\nff{From the perspective of practitioners, there is a growing need for effective approaches to locate performance bugs, memory leaks, and environment-related bugs, which are particularly challenging to identify~\cite{zou2018practitioners}. Practitioners prefer solutions that operate at finer granularity levels, such as method, statement, or block~\cite{kochhar2016practitioners}. Additionally, Kochhar et al.~\cite{kochhar2016practitioners} found that practitioners have high thresholds for tool adoption: 73.58\% consider inspecting more than five program elements unacceptable, and nearly 98\% reject tools requiring inspection of over ten elements. Future research should focus on developing efficient and practical approaches that address these challenges while meeting practitioners' expectations for precision and usability.}

\textbf{Application of LLMs.} Since the introduction of LLMs, the field of software engineering has undergone revolutionary transformations. LLMs present substantial challenges alongside vast opportunities for research in software engineering, including code generation, software testing, and program repair.
\nff{While some studies have started exploring the use of LLMs for IRBL, such as BERT~\cite{mohsen2023enhancingpbl, ciborowska2023too, ciborowska2022fastfbl-bert}, CodeBERT~\cite{zhou2024multimacl-irfl}, CodeT5~\cite{han2024bjenet}, and CodeSage~\cite{chakraborty2024blaze}, significant potential for further investigation remains. Future research should focus on effectively leveraging popular LLMs like GPT and LLaMA, as well as exploring prompt engineering and agent-based technologies, to enhance IRBL in software management. Furthermore, several benchmarks could be explored to evaluate the performance of LLMs in this domain~\cite{jimenez2023swe}. Given that using LLMs for IRBL can be resource-intensive, particularly when processing entire code repositories~\cite{tao2024magis}, and considering that some LLMs have token limitations that hinder their ability to handle large files, future work should prioritize optimizing this process to improve both efficiency and scalability.}

\textbf{Multilingual Bug Localization.} Existing IRBL approaches primarily focus on bug localization within the same language context, where bug reports, identifiers, and comments in source code files are typically written in English. However, in non-English-speaking countries, bug reports are often written in native languages. Consequently, existing approaches initially translate non-English texts into English~\cite{xia2014cross}. This translation process, however, can introduce biases that negatively affect bug localization performance. Recently, LLMs (such as ChatGPT) have shown remarkable performance across various languages and tasks, indicating promising potential for their application in Multilingual bug localization.

\section{Threats to validity}\label{sec:threats}
In this section, we conclude the threats to validity from four parts: construction, conclusion, internal and external threats.

\noindent\textbf{Construction Validity.}
The first threat to the construction validity comes from the search and selection of the primary studies. We follow Kitchenham et al.~\cite{kitchenham2004procedures} guidelines to perform the process. We carefully select our search terms by examining related work and queries widely used software engineering databases used in surveys. Moreover, to comprehensively retrieve articles as much as possible, we also adopt backward and forward snowballing using Google Scholar. Nevertheless, we believe that an adequate set of primary studies was collected for this study.

Another threat to the construction validity is the application of the inclusion and exclusion criteria, which is subject to researcher' bias. The authors create a list of inclusion and exclusion criteria, and independently applied inclusion and exclusion criteria to each candidate paper. A joint voting mechanism is used to mitigate the risk of ambiguous interpretations. The two authors will have a discussion if the two authors hold different opinions towards one paper.

\noindent\textbf{Conclusion Validity.}
Our reported results are directly derived from the data; however, variations in perspectives during the study selection process may lead other researchers to present dissimilar results. To address this potential threat, the authors conducted individual assessments and analyses of the papers.

\noindent\textbf{Internal Validity.}
The study selection process, which comprises inclusion/exclusion criteria, is exposed to the researchers' bias. The inclusion/exclusion criteria was ultimately determined through thorough discussions to reach a consensus. In the final selection procedure of primary studies, we made sure that all the authors were fully involved. In case of disagreement, conflicts were resolved through manual discussion.

\noindent\textbf{External Validity}
This study exclusively investigates published works dealing with IRBL issues employing DL techniques. Potential data quality issues that have not been explored may exist. Furthermore, our findings are primarily applicable to IRBL and should not be generalized beyond this domain. It is important to highlight that the outcomes of this study may vary from those in spectrum-based fault localization and other fault localization research.

\section{Conclusion and Future Work}\label{sec:conclusion}

Bug localization is to identify and pinpoint the specific location or locations within a software program's source code where a bug or defect is present. Bug localization aims to assist developers in efficiently locating and addressing the root cause of software issues, reducing the time and effort required for debugging. This process contributes to enhancing the overall software development and maintenance workflow by facilitating a quicker resolution of bugs. IRBL using DL has particularly gained attention in the recent years. In this study, we conduct a systematic literature survey of existing IRBL techniques using DL to draw a picture on the state of the art. We studied 61 primary studies that leverage DL for IRBL, summarized the state of the art, mitigated issues and open challenges. Our study suggests that the integration of DL in IRBL enhances the model's capacity to extract semantic and syntactic information from both bug reports and source code, addressing issues such as lexical gaps, neglect of code structure information, cold-start problems, and more. Future research avenues for IRBL encompass exploring diversity in programming languages, adopting finer-grained granularity, and focusing on real-world applications. \nff{Most importantly, although some studies have started using LLMs for IRBL, there is still a need for more in-depth exploration and thorough investigation in this area.}

In future research, our primary focus will be on addressing the aforementioned challenges. Besides, we intend to conduct an empirical study to compare and evaluate the performance of various IRBL approaches on a common comprehensive and representative benchmark containing recent data.





\begin{acks}
This research is supported by CCF-Huawei Populus Grove Fund, the National Natural Science Foundation of China (No. 62172214), National Research Foundation,
under its Investigatorship Grant (NRF-NRFI08-2022-0002). Any opinions, findings and conclusions
or recommendations expressed in this material are those of the author(s) and do not reflect the
views of National Research Foundation, Singapore. Chuanyi Li is the corresponding author.
\end{acks}


\appendix
\renewcommand{\thesection}{Appendix \Alph{section}}
\section{Additional Material}

\begin{center}
\begin{table}[htbp]%
\caption{List of Digital Libraries.}
\label{tab:digitallibrary}
\centering
\begin{tabular}{m{0.3\linewidth}<{\raggedright}m{0.6\linewidth}<{\raggedright}}
\toprule
Digital   libraries & URLs \\ 
\midrule
IEEE Xplore & http://www.ieee.org/web/publications /xplore/ \\
\rowcolor[HTML]{EFEFEF} 
ACM Digital Library & http://portal.acm.org \\
Science Direct      & http://www.sciencedirect.com/  \\
\rowcolor[HTML]{EFEFEF} 
Springer Link       & https://link.springer.com/    \\
Wiley InterScience  & https://onlinelibrary.wiley.com/   \\
\rowcolor[HTML]{EFEFEF} 
Elsevier  & http://www.elsevier.com       \\
Google Scholar      & https://scholar.google.com    \\
\rowcolor[HTML]{EFEFEF} 
DBLP  & https://dblp.org \\\bottomrule
\end{tabular}
\end{table}
\end{center}

\begin{table}[]
\centering
\caption{Inclusion and Exclusion Criteria.}
\label{tab:criteria}
\begin{tabular}{m{0.1\linewidth}<{\raggedright}m{0.8\linewidth}<{\raggedright}}
\hline
\multicolumn{2}{l}{Inclusion Criteria} \\ \hline
\rowcolor[HTML]{EFEFEF} 
IC1 & \nff{The study focuses on information retrieval-based approaches to bug localization.} \\
IC2 & \nff{The research employs deep learning algorithms (e.g., CNNs, LSTMs, RNNs, transformers) in its methodology. }\\
\rowcolor[HTML]{EFEFEF} 
IC3 & \nff{The paper was published prior to November 2024.} \\
IC4 & \nff{The work is published in a peer-reviewed journal, conference proceedings, or workshop, with the full text accessible, including recent publications available on arXiv.} \\ \hline
\multicolumn{2}{l}{Exclusion Criteria} \\ \hline
\rowcolor[HTML]{EFEFEF} 
EC1 & \nff{Literature not written in English.} \\
EC2 & \nff{The paper is a technical report or thesis.} \\
\rowcolor[HTML]{EFEFEF} 
EC3 & \nff{Duplicate studies or different versions of similar work by the same authors that report identical findings.} \\
EC4 & \nff{Papers that do not specifically target bug localization or that focus solely on general debugging techniques without an information retrieval component.} \\
\rowcolor[HTML]{EFEFEF} 
EC5 & \nff{Studies employing traditional machine learning techniques or non-machine learning approaches (e.g., rule-based systems) that do not incorporate deep learning.} \\
EC6 & \nff{Research primarily centered on other fault localization approaches, such as spectrum-based fault localization, rather than utilizing bug reports for bug localization.} \\
\rowcolor[HTML]{EFEFEF} 
EC7 & \nff{Studies that address bug localization specifically for deep learning models, rather than focusing on deep learning-based techniques for bug localization.} \\
\hline
\end{tabular}
\end{table}

\begin{table}[htpb]
\centering
\caption{The Quality Checklist.}
\label{tab:qualityassessment}
\begin{tabular}{m{0.1\linewidth}<{\raggedright}m{0.8\linewidth}<{\raggedright}}
\hline
\multicolumn{2}{l}{Data Criteria} \\ \hline
\rowcolor[HTML]{EFEFEF} 
DC1       & The paper reported the dataset.       \\
DC2       & Statistics of the dataset must be reported, including size and granularity. \\
\rowcolor[HTML]{EFEFEF}  \hline
\multicolumn{2}{l}{Model Criteria} \\ \hline
MC1       & The overall framework of the model must be presented. \\
\rowcolor[HTML]{EFEFEF}  MC2       & The feature extraction of bug reports and source code must be clearly demonstrated.  \\  \hline
\multicolumn{2}{l}{Evaluation Criteria} \\ \hline
\rowcolor[HTML]{EFEFEF}  EC1       & The evaluation process must be clearly explained.  \\
EC2 & The performance of the model must be reported. \\
\hline
\end{tabular}
\end{table}

\begin{table}[]
\centering
\caption{The Data Extraction Form.}
\label{tab:extractionform}
\begin{tabular}{m{0.4\linewidth}<{\raggedright}m{0.27\linewidth}<{\raggedright}m{0.2\linewidth}<{\raggedright}}
\hline
Field & Input type & Relevant RQ \\ \hline
Paper ID & Auto-fill & metadata \\
Paper title & Free text & metadata \\
Publication year & Number & metadata \\
Publication venue & Free text & metadata\\
Localization scenario & Multiple selection & metadata \\ \hline
Dataset & Free text & RQ1 \\
Granularity of localization & Multiple selection & RQ1\\
Validation approach & Free text & RQ1 \\
Evaluation metrics & Multiple selection & RQ1 \\
Class imbalance & Free text & RQ1 \\
Before-fix version & Free text & RQ1\\ \hline  
DL model & Free text & RQ2 \\
Text representation & Free text & RQ2 \\
Code representation & Free text & RQ2 \\
Other features & Free text & RQ2\\ \hline
Challenges and proposed solutions & Free text & RQ3\\
Future work & Free text & RQ3 \\  \hline
\end{tabular}
\end{table}

\begin{figure}[htbp]
\centering
\includegraphics[width=0.9\textwidth]{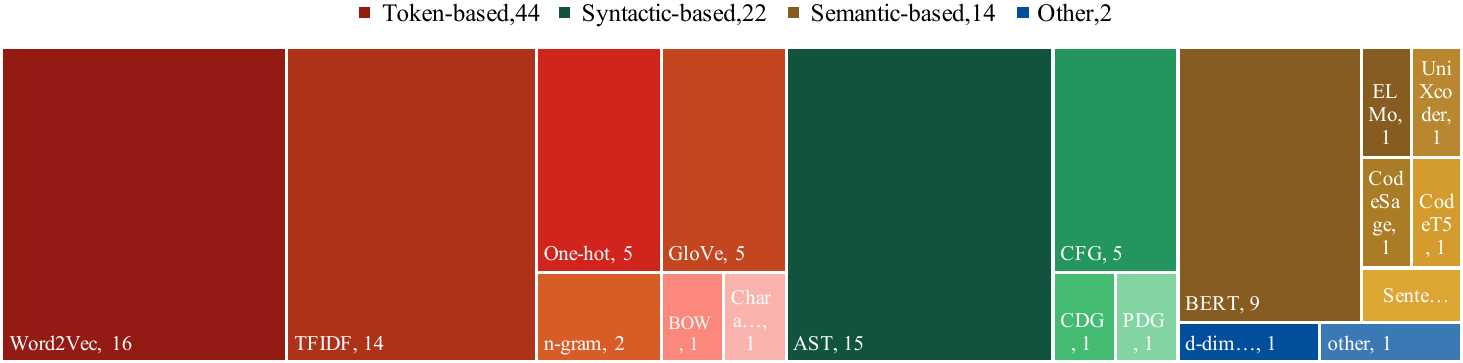}
\caption{Code Representation Approaches.}
\Description[]{}
\label{fig:sunburst}
\end{figure}

\begin{table}[]
  \caption{Other Features Adopted in Primary Studies.}
  \label{tab:otherfeature}
  \begin{tabular}{lp{10cm}}
    \toprule
    Feature& Primary Studies \\ \midrule
    Stack Trace    & \cite{yang2020applying}, \cite{xiao2023bugradar}, \cite{xu2023buglocfront}, \cite{ahmad2023attentivebuglocator}, \cite{yan2023bug}, \cite{ciborowska2023too}, \cite{alsaedi2024two}, \cite{zhang2023enhancing}\\
    Bug Fixing History  & \cite{lam2015combining}, \cite{lam2017bugdnnloc}, \cite{xiao2019improvingdeeploc}, \cite{sangle2020drast}, \cite{wang2020multimd-cnn}, \cite{cao2020bugpecker}, \cite{anh2021imbalanced}, \cite{yang2021locatingmram}, \cite{qi2021dreamloc}, \cite{shi2022semirfl}, \cite{xiao2023bugradar}, \cite{xu2023buglocfront}, \cite{Al-Aidaroosimpact}, \cite{alsaedi2024two}, \cite{zhang2023enhancing} \\
    Collaborative Filtering Score & \cite{lam2017bugdnnloc}, \cite{xiao2019improvingdeeploc} \cite{sangle2020drast}, \cite{cao2020bugpecker}, \cite{yang2021locatingmram}, \cite{qi2021dreamloc}, \cite{shi2022semirfl},\cite{xu2023buglocfront}, \cite{zhang2023enhancing}   \\
    Class Name     & \cite{lam2017bugdnnloc}, \cite{xiao2019improvingdeeploc}, \cite{sangle2020drast}, \cite{shi2022semirfl} \\
    Code Comments  & \cite{lam2015combining}, \cite{lam2017bugdnnloc} \cite{huang2022sbuglocater}   \\
    Others & \cite{xiao2018machinebugtranslator}, \cite{liu2019convolutionalsls-cnn}, \cite{zhu2022bl-gan}, \cite{chandramohan2024supporting}, \cite{alsaedi2024two}, \cite{zhang2023enhancing} \\ 
  \bottomrule
\end{tabular}
\end{table}

\begin{figure}[!t]
\centering
\includegraphics[width=0.5\textwidth]{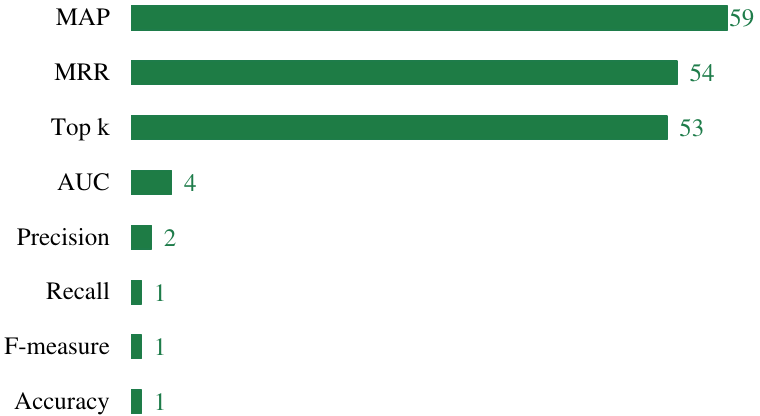}
\caption{Evaluation Metrics.}
\Description[]{}
\label{fig:evaluation}
\end{figure}

\begin{table*}[htbp]
\caption{Overview of open sourced datasets.}
\label{tab:dataset}
\centering
\resizebox{\textwidth}{!}{
\begin{tabular}{llcc}
\hline
Primary Study & {\color[HTML]{000000} Dataset}      & Language & Year of Research \\ \hline
\rowcolor[HTML]{EFEFEF} CNN-Forest\cite{xiao2018bugcnn_forest}    & https://github.com/yanxiao6/BugLocalization-dataset & Java   &  2018  \\
DRAST\cite{sangle2020drast} & https://doi.org/10.5281/zenodo.4153560& C     &     2020\\
\rowcolor[HTML]{EFEFEF} COOBA\cite{zhu2020cooba} & http://dx.doi.org/10.6084/m9.fifigshare.951967      & Java     & 2020  \\
DreamLoc\cite{qi2021dreamloc}      & https://github.com/qibinhang/dream\_loc   & Java     &   2021 \\
\rowcolor[HTML]{EFEFEF} BLESER~\cite{zou2021bleser} & https://github.com/rjust/Defects4J & Java  & 2021 \\
 FLIM~\cite{liang2022modelingflim} & https://github.com/hongliangliang/flim& Java     &   2022 \\
\rowcolor[HTML]{EFEFEF} HGW-SFO-CDNN~\cite{ali2023automatedhgwsfocdnn}  & https://github.com/AliWaqas27/paper\_code.git       & Java     &    2023   \\
 BRS\_BL~\cite{zhang2023enhancing} & https://github.com/Zhang612-alt/BRS\_BL-datasets& Java     &   2023 \\
 BLAZE~\cite{chakraborty2024blaze} & https://zenodo.org/records/11553913 & C++, Go, Java, JavaScript, Python &2024\\
\hline
\end{tabular}}
\end{table*}

\begin{figure}[htbp]
\centering
\includegraphics[width=0.7\textwidth]{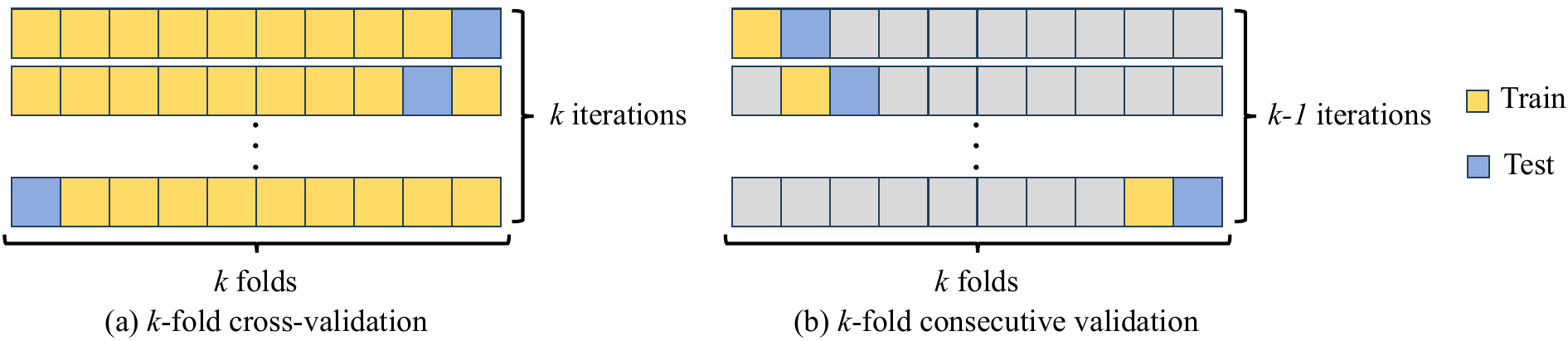}
\caption{Illustration of validation strategies. \nff{(a) \textit{k}-fold cross-validation randomly divides the dataset into \textit{k} equal folds, where each fold serves as the test set once, while the remaining folds form the training set. The process is repeated \textit{k} times (iterations), ensuring that each fold is tested and trained equally. (b) \textit{k}-fold consecutive validation also divides the data into \textit{k} folds, but in a consecutive manner. It preserves the temporal order of the data, with the training set starting from the initial portion of the dataset and testing performed on subsequent consecutive portions.}}
\Description[]{}
\label{fig:validation}
\end{figure}

\begin{table}[htbp]
\caption{\nff{Average performance metrics as reported in the original studies for each approach. Values highlighted in \textbf{bold} indicate MAP, MRR, Top 1 $\geq$ 50\%, Top 5 $\geq$ 70\%, and Top 10 $\geq$ 80\%. ``-'' indicates that the original study did not report.}}
\label{tab:performance}
\resizebox{\textwidth}{!}{
\begin{tabular}{|lccccc|l|lccccc|}
\cline{1-6} \cline{8-13}
\textbf{Approach (\%)}      & \textbf{MAP}    & \textbf{MRR}    & \textbf{Top 1}  & \textbf{Top 5}  & \textbf{Top 10} &       & \textbf{Approach} (\%)      & \textbf{MAP}         & \textbf{MRR}    & \textbf{Top 1}  & \textbf{Top 5}  & \textbf{Top 10} \\ \cline{1-6} \cline{8-13} 
\multicolumn{6}{|c|}{\textbf{Within-project}}      &       &         FBL-BERT (changeset-file)~\cite{ciborowska2022fastfbl-bert}      &             34.8         &   9.7     &      29.3  &  13.8 & -        \\ \cline{1-6}
HyLoc~\cite{lam2015combining}              & 36.0  & 46.8  & 36.0  & 59.6  & 68.9  &       &  FBL-BERT (hunk)~\cite{ciborowska2022fastfbl-bert} &    35.5         &   10.7     &      29.6  &  14.9 & -    \\
NP-CNN~\cite{huo2016learningnp-cnn}            & \textbf{55.7}  &   -     &   -     &   -     &    -    &       & BL-GAN~\cite{zhu2022bl-gan}        & 35.8       & -      & 37.0  & 56.6  & 68.8  \\
DeepLocator~\cite{xiao2017improvingdeeplocator}   & 43.0  &  -      &   -     &    -    &   -     &       & MLA~\cite{ma2022learningmla}           & \textbf{50.6}       & \textbf{60.5}  &  -      &  -      &    -    \\
DNNLOC~\cite{lam2017bugdnnloc}        & 36.0  & 46.8  & 41.4  & 65.1  & 74.9  &       & Bugradar~\cite{xiao2023bugradar}      & 41.9       & 49.7  & 37.9  & 63.5  & 73.4  \\
LS-CNN~\cite{huo2017enhancingls-cnn}        & \textbf{55.6}  &   -     &   -     &  -      & \textbf{86.9}  &       & sgAttention~\cite{ma2023capturingsgattention}   & \textbf{66.5}       & \textbf{77.8}  &   -     &   -     &   -     \\
BugTranslator~\cite{xiao2018machinebugtranslator} & 34.7  & 41.0  &  -      &   -     &   -     &       & S-BugLocator~\cite{huang2022sbuglocater}  & 45.8       & \textbf{54.3}  & 46.2  & \textbf{74.2}  & \textbf{81.9}  \\
CNN\_Forest~\cite{xiao2018bugcnn_forest}   & 44.7  & \textbf{53.3}  &   -     &   -     &   -     &       & PBL~\cite{mohsen2023enhancingpbl}           & -           & -      & -      & -      & -      \\
Loyola et al.~\cite{loyola2018bug} & \textbf{51.0}  & \textbf{53.4}  & \textbf{63.6}  & \textbf{78.3}  &   -     &     -  & HMCBL (commit)~\cite{du2023prehmcbl}   &  10.4 & 32.9       &   28.4     &  16.0      &    14.0    \\
Xiao et al.~\cite{xiao2018improving}   & 35.7  & 41.3  &   -     &    -    &    -    &       &     HMCBL (method)~\cite{du2023prehmcbl}   &    13.1         &   41.7     &   37.2     &   20.2     &   19.5     \\
Polisetty et al.~\cite{polisetty2019usefulness}              & 27.2  & 28.6  &   -     &  -      &    -    &       &    HMCBL (hunk)~\cite{du2023prehmcbl}   &     14.3        &   40.8     &   35.9     &   24.1     & 21.3   \\
DeepLoc\cite{xiao2019improvingdeeploc}       & 45.0  & \textbf{53.6}  &    -    & -       &   -     &       & LocFront\cite{xu2023buglocfront}      & 48.8       & \textbf{57.2}  & 45.2  & \textbf{71.8}  & 78.9  \\
TRANP-CNN\cite{huo2019deeptranp}     & 46.7  & \textbf{56.0}  & 47.6  & 64.4  & 73.4  &       & HGW-SFO-CDNN~\cite{ali2023automatedhgwsfocdnn}  & 42.1       & 49.3  &    -    &   -     &  -      \\
SLS-CNN~\cite{liu2019convolutionalsls-cnn}       & 44.0  & \textbf{52.8}  & 43.7  & 67.1  & 76.2  &       & AttentiveBugLocator~\cite{ahmad2023attentivebuglocator}           & 48.4       & \textbf{53.2}  & \textbf{59.5}  & 76.0  & 82.3  \\
CAST~\cite{liang2019deepcast}          & 44.7  & \textbf{53.0}  & 45.3  & \textbf{72.9}  & 74.5  &       & AI-Aidaroos~\cite{Al-Aidaroosimpact}      & 38.5       & 49.5  & 41.0  & 58.5  & 70.0  \\
DependLoc~\cite{yuan2020dependloc}     & 40.2  & 47.6  &   -     &   -     & 73.8  &       & Ciborowska et al.~\cite{ciborowska2023too}              & 36.8       & 14.9  & 26.9  & 19.2  & 18.2  \\
Yang et al.~\cite{yang2020applying}   & -      & -      & -      & -      & -      &       & RLocator~\cite{chakraborty2024rlocator}      & 43.2       & 46.8  & 34.0  & 55.7  & 65.3  \\
DRAST~\cite{sangle2020drast}        & \textbf{63.7}  & \textbf{70.4}  & \textbf{69.4}  & \textbf{71.6}  & 72.6  &       & bjXnet~\cite{han2023bjxnet}        & \textbf{56.0 }      & \textbf{65.0}  & \textbf{51.7}  & \textbf{80.7}  & \textbf{88.7}  \\
CG-CNN~\cite{huo2020controlcg-cnn}        & 48.4  & \textbf{56.5}  &   -     &    -    &   -     &       & Yan et al.~\cite{yan2023bug}      & \textbf{50.2}       & \textbf{57.5}  & \textbf{54.4}  & \textbf{78.0}  & \textbf{87.1}  \\
MD-CNN~\cite{wang2020multimd-cnn}        & 44.0  & 49.2  & 41.2  & 66.3  & 77.3  &       & Alsaedi et al. (class)~\cite{alsaedi2024two}        & 26.7       & 39.1  & 29.7  & 49.9  & 59.7  \\
DBL~\cite{jiang2020deepdbl}           & 49.4  & \textbf{57.8}  &   -     & \textbf{77.2}  &   -     &       & Alsaedi et al. (method)~\cite{alsaedi2024two} &- &     -   &  -      & 1.7   & 3.0   \\
KGBugLocator~\cite{zhang2020exploitingkgbuglocator}  & \textbf{55.8}  & \textbf{52.0}  & 45.9  & \textbf{73.5}  & \textbf{81.3}  &       & BRS\_BL~\cite{zhang2023enhancing}       & \textbf{52.5}       & \textbf{61.5}  & 48.8  & \textbf{78.8}  & \textbf{87.5}  \\
BugPecker (Method)~\cite{cao2020bugpecker}     & 21.2  & 23.4  & 19.7  & 23.0  & 29.4  &       & MSP-COA-IDN-LSTM~\cite{ali2024softwaremsp-coa-idn-lstm}              & 48.7       & \textbf{50.1}  &  -      &   -     &  -      \\
Yang et al.~\cite{yang2021utilizing}   & -      & -      & -      & -      & -      &       & bjEnet~\cite{han2024bjenet}        & \textbf{58.0}       & \textbf{65.0}  & \textbf{53.0}  & \textbf{80.0} & \textbf{91.0}  \\
ImbalancedBugLoc~\cite{anh2021imbalanced} & \textbf{50.0}  & \textbf{58.2}  & 45.9  & \textbf{74.1}  & \textbf{82.2}  &       & IBL(commit)~\cite{zhao2024fineibl}   & 6.3        & 21.9  & 25.0  & 11.0  & 12.0  \\
MRAM (Method)~\cite{yang2021locatingmram} & 6.1   & 7.6   & 5.1   & 8.9   & 12.6  &       & MACL-IRFL~\cite{zhou2024multimacl-irfl}     & \textbf{63.5}       & \textbf{71.0 } & \textbf{60.7}  & \textbf{83.5}  & \textbf{90.1 } \\ \cline{8-13} 
DreamLoc~\cite{qi2021dreamloc}      & 48.0  & \textbf{57.0}  & 45.5  & \textbf{71.5}  & \textbf{80.0}  & & \multicolumn{6}{c|}{\textbf{Cross-project}}              \\ \cline{8-13} 
DEMOB\cite{zhu2021deepdemob}         & \textbf{54.8}  &  -      & \textbf{58.0}  & \textbf{80.3}  & \textbf{87.7 } &       & Loyola et al. & 27.7       & 29.6  &  -      &  -      &  -      \\
BLESER (Method)~\cite{zou2021bleser}   &   29.4     &   31.1     &   29.2     &  33.2      &   33.2     &       & TRANP-CNN     & 24.8       & 33.7  & 22.7  & 43.9  & 56.9  \\
SBugLocater~\cite{huang2022sbuglocater}   & 48.1  & \textbf{57.6}  & 46.2  & \textbf{71.4}  & \textbf{81.4}  &       & COOBA~\cite{zhu2020cooba}         & 35.0       &  -      &   -     &     -   & -       \\
CoLoc~\cite{luo2022improvingcoloc}         & \textbf{51.1}  & \textbf{60.2}  & \textbf{53.4}  & \textbf{76.6}  & \textbf{85.3}  &       & MRAM (Method)~\cite{yang2021locatingmram} & 4.1        & 5.5   & 3.6   & 6.2   & 9.6   \\
FLIM~\cite{liang2022modelingflim}          & 41.2  & 48.7  & 37.7  & 62.2  & 71.2  &       & TROBO~\cite{zhu2021trobo}         & 37.3       & -       &     -   &    -    &    -    \\
CGMBL~\cite{chen2022cgmbl}         & \textbf{57.0}  & \textbf{71.3}  & \textbf{68.0}  & \textbf{74.7}  & 77.3  &       & FLIM~\cite{liang2022modelingflim}          & 41.6       & 48.8  & 37.1  & 63.6  & 71.7  \\
cFlow~\cite{ma2022flowingcflow}         & 48.6  & \textbf{59.4}  &   -     &   -     &    -    &       & BL-GAN~\cite{zhu2022bl-gan}        &      -       &   -     &   -     &      -  &   -     \\
SemirFL~\cite{shi2022semirfl}       & \textbf{53.4}  & \textbf{58.7}  & 48.5  & 66.2  & \textbf{80.8}  &       & Chandramohan et al.~\cite{chandramohan2024supporting} &-    & \textbf{64.9}  & \textbf{58.6 } & \textbf{73.6}  & 73.8  \\ \cline{8-13} 
BLoco~\cite{zhu2022enhancingbloco}   &     \textbf{50.0}  & - &   \textbf{50.0}     &   \textbf{75.9}     &  \textbf{85.4}        & & \multicolumn{6}{c|}{\textbf{Cross-language}}             \\ \cline{8-13} 
FBL-BERT (changeset)~\cite{ciborowska2022fastfbl-bert}   &  3.6      & 9.6       &   4.5     &    4.9    &     -   &       & Chandramohan et al.~\cite{chandramohan2024supporting}           & 38.3       & 48.1  & 36.5  & 62.6  & 76.9  \\ 
&&&&&& & Blaze~\cite{chakraborty2024blaze} & 28.7 & 32.7&21.0&46.7&56.0\\\cline{1-6} \cline{8-13}
\end{tabular}}
\end{table}

\begin{table}[htbp]
\scriptsize
\centering
\caption{Key Studies.}
    \label{tab:keystudy}
    \begin{tabularx}{\textwidth}{|l|l|l|X|X|X|}
    \hline
    Approach  & Venue & Year & Methodologies & Contribution & Limitation \\ \hline
    HyLoc     & ASE   & 2015 & Uses 3 DNNs to learn relationships between bug reports and source code, with bug-fixing recency score. & First to use DL to bridge the lexical gap between bug reports and source code. &Does not consider semantic features or code structure information. \\ \hline
    TRANP-CNN & TSE   & 2019 & Deep transfer learning approach for cross-project bug localization, uses word2vec encoding and CNN for feature extraction. & First deep transfer learning model for cross-project bug localization, directly takes bug reports and source code in their raw format as input and outputs the localization result. & Requires 20\% labeled data from target project, not fully solving the cold-start problem. \\ \hline
    BugPecker & ASE   & 2020 & Builds revision graphs from past fixes and uses DNN for bug localization, combining semantic matching, collaborative filtering scores, and bug-fixing history. & First to introduce revision graphs and AST parsing, improving method-level bug localization. & Cold-start problem, and challenges in knowledge transfer between projects.\\ \hline
    FBL-BERT  & ICSE  & 2022 & Applies BERT to localize bug-inducing changesets, improves retrieval speed and quality with special tokens for code modifications. & First to apply BERT for changeset-level bug localization, significantly improves retrieval performance. & Special token-based approach complicates data representation; not robust enough across projects. \\ \hline
    FLIM      & EMSE  & 2022 & Uses CodeBERT to extract function-level semantic features, addresses long code sequences and noisy data, and uses a learning-to-rank model for bug localization. & Proposes a function-level semantic feature extraction framework, improving bug localization accuracy. & Does not consider function structural features (e.g., AST, CFG), limiting its ability to capture full code semantics. \\ \hline
    HMCBL     & FSE   & 2023 & Introduces Semantic Flow Graph (SFG) for code semantics, trains SemanticCodeBERT, and applies HMCBL for bug localization optimization. & Proposes SFG and SemanticCodeBERT, improving bug localization with HMCBL technique. & Complex SFG model with high computational cost, challenging to scale for large codebases. \\ \hline
    RLocator  & TSE   & 2024 & Reinforcement learning-based approach leveraging Markov Decision Process to optimize bug localization, using ElasticSearch to identify relevant files. & First to use reinforcement learning for direct optimization of bug localization evaluation metrics. & Reinforcement learning model complexity and computational cost, difficult to scale for large datasets. \\ \hline
    \end{tabularx}
\end{table}

\begin{table*}[htbp]
  \caption{Model Structures of Primary Study.}
  \label{tab:modelstructure}
  \resizebox{\textwidth}{!}{
  \begin{tabular}{lp{10cm}}
    \toprule
    Model Structure & Primary Study \\ \midrule
    Heterogeneous Model & \cite{huo2016learningnp-cnn}, \cite{huo2017enhancingls-cnn}, \cite{xiao2018bugcnn_forest}, \cite{loyola2018bug}, \cite{xiao2019improvingdeeploc}, \cite{huo2019deeptranp}, \cite{liu2019convolutionalsls-cnn}, \cite{liang2019deepcast}, \cite{yuan2020dependloc}, \cite{yang2020applying}, \cite{huo2020controlcg-cnn}, \cite{jiang2020deepdbl}, \cite{zhu2020cooba}, \cite{zhang2020exploitingkgbuglocator}, \cite{cao2020bugpecker}, \cite{yang2021locatingmram}, \cite{anh2021imbalanced}, \cite{qi2021dreamloc}, \cite{zhu2021deepdemob}, \cite{zou2021bleser}, \cite{ma2022flowingcflow}, \cite{shi2022semirfl}, \cite{zhu2022enhancingbloco}, 
    \cite{xiao2023bugradar}, \cite{ma2023capturingsgattention}, \cite{yong2023decomposings-buglocator}, \cite{du2023prehmcbl}, \cite{ali2023automatedhgwsfocdnn}, \cite{Al-Aidaroosimpact}, \cite{han2023bjxnet}, \cite{ma2022learningmla} \cite{yan2023bug}, \cite{zhao2024fineibl}\\ \cline{2-2}
    Homogeneous Model & \cite{lam2015combining}, \cite{lam2017bugdnnloc}, \cite{polisetty2019usefulness}, \cite{sangle2020drast}, \cite{wang2020multimd-cnn}, \cite{yang2021utilizing}, \cite{huang2022sbuglocater}, \cite{ciborowska2022fastfbl-bert}, \cite{ahmad2023attentivebuglocator}, 
    \cite{ciborowska2023too}, \cite{chakraborty2024rlocator}, \cite{ali2024softwaremsp-coa-idn-lstm}, \cite{zhou2024multimacl-irfl} \\ \cline{2-2}
    Relevance Matching Model & \cite{luo2022improvingcoloc}, \cite{liang2022modelingflim}, \cite{mohsen2023enhancingpbl}, \cite{chandramohan2024supporting}, \cite{alsaedi2024two},\cite{chakraborty2024blaze} \\ \cline{2-2}
    Others & \cite{xiao2017improvingdeeplocator}, \cite{xiao2018machinebugtranslator}, \cite{xiao2018improving}, \cite{zhu2021trobo}, \cite{chen2022cgmbl}, \cite{zhu2022bl-gan}, \cite{xu2023buglocfront}, \cite{zhang2023enhancing}, \cite{han2024bjenet} \\
  \bottomrule
\end{tabular}}
\end{table*}





\begin{thebibliography}{195}


\ifx \showCODEN    \undefined \def \showCODEN     #1{\unskip}     \fi
\ifx \showDOI      \undefined \def \showDOI       #1{#1}\fi
\ifx \showISBNx    \undefined \def \showISBNx     #1{\unskip}     \fi
\ifx \showISBNxiii \undefined \def \showISBNxiii  #1{\unskip}     \fi
\ifx \showISSN     \undefined \def \showISSN      #1{\unskip}     \fi
\ifx \showLCCN     \undefined \def \showLCCN      #1{\unskip}     \fi
\ifx \shownote     \undefined \def \shownote      #1{#1}          \fi
\ifx \showarticletitle \undefined \def \showarticletitle #1{#1}   \fi
\ifx \showURL      \undefined \def \showURL       {\relax}        \fi
\providecommand\bibfield[2]{#2}
\providecommand\bibinfo[2]{#2}
\providecommand\natexlab[1]{#1}
\providecommand\showeprint[2][]{arXiv:#2}

\bibitem[Abreu et~al\mbox{.}(2007)]%
        {abreu2007accuracy}
\bibfield{author}{\bibinfo{person}{Rui Abreu}, \bibinfo{person}{Peter Zoeteweij}, {and} \bibinfo{person}{Arjan~JC Van~Gemund}.} \bibinfo{year}{2007}\natexlab{}.
\newblock \showarticletitle{On the accuracy of spectrum-based fault localization}. In \bibinfo{booktitle}{\emph{TAICPART-MUTATION 2007}}. IEEE, \bibinfo{pages}{89--98}.
\newblock


\bibitem[Agarwal and Agrawal(2014)]%
        {agarwal2014fault}
\bibfield{author}{\bibinfo{person}{Pragya Agarwal} {and} \bibinfo{person}{Arun~Prakash Agrawal}.} \bibinfo{year}{2014}\natexlab{}.
\newblock \showarticletitle{Fault-localization techniques for software systems: A literature review}.
\newblock \bibinfo{journal}{\emph{ACM SIGSOFT Software Engineering Notes}} \bibinfo{volume}{39}, \bibinfo{number}{5} (\bibinfo{year}{2014}), \bibinfo{pages}{1--8}.
\newblock


\bibitem[Agrawal et~al\mbox{.}(1995)]%
        {agrawal1995fault}
\bibfield{author}{\bibinfo{person}{Hiralal Agrawal}, \bibinfo{person}{Joseph~R Horgan}, \bibinfo{person}{Saul London}, {and} \bibinfo{person}{W~Eric Wong}.} \bibinfo{year}{1995}\natexlab{}.
\newblock \showarticletitle{Fault localization using execution slices and dataflow tests}. In \bibinfo{booktitle}{\emph{Proceedings of Sixth ISSRE}}. IEEE, \bibinfo{pages}{143--151}.
\newblock


\bibitem[Ahmad et~al\mbox{.}(2023)]%
        {ahmad2023attentivebuglocator}
\bibfield{author}{\bibinfo{person}{Aminu~A Ahmad}, \bibinfo{person}{Lasheng Yu}, \bibinfo{person}{Mohamed Kholief}, {and} \bibinfo{person}{Abba Garba}.} \bibinfo{year}{2023}\natexlab{}.
\newblock \showarticletitle{AttentiveBugLocator: A Bug Localization Model using Attention-based SemanticFeatures and Information Retrieval}.
\newblock \bibinfo{journal}{\emph{Research Square}} (\bibinfo{year}{2023}).
\newblock


\bibitem[Akbar and Kak(2020)]%
        {akbar2020large}
\bibfield{author}{\bibinfo{person}{Shayan~A Akbar} {and} \bibinfo{person}{Avinash~C Kak}.} \bibinfo{year}{2020}\natexlab{}.
\newblock \showarticletitle{A large-scale comparative evaluation of IR-based tools for bug localization}. In \bibinfo{booktitle}{\emph{Proceedings of the 17th MSR}}. \bibinfo{pages}{21--31}.
\newblock


\bibitem[Al-Aidaroos and Bamzahem(2023)]%
        {Al-Aidaroosimpact}
\bibfield{author}{\bibinfo{person}{Ahmed~Sheikh Al-Aidaroos} {and} \bibinfo{person}{Sara~Mohammed Bamzahem}.} \bibinfo{year}{2023}\natexlab{}.
\newblock \showarticletitle{The Impact of GloVe and Word2Vec Word-Embedding Technologies on Bug Localization with Convolutional Neural Network}.
\newblock \bibinfo{journal}{\emph{IJSEA}} (\bibinfo{year}{2023}), \bibinfo{pages}{108--111}.
\newblock


\bibitem[Ali et~al\mbox{.}(2024)]%
        {ali2024softwaremsp-coa-idn-lstm}
\bibfield{author}{\bibinfo{person}{Waqas Ali}, \bibinfo{person}{Lili Bo}, \bibinfo{person}{Xiaobing Sun}, \bibinfo{person}{Xiaoxue Wu}, \bibinfo{person}{Aakash Ali}, {and} \bibinfo{person}{Ying Wei}.} \bibinfo{year}{2024}\natexlab{}.
\newblock \showarticletitle{Software bug localization based on optimized and ensembled deep learning models}.
\newblock \bibinfo{journal}{\emph{Journal of Software: Evolution and Process}} \bibinfo{volume}{36}, \bibinfo{number}{8} (\bibinfo{year}{2024}), \bibinfo{pages}{e2654}.
\newblock


\bibitem[Ali et~al\mbox{.}(2023)]%
        {ali2023automatedhgwsfocdnn}
\bibfield{author}{\bibinfo{person}{Waqas Ali}, \bibinfo{person}{Lili Bo}, \bibinfo{person}{Xiaobing Sun}, \bibinfo{person}{Xiaoxue Wu}, \bibinfo{person}{Saifullah Memon}, \bibinfo{person}{Saima Siraj}, {and} \bibinfo{person}{Ann~Suwaree Ashton}.} \bibinfo{year}{2023}\natexlab{}.
\newblock \showarticletitle{Automated Software Bug Localization enabled by Meta-heuristic-based Convolutional Neural Network and Improved Deep Neural Network}.
\newblock \bibinfo{journal}{\emph{Expert Systems with Applications}} (\bibinfo{year}{2023}), \bibinfo{pages}{120562}.
\newblock


\bibitem[Alsaedi et~al\mbox{.}(2024)]%
        {alsaedi2024two}
\bibfield{author}{\bibinfo{person}{Shatha Alsaedi}, \bibinfo{person}{Ahmed~AA Gad-Elrab}, \bibinfo{person}{Amin Noaman}, {and} \bibinfo{person}{Fathy Eassa}.} \bibinfo{year}{2024}\natexlab{}.
\newblock \showarticletitle{Two-Level Information-Retrieval-Based Model for Bug Localization Based on Bug Reports}.
\newblock \bibinfo{journal}{\emph{Electronics}} \bibinfo{volume}{13}, \bibinfo{number}{2} (\bibinfo{year}{2024}), \bibinfo{pages}{321}.
\newblock


\bibitem[Anh and Luyen(2021)]%
        {anh2021imbalanced}
\bibfield{author}{\bibinfo{person}{Bui Thi~Mai Anh} {and} \bibinfo{person}{Nguyen~Viet Luyen}.} \bibinfo{year}{2021}\natexlab{}.
\newblock \showarticletitle{An Imbalanced Deep Learning Model for Bug Localization}. In \bibinfo{booktitle}{\emph{2021 28th APSEC Workshops}}. IEEE, \bibinfo{pages}{32--40}.
\newblock


\bibitem[B.~Le et~al\mbox{.}(2016)]%
        {b2016learning}
\bibfield{author}{\bibinfo{person}{Tien-Duy B.~Le}, \bibinfo{person}{David Lo}, \bibinfo{person}{Claire Le~Goues}, {and} \bibinfo{person}{Lars Grunske}.} \bibinfo{year}{2016}\natexlab{}.
\newblock \showarticletitle{A learning-to-rank based fault localization approach using likely invariants}. In \bibinfo{booktitle}{\emph{25th ISSTA}}. \bibinfo{pages}{177--188}.
\newblock


\bibitem[Bachmann and Bernstein(2009)]%
        {bachmann2009software}
\bibfield{author}{\bibinfo{person}{Adrian Bachmann} {and} \bibinfo{person}{Abraham Bernstein}.} \bibinfo{year}{2009}\natexlab{}.
\newblock \showarticletitle{Software process data quality and characteristics: a historical view on open and closed source projects}. In \bibinfo{booktitle}{\emph{Proceedings of the joint IWPSE and Evol workshops}}. \bibinfo{pages}{119--128}.
\newblock


\bibitem[Bahdanau et~al\mbox{.}(2014)]%
        {bahdanau2014neural}
\bibfield{author}{\bibinfo{person}{Dzmitry Bahdanau}, \bibinfo{person}{Kyunghyun Cho}, {and} \bibinfo{person}{Yoshua Bengio}.} \bibinfo{year}{2014}\natexlab{}.
\newblock \showarticletitle{Neural machine translation by jointly learning to align and translate}.
\newblock \bibinfo{journal}{\emph{arXiv:1409.0473}} (\bibinfo{year}{2014}).
\newblock


\bibitem[Baliyan et~al\mbox{.}(2021)]%
        {baliyan2021multilingual}
\bibfield{author}{\bibinfo{person}{Anupam Baliyan}, \bibinfo{person}{Akshit Batra}, {and} \bibinfo{person}{Sunil~Pratap Singh}.} \bibinfo{year}{2021}\natexlab{}.
\newblock \showarticletitle{Multilingual sentiment analysis using RNN-LSTM and neural machine translation}. In \bibinfo{booktitle}{\emph{2021 8th INDIACom}}. IEEE, \bibinfo{pages}{710--713}.
\newblock


\bibitem[Benton et~al\mbox{.}(2019)]%
        {benton2019defexts}
\bibfield{author}{\bibinfo{person}{Samuel Benton}, \bibinfo{person}{Ali Ghanbari}, {and} \bibinfo{person}{Lingming Zhang}.} \bibinfo{year}{2019}\natexlab{}.
\newblock \showarticletitle{Defexts: A curated dataset of reproducible real-world bugs for modern jvm languages}. In \bibinfo{booktitle}{\emph{2019 IEEE/ACM 41st ICSE-Companion}}. IEEE, \bibinfo{pages}{47--50}.
\newblock


\bibitem[Bettenburg et~al\mbox{.}(2007)]%
        {bettenburg2007quality}
\bibfield{author}{\bibinfo{person}{Nicolas Bettenburg}, \bibinfo{person}{Sascha Just}, \bibinfo{person}{Adrian Schr{\"o}ter}, \bibinfo{person}{Cathrin Wei{\ss}}, \bibinfo{person}{Rahul Premraj}, {and} \bibinfo{person}{Thomas Zimmermann}.} \bibinfo{year}{2007}\natexlab{}.
\newblock \showarticletitle{Quality of bug reports in eclipse}. In \bibinfo{booktitle}{\emph{Proceedings of the 2007 OOPSLA workshop on eclipse technology eXchange}}. \bibinfo{pages}{21--25}.
\newblock


\bibitem[Bettenburg et~al\mbox{.}(2008a)]%
        {bettenburg2008makes}
\bibfield{author}{\bibinfo{person}{Nicolas Bettenburg}, \bibinfo{person}{Sascha Just}, \bibinfo{person}{Adrian Schr{\"o}ter}, \bibinfo{person}{Cathrin Weiss}, \bibinfo{person}{Rahul Premraj}, {and} \bibinfo{person}{Thomas Zimmermann}.} \bibinfo{year}{2008}\natexlab{a}.
\newblock \showarticletitle{What makes a good bug report?}. In \bibinfo{booktitle}{\emph{Proceedings of the 16th ACM SIGSOFT FSE}}. \bibinfo{pages}{308--318}.
\newblock


\bibitem[Bettenburg et~al\mbox{.}(2008b)]%
        {bettenburg2008extracting}
\bibfield{author}{\bibinfo{person}{Nicolas Bettenburg}, \bibinfo{person}{Rahul Premraj}, \bibinfo{person}{Thomas Zimmermann}, {and} \bibinfo{person}{Sunghun Kim}.} \bibinfo{year}{2008}\natexlab{b}.
\newblock \showarticletitle{Extracting structural information from bug reports}. In \bibinfo{booktitle}{\emph{Proceedings of the 2008 MSR}}. \bibinfo{pages}{27--30}.
\newblock


\bibitem[Bhandari et~al\mbox{.}(2023)]%
        {bhandari2023data}
\bibfield{author}{\bibinfo{person}{Kirti Bhandari}, \bibinfo{person}{Kuldeep Kumar}, {and} \bibinfo{person}{Amrit~Lal Sangal}.} \bibinfo{year}{2023}\natexlab{}.
\newblock \showarticletitle{Data quality issues in software fault prediction: a systematic literature review}.
\newblock \bibinfo{journal}{\emph{Artificial Intelligence Review}} \bibinfo{volume}{56}, \bibinfo{number}{8} (\bibinfo{year}{2023}), \bibinfo{pages}{7839--7908}.
\newblock


\bibitem[Brudaru and Zeller(2008)]%
        {10.1145/1454247.1454257}
\bibfield{author}{\bibinfo{person}{Irina~Ioana Brudaru} {and} \bibinfo{person}{Andreas Zeller}.} \bibinfo{year}{2008}\natexlab{}.
\newblock \showarticletitle{What is the Long-Term Impact of Changes?}. In \bibinfo{booktitle}{\emph{RSSE'08}} (Atlanta, Georgia). \bibinfo{publisher}{Association for Computing Machinery}, \bibinfo{address}{New York, NY, USA}, \bibinfo{pages}{30–32}.
\newblock
\showISBNx{9781605582283}
\urldef\tempurl%
\url{https://doi.org/10.1145/1454247.1454257}
\showDOI{\tempurl}


\bibitem[Cao et~al\mbox{.}(2020)]%
        {cao2020bugpecker}
\bibfield{author}{\bibinfo{person}{Junming Cao}, \bibinfo{person}{Shouliang Yang}, \bibinfo{person}{Wenhui Jiang}, \bibinfo{person}{Hushuang Zeng}, \bibinfo{person}{Beijun Shen}, {and} \bibinfo{person}{Hao Zhong}.} \bibinfo{year}{2020}\natexlab{}.
\newblock \showarticletitle{Bugpecker: Locating faulty methods with deep learning on revision graphs}. In \bibinfo{booktitle}{\emph{Proceedings of the 35th IEEE/ACM ASE}}. \bibinfo{pages}{1214--1218}.
\newblock


\bibitem[Catal et~al\mbox{.}(2022)]%
        {catal2022applications}
\bibfield{author}{\bibinfo{person}{Cagatay Catal}, \bibinfo{person}{G{\"o}rkem Giray}, {and} \bibinfo{person}{Bedir Tekinerdogan}.} \bibinfo{year}{2022}\natexlab{}.
\newblock \showarticletitle{Applications of deep learning for mobile malware detection: A systematic literature review}.
\newblock \bibinfo{journal}{\emph{Neural Computing and Applications}} (\bibinfo{year}{2022}), \bibinfo{pages}{1--26}.
\newblock


\bibitem[Chakraborty et~al\mbox{.}(2024a)]%
        {chakraborty2024blaze}
\bibfield{author}{\bibinfo{person}{Partha Chakraborty}, \bibinfo{person}{Mahmoud Alfadel}, {and} \bibinfo{person}{Meiyappan Nagappan}.} \bibinfo{year}{2024}\natexlab{a}.
\newblock \showarticletitle{BLAZE: Cross-Language and Cross-Project Bug Localization via Dynamic Chunking and Hard Example Learning}.
\newblock \bibinfo{journal}{\emph{arXiv:2407.17631}} (\bibinfo{year}{2024}).
\newblock


\bibitem[Chakraborty et~al\mbox{.}(2024b)]%
        {chakraborty2024rlocator}
\bibfield{author}{\bibinfo{person}{Partha Chakraborty}, \bibinfo{person}{Mahmoud Alfadel}, {and} \bibinfo{person}{Meiyappan Nagappan}.} \bibinfo{year}{2024}\natexlab{b}.
\newblock \showarticletitle{RLocator: Reinforcement learning for bug localization}.
\newblock \bibinfo{journal}{\emph{IEEE TSE}} (\bibinfo{year}{2024}).
\newblock


\bibitem[Chandramohan et~al\mbox{.}(2024)]%
        {chandramohan2024supporting}
\bibfield{author}{\bibinfo{person}{Mahinthan Chandramohan}, \bibinfo{person}{Dai~Quoc Nguyen}, \bibinfo{person}{Padmanabhan Krishnan}, {and} \bibinfo{person}{Jovan Jancic}.} \bibinfo{year}{2024}\natexlab{}.
\newblock \showarticletitle{Supporting Cross-language Cross-project Bug Localization Using Pre-trained Language Models}.
\newblock \bibinfo{journal}{\emph{arXiv:2407.02732}} (\bibinfo{year}{2024}).
\newblock


\bibitem[Chawla et~al\mbox{.}(2002)]%
        {chawla2002smote}
\bibfield{author}{\bibinfo{person}{Nitesh~V Chawla}, \bibinfo{person}{Kevin~W Bowyer}, \bibinfo{person}{Lawrence~O Hall}, {and} \bibinfo{person}{W~Philip Kegelmeyer}.} \bibinfo{year}{2002}\natexlab{}.
\newblock \showarticletitle{SMOTE: synthetic minority over-sampling technique}.
\newblock \bibinfo{journal}{\emph{Journal of artificial intelligence research}}  \bibinfo{volume}{16} (\bibinfo{year}{2002}), \bibinfo{pages}{321--357}.
\newblock


\bibitem[Chen et~al\mbox{.}(2022)]%
        {chen2022cgmbl}
\bibfield{author}{\bibinfo{person}{Hao Chen}, \bibinfo{person}{Haiyang Yang}, \bibinfo{person}{Zilun Yan}, \bibinfo{person}{Li Kuang}, {and} \bibinfo{person}{Lingyan Zhang}.} \bibinfo{year}{2022}\natexlab{}.
\newblock \showarticletitle{CGMBL: Combining GAN and Method Name for Bug Localization}. In \bibinfo{booktitle}{\emph{2022 IEEE 22nd QRS}}. IEEE, \bibinfo{pages}{231--241}.
\newblock


\bibitem[Ciborowska and Damevski(2022)]%
        {ciborowska2022fastfbl-bert}
\bibfield{author}{\bibinfo{person}{Agnieszka Ciborowska} {and} \bibinfo{person}{Kostadin Damevski}.} \bibinfo{year}{2022}\natexlab{}.
\newblock \showarticletitle{Fast changeset-based bug localization with BERT}. In \bibinfo{booktitle}{\emph{Proceedings of the 44th ICSE}}. \bibinfo{pages}{946--957}.
\newblock


\bibitem[Ciborowska and Damevski(2023)]%
        {ciborowska2023too}
\bibfield{author}{\bibinfo{person}{Agnieszka Ciborowska} {and} \bibinfo{person}{Kostadin Damevski}.} \bibinfo{year}{2023}\natexlab{}.
\newblock \showarticletitle{Too Few Bug Reports? Exploring Data Augmentation for Improved Changeset-based Bug Localization}.
\newblock \bibinfo{journal}{\emph{arXiv:2305.16430}} (\bibinfo{year}{2023}).
\newblock


\bibitem[Croft et~al\mbox{.}(2022)]%
        {croft2022data}
\bibfield{author}{\bibinfo{person}{Roland Croft}, \bibinfo{person}{Yongzheng Xie}, {and} \bibinfo{person}{Muhammad~Ali Babar}.} \bibinfo{year}{2022}\natexlab{}.
\newblock \showarticletitle{Data preparation for software vulnerability prediction: A systematic literature review}.
\newblock \bibinfo{journal}{\emph{IEEE TSE}} \bibinfo{volume}{49}, \bibinfo{number}{3} (\bibinfo{year}{2022}), \bibinfo{pages}{1044--1063}.
\newblock


\bibitem[Dallmeier and Zimmermann(2007)]%
        {dallmeier2007extraction}
\bibfield{author}{\bibinfo{person}{Valentin Dallmeier} {and} \bibinfo{person}{Thomas Zimmermann}.} \bibinfo{year}{2007}\natexlab{}.
\newblock \showarticletitle{Extraction of bug localization benchmarks from history}. In \bibinfo{booktitle}{\emph{Proceedings of the 22nd IEEE/ACM ASE}}. \bibinfo{pages}{433--436}.
\newblock


\bibitem[Davies and Roper(2013)]%
        {davies2013bug}
\bibfield{author}{\bibinfo{person}{Steven Davies} {and} \bibinfo{person}{Marc Roper}.} \bibinfo{year}{2013}\natexlab{}.
\newblock \showarticletitle{Bug localisation through diverse sources of information}. In \bibinfo{booktitle}{\emph{2013 IEEE ISSREW}}. IEEE, \bibinfo{pages}{126--131}.
\newblock


\bibitem[Davies et~al\mbox{.}(2012)]%
        {davies2012using}
\bibfield{author}{\bibinfo{person}{Steven Davies}, \bibinfo{person}{Marc Roper}, {and} \bibinfo{person}{Murray Wood}.} \bibinfo{year}{2012}\natexlab{}.
\newblock \showarticletitle{Using bug report similarity to enhance bug localisation}. In \bibinfo{booktitle}{\emph{2012 19th Working Conference on Reverse Engineering}}. IEEE, \bibinfo{pages}{125--134}.
\newblock


\bibitem[Devlin et~al\mbox{.}(2018)]%
        {devlin2018bert}
\bibfield{author}{\bibinfo{person}{Jacob Devlin}, \bibinfo{person}{Ming-Wei Chang}, \bibinfo{person}{Kenton Lee}, {and} \bibinfo{person}{Kristina Toutanova}.} \bibinfo{year}{2018}\natexlab{}.
\newblock \showarticletitle{Bert: Pre-training of deep bidirectional transformers for language understanding}.
\newblock \bibinfo{journal}{\emph{arXiv:1810.04805}} (\bibinfo{year}{2018}).
\newblock


\bibitem[Douzas et~al\mbox{.}(2018)]%
        {douzas2018improving}
\bibfield{author}{\bibinfo{person}{Georgios Douzas}, \bibinfo{person}{Fernando Bacao}, {and} \bibinfo{person}{Felix Last}.} \bibinfo{year}{2018}\natexlab{}.
\newblock \showarticletitle{Improving imbalanced learning through a heuristic oversampling method based on k-means and SMOTE}.
\newblock \bibinfo{journal}{\emph{Information Sciences}}  \bibinfo{volume}{465} (\bibinfo{year}{2018}), \bibinfo{pages}{1--20}.
\newblock


\bibitem[Du and Yu(2023)]%
        {du2023prehmcbl}
\bibfield{author}{\bibinfo{person}{Yali Du} {and} \bibinfo{person}{Zhongxing Yu}.} \bibinfo{year}{2023}\natexlab{}.
\newblock \showarticletitle{Pre-training Code Representation with Semantic Flow Graph for Effective Bug Localization}. In \bibinfo{booktitle}{\emph{Proceedings of the 31st ACM FSE}}. \bibinfo{pages}{579--591}.
\newblock


\bibitem[Feng et~al\mbox{.}(2020)]%
        {feng2020codebert}
\bibfield{author}{\bibinfo{person}{Zhangyin Feng}, \bibinfo{person}{Daya Guo}, \bibinfo{person}{Duyu Tang}, \bibinfo{person}{Nan Duan}, \bibinfo{person}{Xiaocheng Feng}, \bibinfo{person}{Ming Gong}, \bibinfo{person}{Linjun Shou}, \bibinfo{person}{Bing Qin}, \bibinfo{person}{Ting Liu}, \bibinfo{person}{Daxin Jiang}, {et~al\mbox{.}}} \bibinfo{year}{2020}\natexlab{}.
\newblock \showarticletitle{Codebert: A pre-trained model for programming and natural languages}.
\newblock \bibinfo{journal}{\emph{arXiv:2002.08155}} (\bibinfo{year}{2020}).
\newblock


\bibitem[Font and Costa-Jussa(2019)]%
        {font2019equalizing}
\bibfield{author}{\bibinfo{person}{Joel~Escud{\'e} Font} {and} \bibinfo{person}{Marta~R Costa-Jussa}.} \bibinfo{year}{2019}\natexlab{}.
\newblock \showarticletitle{Equalizing gender biases in neural machine translation with word embeddings techniques}.
\newblock \bibinfo{journal}{\emph{arXiv:1901.03116}} (\bibinfo{year}{2019}).
\newblock


\bibitem[Giray et~al\mbox{.}(2023)]%
        {giray2023use}
\bibfield{author}{\bibinfo{person}{G{\"o}rkem Giray}, \bibinfo{person}{Kwabena~Ebo Bennin}, \bibinfo{person}{{\"O}mer K{\"o}ksal}, \bibinfo{person}{{\"O}nder Babur}, {and} \bibinfo{person}{Bedir Tekinerdogan}.} \bibinfo{year}{2023}\natexlab{}.
\newblock \showarticletitle{On the use of deep learning in software defect prediction}.
\newblock \bibinfo{journal}{\emph{Journal of Systems and Software}}  \bibinfo{volume}{195} (\bibinfo{year}{2023}), \bibinfo{pages}{111537}.
\newblock


\bibitem[Guo et~al\mbox{.}(2022)]%
        {guo2022unixcoder}
\bibfield{author}{\bibinfo{person}{Daya Guo}, \bibinfo{person}{Shuai Lu}, \bibinfo{person}{Nan Duan}, \bibinfo{person}{Yanlin Wang}, \bibinfo{person}{Ming Zhou}, {and} \bibinfo{person}{Jian Yin}.} \bibinfo{year}{2022}\natexlab{}.
\newblock \showarticletitle{Unixcoder: Unified cross-modal pre-training for code representation}.
\newblock \bibinfo{journal}{\emph{arXiv:2203.03850}} (\bibinfo{year}{2022}).
\newblock


\bibitem[Guo et~al\mbox{.}(2020)]%
        {guo2020graphcodebert}
\bibfield{author}{\bibinfo{person}{Daya Guo}, \bibinfo{person}{Shuo Ren}, \bibinfo{person}{Shuai Lu}, \bibinfo{person}{Zhangyin Feng}, \bibinfo{person}{Duyu Tang}, \bibinfo{person}{Shujie Liu}, \bibinfo{person}{Long Zhou}, \bibinfo{person}{Nan Duan}, \bibinfo{person}{Alexey Svyatkovskiy}, \bibinfo{person}{Shengyu Fu}, {et~al\mbox{.}}} \bibinfo{year}{2020}\natexlab{}.
\newblock \showarticletitle{Graphcodebert: Pre-training code representations with data flow}.
\newblock \bibinfo{journal}{\emph{arXiv:2009.08366}} (\bibinfo{year}{2020}).
\newblock


\bibitem[Hall et~al\mbox{.}(2011)]%
        {hall2011systematic}
\bibfield{author}{\bibinfo{person}{Tracy Hall}, \bibinfo{person}{Sarah Beecham}, \bibinfo{person}{David Bowes}, \bibinfo{person}{David Gray}, {and} \bibinfo{person}{Steve Counsell}.} \bibinfo{year}{2011}\natexlab{}.
\newblock \showarticletitle{A systematic literature review on fault prediction performance in software engineering}.
\newblock \bibinfo{journal}{\emph{IEEE TSE}} \bibinfo{volume}{38}, \bibinfo{number}{6} (\bibinfo{year}{2011}), \bibinfo{pages}{1276--1304}.
\newblock


\bibitem[Han et~al\mbox{.}(2024)]%
        {han2024bjenet}
\bibfield{author}{\bibinfo{person}{Jiaxuan Han}, \bibinfo{person}{Cheng Huang}, {and} \bibinfo{person}{Jiayong Liu}.} \bibinfo{year}{2024}\natexlab{}.
\newblock \showarticletitle{bjEnet: a fast and accurate software bug localization method in natural language semantic space}.
\newblock \bibinfo{journal}{\emph{Software Quality Journal}} (\bibinfo{year}{2024}), \bibinfo{pages}{1--24}.
\newblock


\bibitem[Han et~al\mbox{.}(2023)]%
        {han2023bjxnet}
\bibfield{author}{\bibinfo{person}{Jiaxuan Han}, \bibinfo{person}{Cheng Huang}, \bibinfo{person}{Siqi Sun}, \bibinfo{person}{Zhonglin Liu}, {and} \bibinfo{person}{Jiayong Liu}.} \bibinfo{year}{2023}\natexlab{}.
\newblock \showarticletitle{bjXnet: an improved bug localization model based on code property graph and attention mechanism}.
\newblock \bibinfo{journal}{\emph{Automated Software Engineering}} \bibinfo{volume}{30}, \bibinfo{number}{1} (\bibinfo{year}{2023}), \bibinfo{pages}{12}.
\newblock


\bibitem[Harrold et~al\mbox{.}(2000)]%
        {harrold2000empirical}
\bibfield{author}{\bibinfo{person}{Mary~Jean Harrold}, \bibinfo{person}{Gregg Rothermel}, \bibinfo{person}{Kent Sayre}, \bibinfo{person}{Rui Wu}, {and} \bibinfo{person}{Liu Yi}.} \bibinfo{year}{2000}\natexlab{}.
\newblock \showarticletitle{An empirical investigation of the relationship between spectra differences and regression faults}.
\newblock \bibinfo{journal}{\emph{Software Testing, Verification and Reliability}} \bibinfo{volume}{10}, \bibinfo{number}{3} (\bibinfo{year}{2000}), \bibinfo{pages}{171--194}.
\newblock


\bibitem[He et~al\mbox{.}(2008)]%
        {he2008adasyn}
\bibfield{author}{\bibinfo{person}{Haibo He}, \bibinfo{person}{Yang Bai}, \bibinfo{person}{Edwardo~A Garcia}, {and} \bibinfo{person}{Shutao Li}.} \bibinfo{year}{2008}\natexlab{}.
\newblock \showarticletitle{ADASYN: Adaptive synthetic sampling approach for imbalanced learning}. In \bibinfo{booktitle}{\emph{2008 IEEE international joint conference on neural networks}}. Ieee, \bibinfo{pages}{1322--1328}.
\newblock


\bibitem[Herzig et~al\mbox{.}(2013)]%
        {herzig2013s}
\bibfield{author}{\bibinfo{person}{Kim Herzig}, \bibinfo{person}{Sascha Just}, {and} \bibinfo{person}{Andreas Zeller}.} \bibinfo{year}{2013}\natexlab{}.
\newblock \showarticletitle{It's not a bug, it's a feature: how misclassification impacts bug prediction}. In \bibinfo{booktitle}{\emph{2013 35th ICSE}}. IEEE, \bibinfo{pages}{392--401}.
\newblock


\bibitem[Herzig and Zeller(2013)]%
        {herzig2013impact}
\bibfield{author}{\bibinfo{person}{Kim Herzig} {and} \bibinfo{person}{Andreas Zeller}.} \bibinfo{year}{2013}\natexlab{}.
\newblock \showarticletitle{The impact of tangled code changes}. In \bibinfo{booktitle}{\emph{2013 10th MSR}}. IEEE, \bibinfo{pages}{121--130}.
\newblock


\bibitem[Hochreiter(1997)]%
        {hochreiter1997long}
\bibfield{author}{\bibinfo{person}{S Hochreiter}.} \bibinfo{year}{1997}\natexlab{}.
\newblock \showarticletitle{Long Short-term Memory}.
\newblock \bibinfo{journal}{\emph{Neural Computation MIT-Press}} (\bibinfo{year}{1997}).
\newblock


\bibitem[Hosseini et~al\mbox{.}(2017)]%
        {hosseini2017systematic}
\bibfield{author}{\bibinfo{person}{Seyedrebvar Hosseini}, \bibinfo{person}{Burak Turhan}, {and} \bibinfo{person}{Dimuthu Gunarathna}.} \bibinfo{year}{2017}\natexlab{}.
\newblock \showarticletitle{A systematic literature review and meta-analysis on cross project defect prediction}.
\newblock \bibinfo{journal}{\emph{IEEE TSE}} \bibinfo{volume}{45}, \bibinfo{number}{2} (\bibinfo{year}{2017}), \bibinfo{pages}{111--147}.
\newblock


\bibitem[Huang et~al\mbox{.}(2022)]%
        {huang2022sbuglocater}
\bibfield{author}{\bibinfo{person}{Xuxiang Huang}, \bibinfo{person}{Chen Xiang}, \bibinfo{person}{Hua Li}, {and} \bibinfo{person}{Peng He}.} \bibinfo{year}{2022}\natexlab{}.
\newblock \showarticletitle{SBugLocater: Bug Localization Based on Deep Matching and Information Retrieval}.
\newblock \bibinfo{journal}{\emph{Mathematical Problems in Engineering}}  \bibinfo{volume}{2022} (\bibinfo{year}{2022}).
\newblock


\bibitem[Hunt et~al\mbox{.}(2008)]%
        {4526688}
\bibfield{author}{\bibinfo{person}{Bob Hunt}, \bibinfo{person}{Bryn Turner}, {and} \bibinfo{person}{Karen McRitchie}.} \bibinfo{year}{2008}\natexlab{}.
\newblock \showarticletitle{Software Maintenance Implications on Cost and Schedule}. In \bibinfo{booktitle}{\emph{2008 IEEE Aerospace Conference}}. \bibinfo{pages}{1--6}.
\newblock
\urldef\tempurl%
\url{https://doi.org/10.1109/AERO.2008.4526688}
\showDOI{\tempurl}


\bibitem[Huo and Li(2017)]%
        {huo2017enhancingls-cnn}
\bibfield{author}{\bibinfo{person}{Xuan Huo} {and} \bibinfo{person}{Ming Li}.} \bibinfo{year}{2017}\natexlab{}.
\newblock \showarticletitle{Enhancing the Unified Features to Locate Buggy Files by Exploiting the Sequential Nature of Source Code.}. In \bibinfo{booktitle}{\emph{IJCAI}}. \bibinfo{pages}{1909--1915}.
\newblock


\bibitem[Huo et~al\mbox{.}(2020)]%
        {huo2020controlcg-cnn}
\bibfield{author}{\bibinfo{person}{Xuan Huo}, \bibinfo{person}{Ming Li}, {and} \bibinfo{person}{Zhi-Hua Zhou}.} \bibinfo{year}{2020}\natexlab{}.
\newblock \showarticletitle{Control flow graph embedding based on multi-instance decomposition for bug localization}. In \bibinfo{booktitle}{\emph{AAAI}}, Vol.~\bibinfo{volume}{34}. \bibinfo{pages}{4223--4230}.
\newblock


\bibitem[Huo et~al\mbox{.}(2016)]%
        {huo2016learningnp-cnn}
\bibfield{author}{\bibinfo{person}{Xuan Huo}, \bibinfo{person}{Ming Li}, \bibinfo{person}{Zhi-Hua Zhou}, {et~al\mbox{.}}} \bibinfo{year}{2016}\natexlab{}.
\newblock \showarticletitle{Learning unified features from natural and programming languages for locating buggy source code.}. In \bibinfo{booktitle}{\emph{IJCAI}}, Vol.~\bibinfo{volume}{16}. \bibinfo{pages}{1606--1612}.
\newblock


\bibitem[Huo et~al\mbox{.}(2019)]%
        {huo2019deeptranp}
\bibfield{author}{\bibinfo{person}{Xuan Huo}, \bibinfo{person}{Ferdian Thung}, \bibinfo{person}{Ming Li}, \bibinfo{person}{David Lo}, {and} \bibinfo{person}{Shu-Ting Shi}.} \bibinfo{year}{2019}\natexlab{}.
\newblock \showarticletitle{Deep transfer bug localization}.
\newblock \bibinfo{journal}{\emph{IEEE TSE}} \bibinfo{volume}{47}, \bibinfo{number}{7} (\bibinfo{year}{2019}), \bibinfo{pages}{1368--1380}.
\newblock


\bibitem[Iqbal et~al\mbox{.}(2020)]%
        {iqbal2020determining}
\bibfield{author}{\bibinfo{person}{Shahid Iqbal}, \bibinfo{person}{Rashid Naseem}, \bibinfo{person}{Salman Jan}, \bibinfo{person}{Sami Alshmrany}, \bibinfo{person}{Muhammad Yasar}, {and} \bibinfo{person}{Arshad Ali}.} \bibinfo{year}{2020}\natexlab{}.
\newblock \showarticletitle{Determining bug prioritization using feature reduction and clustering with classification}.
\newblock \bibinfo{journal}{\emph{IEEE Access}}  \bibinfo{volume}{8} (\bibinfo{year}{2020}), \bibinfo{pages}{215661--215678}.
\newblock


\bibitem[Jarman et~al\mbox{.}(2021)]%
        {jarman2021legion}
\bibfield{author}{\bibinfo{person}{Darryl Jarman}, \bibinfo{person}{Jeffrey Berry}, \bibinfo{person}{Riley Smith}, \bibinfo{person}{Ferdian Thung}, {and} \bibinfo{person}{David Lo}.} \bibinfo{year}{2021}\natexlab{}.
\newblock \showarticletitle{Legion: Massively composing rankers for improved bug localization at adobe}.
\newblock \bibinfo{journal}{\emph{IEEE TSE}} \bibinfo{volume}{48}, \bibinfo{number}{8} (\bibinfo{year}{2021}), \bibinfo{pages}{3010--3024}.
\newblock


\bibitem[Jiang et~al\mbox{.}(2020)]%
        {jiang2020deepdbl}
\bibfield{author}{\bibinfo{person}{Bo Jiang}, \bibinfo{person}{Pengfei Liu}, {and} \bibinfo{person}{Jie Xu}.} \bibinfo{year}{2020}\natexlab{}.
\newblock \showarticletitle{A deep learning approach to locate buggy files}. In \bibinfo{booktitle}{\emph{11th DESSERT}}. IEEE.
\newblock


\bibitem[Jimenez et~al\mbox{.}(2023)]%
        {jimenez2023swe}
\bibfield{author}{\bibinfo{person}{Carlos~E Jimenez}, \bibinfo{person}{John Yang}, \bibinfo{person}{Alexander Wettig}, \bibinfo{person}{Shunyu Yao}, \bibinfo{person}{Kexin Pei}, \bibinfo{person}{Ofir Press}, {and} \bibinfo{person}{Karthik Narasimhan}.} \bibinfo{year}{2023}\natexlab{}.
\newblock \showarticletitle{Swe-bench: Can language models resolve real-world github issues?}
\newblock \bibinfo{journal}{\emph{arXiv:2310.06770}} (\bibinfo{year}{2023}).
\newblock


\bibitem[Just et~al\mbox{.}(2014)]%
        {just2014defects4j}
\bibfield{author}{\bibinfo{person}{Ren{\'e} Just}, \bibinfo{person}{Darioush Jalali}, {and} \bibinfo{person}{Michael~D Ernst}.} \bibinfo{year}{2014}\natexlab{}.
\newblock \showarticletitle{Defects4J: A database of existing faults to enable controlled testing studies for Java programs}. In \bibinfo{booktitle}{\emph{Proceedings of the 2014 ISSTA}}. \bibinfo{pages}{437--440}.
\newblock


\bibitem[Kalchbrenner et~al\mbox{.}(2014)]%
        {kalchbrenner2014convolutional}
\bibfield{author}{\bibinfo{person}{Nal Kalchbrenner}, \bibinfo{person}{Edward Grefenstette}, {and} \bibinfo{person}{Phil Blunsom}.} \bibinfo{year}{2014}\natexlab{}.
\newblock \showarticletitle{A convolutional neural network for modelling sentences}.
\newblock \bibinfo{journal}{\emph{arXiv:1404.2188}} (\bibinfo{year}{2014}).
\newblock


\bibitem[Keele et~al\mbox{.}(2007)]%
        {keele2007guidelines}
\bibfield{author}{\bibinfo{person}{Staffs Keele} {et~al\mbox{.}}} \bibinfo{year}{2007}\natexlab{}.
\newblock \bibinfo{title}{Guidelines for performing systematic literature reviews in software engineering}.
\newblock
\newblock


\bibitem[Kim et~al\mbox{.}(2007)]%
        {kim2007predicting}
\bibfield{author}{\bibinfo{person}{Sunghun Kim}, \bibinfo{person}{Thomas Zimmermann}, \bibinfo{person}{E~James Whitehead~Jr}, {and} \bibinfo{person}{Andreas Zeller}.} \bibinfo{year}{2007}\natexlab{}.
\newblock \showarticletitle{Predicting faults from cached history}. In \bibinfo{booktitle}{\emph{29th ICSE}}. IEEE, \bibinfo{pages}{489--498}.
\newblock


\bibitem[Kipf and Welling(2016)]%
        {kipf2016semi}
\bibfield{author}{\bibinfo{person}{Thomas~N Kipf} {and} \bibinfo{person}{Max Welling}.} \bibinfo{year}{2016}\natexlab{}.
\newblock \showarticletitle{Semi-supervised classification with graph convolutional networks}.
\newblock \bibinfo{journal}{\emph{arXiv:1609.02907}} (\bibinfo{year}{2016}).
\newblock


\bibitem[Kitchenham(2004)]%
        {kitchenham2004procedures}
\bibfield{author}{\bibinfo{person}{Barbara Kitchenham}.} \bibinfo{year}{2004}\natexlab{}.
\newblock \showarticletitle{Procedures for performing systematic reviews}.
\newblock \bibinfo{journal}{\emph{Keele, UK, Keele University}}  \bibinfo{volume}{33} (\bibinfo{year}{2004}), \bibinfo{pages}{1--26}.
\newblock


\bibitem[Kochhar et~al\mbox{.}(2014a)]%
        {kochhar2014s}
\bibfield{author}{\bibinfo{person}{Pavneet~Singh Kochhar}, \bibinfo{person}{Tien-Duy~B Le}, {and} \bibinfo{person}{David Lo}.} \bibinfo{year}{2014}\natexlab{a}.
\newblock \showarticletitle{It's not a bug, it's a feature: Does misclassification affect bug localization?}. In \bibinfo{booktitle}{\emph{Proceedings of the 11th MSR}}. \bibinfo{pages}{296--299}.
\newblock


\bibitem[Kochhar et~al\mbox{.}(2014b)]%
        {kochhar2014potential}
\bibfield{author}{\bibinfo{person}{Pavneet~Singh Kochhar}, \bibinfo{person}{Yuan Tian}, {and} \bibinfo{person}{David Lo}.} \bibinfo{year}{2014}\natexlab{b}.
\newblock \showarticletitle{Potential biases in bug localization: Do they matter?}. In \bibinfo{booktitle}{\emph{Proceedings of the 29th ACM/IEEE ASE}}. \bibinfo{pages}{803--814}.
\newblock


\bibitem[Kochhar et~al\mbox{.}(2016)]%
        {kochhar2016practitioners}
\bibfield{author}{\bibinfo{person}{Pavneet~Singh Kochhar}, \bibinfo{person}{Xin Xia}, \bibinfo{person}{David Lo}, {and} \bibinfo{person}{Shanping Li}.} \bibinfo{year}{2016}\natexlab{}.
\newblock \showarticletitle{Practitioners' expectations on automated fault localization}. In \bibinfo{booktitle}{\emph{Proceedings of the 25th ISSTA}}. \bibinfo{pages}{165--176}.
\newblock


\bibitem[Kotsiantis et~al\mbox{.}(2006)]%
        {kotsiantis2006handling}
\bibfield{author}{\bibinfo{person}{Sotiris Kotsiantis}, \bibinfo{person}{Dimitris Kanellopoulos}, \bibinfo{person}{Panayiotis Pintelas}, {et~al\mbox{.}}} \bibinfo{year}{2006}\natexlab{}.
\newblock \showarticletitle{Handling imbalanced datasets: A review}.
\newblock \bibinfo{journal}{\emph{GESTS international transactions on computer science and engineering}} \bibinfo{volume}{30}, \bibinfo{number}{1} (\bibinfo{year}{2006}), \bibinfo{pages}{25--36}.
\newblock


\bibitem[Kuhn et~al\mbox{.}(2007)]%
        {kuhn2007semantic}
\bibfield{author}{\bibinfo{person}{Adrian Kuhn}, \bibinfo{person}{St{\'e}phane Ducasse}, {and} \bibinfo{person}{Tudor G{\^\i}rba}.} \bibinfo{year}{2007}\natexlab{}.
\newblock \showarticletitle{Semantic clustering: Identifying topics in source code}.
\newblock \bibinfo{journal}{\emph{IST}} \bibinfo{volume}{49}, \bibinfo{number}{3} (\bibinfo{year}{2007}), \bibinfo{pages}{230--243}.
\newblock


\bibitem[Lam et~al\mbox{.}(2015)]%
        {lam2015combining}
\bibfield{author}{\bibinfo{person}{An~Ngoc Lam}, \bibinfo{person}{Anh~Tuan Nguyen}, \bibinfo{person}{Hoan~Anh Nguyen}, {and} \bibinfo{person}{Tien~N Nguyen}.} \bibinfo{year}{2015}\natexlab{}.
\newblock \showarticletitle{Combining deep learning with information retrieval to localize buggy files for bug reports (n)}. In \bibinfo{booktitle}{\emph{2015 30th IEEE/ACM ASE}}. IEEE, \bibinfo{pages}{476--481}.
\newblock


\bibitem[Lam et~al\mbox{.}(2017)]%
        {lam2017bugdnnloc}
\bibfield{author}{\bibinfo{person}{An~Ngoc Lam}, \bibinfo{person}{Anh~Tuan Nguyen}, \bibinfo{person}{Hoan~Anh Nguyen}, {and} \bibinfo{person}{Tien~N Nguyen}.} \bibinfo{year}{2017}\natexlab{}.
\newblock \showarticletitle{Bug localization with combination of deep learning and information retrieval}. In \bibinfo{booktitle}{\emph{2017 IEEE/ACM 25th ICPC}}. IEEE, \bibinfo{pages}{218--229}.
\newblock


\bibitem[Lan et~al\mbox{.}(2019)]%
        {lan2019albert}
\bibfield{author}{\bibinfo{person}{Zhenzhong Lan}, \bibinfo{person}{Mingda Chen}, \bibinfo{person}{Sebastian Goodman}, \bibinfo{person}{Kevin Gimpel}, \bibinfo{person}{Piyush Sharma}, {and} \bibinfo{person}{Radu Soricut}.} \bibinfo{year}{2019}\natexlab{}.
\newblock \showarticletitle{Albert: A lite bert for self-supervised learning of language representations}.
\newblock \bibinfo{journal}{\emph{arXiv:1909.11942}} (\bibinfo{year}{2019}).
\newblock


\bibitem[LeCun et~al\mbox{.}(2015)]%
        {lecun2015deep}
\bibfield{author}{\bibinfo{person}{Yann LeCun}, \bibinfo{person}{Yoshua Bengio}, {and} \bibinfo{person}{Geoffrey Hinton}.} \bibinfo{year}{2015}\natexlab{}.
\newblock \showarticletitle{Deep learning}.
\newblock \bibinfo{journal}{\emph{nature}} \bibinfo{volume}{521}, \bibinfo{number}{7553} (\bibinfo{year}{2015}), \bibinfo{pages}{436--444}.
\newblock


\bibitem[LeCun et~al\mbox{.}(1998)]%
        {lecun1998gradient}
\bibfield{author}{\bibinfo{person}{Yann LeCun}, \bibinfo{person}{L{\'e}on Bottou}, \bibinfo{person}{Yoshua Bengio}, {and} \bibinfo{person}{Patrick Haffner}.} \bibinfo{year}{1998}\natexlab{}.
\newblock \showarticletitle{Gradient-based learning applied to document recognition}.
\newblock \bibinfo{journal}{\emph{Proc. IEEE}} \bibinfo{volume}{86}, \bibinfo{number}{11} (\bibinfo{year}{1998}), \bibinfo{pages}{2278--2324}.
\newblock


\bibitem[Lee et~al\mbox{.}(2018)]%
        {lee2018bench4bl}
\bibfield{author}{\bibinfo{person}{Jaekwon Lee}, \bibinfo{person}{Dongsun Kim}, \bibinfo{person}{Tegawend{\'e}~F Bissyand{\'e}}, \bibinfo{person}{Woosung Jung}, {and} \bibinfo{person}{Yves Le~Traon}.} \bibinfo{year}{2018}\natexlab{}.
\newblock \showarticletitle{Bench4bl: reproducibility study on the performance of ir-based bug localization}. In \bibinfo{booktitle}{\emph{27th ISSTA}}. \bibinfo{pages}{61--72}.
\newblock


\bibitem[Lewis et~al\mbox{.}(2013)]%
        {lewis2013does}
\bibfield{author}{\bibinfo{person}{Chris Lewis}, \bibinfo{person}{Zhongpeng Lin}, \bibinfo{person}{Caitlin Sadowski}, \bibinfo{person}{Xiaoyan Zhu}, \bibinfo{person}{Rong Ou}, {and} \bibinfo{person}{E~James Whitehead}.} \bibinfo{year}{2013}\natexlab{}.
\newblock \showarticletitle{Does bug prediction support human developers? findings from a google case study}. In \bibinfo{booktitle}{\emph{2013 35th ICSE}}. IEEE, \bibinfo{pages}{372--381}.
\newblock


\bibitem[Li et~al\mbox{.}(2022)]%
        {li2022empirical}
\bibfield{author}{\bibinfo{person}{Wei Li}, \bibinfo{person}{Qingan Li}, \bibinfo{person}{Yunlong Ming}, \bibinfo{person}{Weijiao Dai}, \bibinfo{person}{Shi Ying}, {and} \bibinfo{person}{Mengting Yuan}.} \bibinfo{year}{2022}\natexlab{}.
\newblock \showarticletitle{An empirical study of the effectiveness of IR-based bug localization for large-scale industrial projects}.
\newblock \bibinfo{journal}{\emph{EMSE}} \bibinfo{volume}{27}, \bibinfo{number}{2} (\bibinfo{year}{2022}), \bibinfo{pages}{47}.
\newblock


\bibitem[Li et~al\mbox{.}(2019)]%
        {li2019deepfl}
\bibfield{author}{\bibinfo{person}{Xia Li}, \bibinfo{person}{Wei Li}, \bibinfo{person}{Yuqun Zhang}, {and} \bibinfo{person}{Lingming Zhang}.} \bibinfo{year}{2019}\natexlab{}.
\newblock \showarticletitle{Deepfl: Integrating multiple fault diagnosis dimensions for deep fault localization}. In \bibinfo{booktitle}{\emph{28th ISSTA}}. \bibinfo{pages}{169--180}.
\newblock


\bibitem[Li et~al\mbox{.}(2015)]%
        {li2015gated}
\bibfield{author}{\bibinfo{person}{Yujia Li}, \bibinfo{person}{Daniel Tarlow}, \bibinfo{person}{Marc Brockschmidt}, {and} \bibinfo{person}{Richard Zemel}.} \bibinfo{year}{2015}\natexlab{}.
\newblock \showarticletitle{Gated graph sequence neural networks}.
\newblock \bibinfo{journal}{\emph{arXiv:1511.05493}} (\bibinfo{year}{2015}).
\newblock


\bibitem[Liang et~al\mbox{.}(2022)]%
        {liang2022modelingflim}
\bibfield{author}{\bibinfo{person}{Hongliang Liang}, \bibinfo{person}{Dengji Hang}, {and} \bibinfo{person}{Xiangyu Li}.} \bibinfo{year}{2022}\natexlab{}.
\newblock \showarticletitle{Modeling function-level interactions for file-level bug localization}.
\newblock \bibinfo{journal}{\emph{EMSE}} \bibinfo{volume}{27}, \bibinfo{number}{7} (\bibinfo{year}{2022}), \bibinfo{pages}{186}.
\newblock


\bibitem[Liang et~al\mbox{.}(2019)]%
        {liang2019deepcast}
\bibfield{author}{\bibinfo{person}{Hongliang Liang}, \bibinfo{person}{Lu Sun}, \bibinfo{person}{Meilin Wang}, {and} \bibinfo{person}{Yuxing Yang}.} \bibinfo{year}{2019}\natexlab{}.
\newblock \showarticletitle{Deep learning with customized abstract syntax tree for bug localization}.
\newblock \bibinfo{journal}{\emph{IEEE Access}}  \bibinfo{volume}{7} (\bibinfo{year}{2019}), \bibinfo{pages}{116309--116320}.
\newblock


\bibitem[Liu et~al\mbox{.}(2019)]%
        {liu2019convolutionalsls-cnn}
\bibfield{author}{\bibinfo{person}{Guangliang Liu}, \bibinfo{person}{Yang Lu}, \bibinfo{person}{Ke Shi}, \bibinfo{person}{Jingfei Chang}, {and} \bibinfo{person}{Xing Wei}.} \bibinfo{year}{2019}\natexlab{}.
\newblock \showarticletitle{Convolutional neural networks-based locating relevant buggy code files for bug reports affected by data imbalance}.
\newblock \bibinfo{journal}{\emph{IEEE Access}}  \bibinfo{volume}{7} (\bibinfo{year}{2019}), \bibinfo{pages}{131304--131316}.
\newblock


\bibitem[Loyola et~al\mbox{.}(2018)]%
        {loyola2018bug}
\bibfield{author}{\bibinfo{person}{Pablo Loyola}, \bibinfo{person}{Kugamoorthy Gajananan}, {and} \bibinfo{person}{Fumiko Satoh}.} \bibinfo{year}{2018}\natexlab{}.
\newblock \showarticletitle{Bug localization by learning to rank and represent bug inducing changes}. In \bibinfo{booktitle}{\emph{Proceedings of the 27th ACM CIKM}}. \bibinfo{pages}{657--665}.
\newblock


\bibitem[Lukins et~al\mbox{.}(2008)]%
        {lukins2008source}
\bibfield{author}{\bibinfo{person}{Stacy~K Lukins}, \bibinfo{person}{Nicholas~A Kraft}, {and} \bibinfo{person}{Letha~H Etzkorn}.} \bibinfo{year}{2008}\natexlab{}.
\newblock \showarticletitle{Source code retrieval for bug localization using latent dirichlet allocation}. In \bibinfo{booktitle}{\emph{2008 15Th working conference on reverse engineering}}. IEEE, \bibinfo{pages}{155--164}.
\newblock


\bibitem[Luo et~al\mbox{.}(2022)]%
        {luo2022improvingcoloc}
\bibfield{author}{\bibinfo{person}{Zhengmao Luo}, \bibinfo{person}{Wenyao Wang}, {and} \bibinfo{person}{Caichun Cen}.} \bibinfo{year}{2022}\natexlab{}.
\newblock \showarticletitle{Improving Bug Localization With Effective Contrastive Learning Representation}.
\newblock \bibinfo{journal}{\emph{IEEE Access}}  \bibinfo{volume}{11} (\bibinfo{year}{2022}), \bibinfo{pages}{32523--32533}.
\newblock


\bibitem[Ma et~al\mbox{.}(2023)]%
        {ma2023capturingsgattention}
\bibfield{author}{\bibinfo{person}{Yi-Fan Ma}, \bibinfo{person}{Yali Du}, {and} \bibinfo{person}{Ming Li}.} \bibinfo{year}{2023}\natexlab{}.
\newblock \showarticletitle{Capturing the long-distance dependency in the control flow graph via structural-guided attention for bug localization}. In \bibinfo{booktitle}{\emph{Proceedings of the Thirty-Second International Joint Conference on Artificial Intelligence, IJCAI}}.
\newblock


\bibitem[Ma and Li(2022a)]%
        {ma2022flowingcflow}
\bibfield{author}{\bibinfo{person}{Yi-Fan Ma} {and} \bibinfo{person}{Ming Li}.} \bibinfo{year}{2022}\natexlab{a}.
\newblock \showarticletitle{The flowing nature matters: feature learning from the control flow graph of source code for bug localization}.
\newblock \bibinfo{journal}{\emph{Machine Learning}} \bibinfo{volume}{111}, \bibinfo{number}{3} (\bibinfo{year}{2022}), \bibinfo{pages}{853--870}.
\newblock


\bibitem[Ma and Li(2022b)]%
        {ma2022learningmla}
\bibfield{author}{\bibinfo{person}{Yi-Fan Ma} {and} \bibinfo{person}{Ming Li}.} \bibinfo{year}{2022}\natexlab{b}.
\newblock \showarticletitle{Learning from the Multi-Level Abstraction of the Control Flow Graph via Alternating Propagation for Bug Localization}. In \bibinfo{booktitle}{\emph{2022 IEEE ICDM}}. IEEE, \bibinfo{pages}{299--308}.
\newblock


\bibitem[Manning et~al\mbox{.}(2008)]%
        {manning2008introduction}
\bibfield{author}{\bibinfo{person}{Christopher~D Manning}, \bibinfo{person}{Prabhakar Raghavan}, {and} \bibinfo{person}{Hinrich Sch{\"u}tze}.} \bibinfo{year}{2008}\natexlab{}.
\newblock \bibinfo{booktitle}{\emph{Introduction to information retrieval}}.
\newblock \bibinfo{publisher}{Cambridge university press}.
\newblock


\bibitem[Marcus et~al\mbox{.}(2004)]%
        {marcus2004information}
\bibfield{author}{\bibinfo{person}{Andrian Marcus}, \bibinfo{person}{Andrey Sergeyev}, \bibinfo{person}{Vaclav Rajlich}, {and} \bibinfo{person}{Jonathan~I Maletic}.} \bibinfo{year}{2004}\natexlab{}.
\newblock \showarticletitle{An information retrieval approach to concept location in source code}. In \bibinfo{booktitle}{\emph{11th working conference on reverse engineering}}. IEEE, \bibinfo{pages}{214--223}.
\newblock


\bibitem[Martinez and Monperrus(2016)]%
        {martinez2016astor}
\bibfield{author}{\bibinfo{person}{Matias Martinez} {and} \bibinfo{person}{Martin Monperrus}.} \bibinfo{year}{2016}\natexlab{}.
\newblock \showarticletitle{Astor: A program repair library for java}. In \bibinfo{booktitle}{\emph{25th ISSTA}}. \bibinfo{pages}{441--444}.
\newblock


\bibitem[Matthew et~al\mbox{.}(1802)]%
        {matthew1802deep}
\bibfield{author}{\bibinfo{person}{E~Peters Matthew}, \bibinfo{person}{N Mark}, \bibinfo{person}{I Mohit}, \bibinfo{person}{G Matt}, \bibinfo{person}{C Christopher}, {and} \bibinfo{person}{L Kenton}.} \bibinfo{year}{1802}\natexlab{}.
\newblock \showarticletitle{Deep contextualized word representations (2018)}.
\newblock \bibinfo{journal}{\emph{arXiv:1802.05365}} (\bibinfo{year}{1802}).
\newblock


\bibitem[Mikolov et~al\mbox{.}(2013a)]%
        {mikolov2013efficientword2vec}
\bibfield{author}{\bibinfo{person}{Tomas Mikolov}, \bibinfo{person}{Kai Chen}, \bibinfo{person}{Greg Corrado}, {and} \bibinfo{person}{Jeffrey Dean}.} \bibinfo{year}{2013}\natexlab{a}.
\newblock \showarticletitle{Efficient estimation of word representations in vector space}.
\newblock \bibinfo{journal}{\emph{arXiv:1301.3781}} (\bibinfo{year}{2013}).
\newblock


\bibitem[Mikolov et~al\mbox{.}(2013b)]%
        {mikolov2013distributed}
\bibfield{author}{\bibinfo{person}{Tomas Mikolov}, \bibinfo{person}{Ilya Sutskever}, \bibinfo{person}{Kai Chen}, \bibinfo{person}{Greg~S Corrado}, {and} \bibinfo{person}{Jeff Dean}.} \bibinfo{year}{2013}\natexlab{b}.
\newblock \showarticletitle{Distributed representations of words and phrases and their compositionality}.
\newblock \bibinfo{journal}{\emph{Advances in neural information processing systems}}  \bibinfo{volume}{26} (\bibinfo{year}{2013}).
\newblock


\bibitem[Mohsen et~al\mbox{.}(2023)]%
        {mohsen2023enhancingpbl}
\bibfield{author}{\bibinfo{person}{Amr~Mansour Mohsen}, \bibinfo{person}{Hesham Hassan}, \bibinfo{person}{Khaled Wassif}, \bibinfo{person}{Ramadan Moawad}, {and} \bibinfo{person}{Soha Makady}.} \bibinfo{year}{2023}\natexlab{}.
\newblock \showarticletitle{Enhancing Bug Localization using Phase-Based Approach}.
\newblock \bibinfo{journal}{\emph{IEEE Access}} (\bibinfo{year}{2023}).
\newblock


\bibitem[Moon et~al\mbox{.}(2014)]%
        {moon2014ask}
\bibfield{author}{\bibinfo{person}{Seokhyeon Moon}, \bibinfo{person}{Yunho Kim}, \bibinfo{person}{Moonzoo Kim}, {and} \bibinfo{person}{Shin Yoo}.} \bibinfo{year}{2014}\natexlab{}.
\newblock \showarticletitle{Ask the mutants: Mutating faulty programs for fault localization}. In \bibinfo{booktitle}{\emph{2014 IEEE Seventh ICST}}. IEEE, \bibinfo{pages}{153--162}.
\newblock


\bibitem[Moreno et~al\mbox{.}(2014)]%
        {moreno2014use}
\bibfield{author}{\bibinfo{person}{Laura Moreno}, \bibinfo{person}{John~Joseph Treadway}, \bibinfo{person}{Andrian Marcus}, {and} \bibinfo{person}{Wuwei Shen}.} \bibinfo{year}{2014}\natexlab{}.
\newblock \showarticletitle{On the use of stack traces to improve text retrieval-based bug localization}. In \bibinfo{booktitle}{\emph{2014 IEEE ICSME}}. IEEE, \bibinfo{pages}{151--160}.
\newblock


\bibitem[Moreo et~al\mbox{.}(2021)]%
        {moreo2021word}
\bibfield{author}{\bibinfo{person}{Alejandro Moreo}, \bibinfo{person}{Andrea Esuli}, {and} \bibinfo{person}{Fabrizio Sebastiani}.} \bibinfo{year}{2021}\natexlab{}.
\newblock \showarticletitle{Word-class embeddings for multiclass text classification}.
\newblock \bibinfo{journal}{\emph{Data Mining and Knowledge Discovery}}  \bibinfo{volume}{35} (\bibinfo{year}{2021}), \bibinfo{pages}{911--963}.
\newblock


\bibitem[Murali et~al\mbox{.}(2021)]%
        {murali2021industry}
\bibfield{author}{\bibinfo{person}{Vijayaraghavan Murali}, \bibinfo{person}{Lee Gross}, \bibinfo{person}{Rebecca Qian}, {and} \bibinfo{person}{Satish Chandra}.} \bibinfo{year}{2021}\natexlab{}.
\newblock \showarticletitle{Industry-scale ir-based bug localization: A perspective from facebook}. In \bibinfo{booktitle}{\emph{2021 IEEE/ACM 43rd ICSE-SEIP}}. IEEE, \bibinfo{pages}{188--197}.
\newblock


\bibitem[Ng(2024)]%
        {andrew}
\bibfield{author}{\bibinfo{person}{Andrew Ng}.} \bibinfo{year}{2024}\natexlab{}.
\newblock \bibinfo{booktitle}{\emph{UNBIGGEN AI}}.
\newblock
\urldef\tempurl%
\url{https://spectrum.ieee.org/andrew-ng-data-centric-ai}
\showURL{%
\tempurl}


\bibitem[Ni et~al\mbox{.}(2022)]%
        {ni2022best}
\bibfield{author}{\bibinfo{person}{Chao Ni}, \bibinfo{person}{Wei Wang}, \bibinfo{person}{Kaiwen Yang}, \bibinfo{person}{Xin Xia}, \bibinfo{person}{Kui Liu}, {and} \bibinfo{person}{David Lo}.} \bibinfo{year}{2022}\natexlab{}.
\newblock \showarticletitle{The best of both worlds: integrating semantic features with expert features for defect prediction and localization}. In \bibinfo{booktitle}{\emph{Proceedings of the 30th ACM FSE}}. \bibinfo{pages}{672--683}.
\newblock


\bibitem[Nichols(2010)]%
        {nichols2010augmented}
\bibfield{author}{\bibinfo{person}{Brent~D Nichols}.} \bibinfo{year}{2010}\natexlab{}.
\newblock \showarticletitle{Augmented bug localization using past bug information}. In \bibinfo{booktitle}{\emph{Proceedings of the 48th Annual Southeast Regional Conference}}. \bibinfo{pages}{1--6}.
\newblock


\bibitem[Niu et~al\mbox{.}(2023a)]%
        {niu2023rat}
\bibfield{author}{\bibinfo{person}{Feifei Niu}, \bibinfo{person}{Wesley~KG Assun{\c{c}}ao}, \bibinfo{person}{LiGuo Huang}, \bibinfo{person}{Christoph Mayr-Dorn}, \bibinfo{person}{Jidong Ge}, \bibinfo{person}{Bin Luo}, {and} \bibinfo{person}{Alexander Egyed}.} \bibinfo{year}{2023}\natexlab{a}.
\newblock \showarticletitle{Rat: A refactoring-aware traceability model for bug localization}. In \bibinfo{booktitle}{\emph{2023 IEEE/ACM 45th ICSE}}. IEEE, \bibinfo{pages}{196--207}.
\newblock


\bibitem[Niu et~al\mbox{.}(2023b)]%
        {niu2023ablots}
\bibfield{author}{\bibinfo{person}{Feifei Niu}, \bibinfo{person}{Christoph Mayr-Dorn}, \bibinfo{person}{Wesley~KG Assun{\c{c}}{\~a}o}, \bibinfo{person}{LiGuo Huang}, \bibinfo{person}{Jidong Ge}, \bibinfo{person}{Bin Luo}, {and} \bibinfo{person}{Alexander Egyed}.} \bibinfo{year}{2023}\natexlab{b}.
\newblock \showarticletitle{The ABLoTS Approach for Bug Localization: is it replicable and generalizable?}. In \bibinfo{booktitle}{\emph{20th MSR}}. IEEE, \bibinfo{pages}{576--587}.
\newblock


\bibitem[Niu et~al\mbox{.}(2024)]%
        {niu2024extensive}
\bibfield{author}{\bibinfo{person}{Feifei Niu}, \bibinfo{person}{Enshuo Zhang}, \bibinfo{person}{Christoph Mayr-Dorn}, \bibinfo{person}{Wesley Klewerton~Guez Assun{\c{c}}{\~a}o}, \bibinfo{person}{Liguo Huang}, \bibinfo{person}{Jidong Ge}, \bibinfo{person}{Bin Luo}, {and} \bibinfo{person}{Alexander Egyed}.} \bibinfo{year}{2024}\natexlab{}.
\newblock \showarticletitle{An extensive replication study of the ABLoTS approach for bug localization}.
\newblock \bibinfo{journal}{\emph{EMSE}} \bibinfo{volume}{29}, \bibinfo{number}{6} (\bibinfo{year}{2024}), \bibinfo{pages}{1--37}.
\newblock


\bibitem[Pagliardini et~al\mbox{.}(2017)]%
        {pagliardini2017unsupervised}
\bibfield{author}{\bibinfo{person}{Matteo Pagliardini}, \bibinfo{person}{Prakhar Gupta}, {and} \bibinfo{person}{Martin Jaggi}.} \bibinfo{year}{2017}\natexlab{}.
\newblock \showarticletitle{Unsupervised learning of sentence embeddings using compositional n-gram features}.
\newblock \bibinfo{journal}{\emph{arXiv:1703.02507}} (\bibinfo{year}{2017}).
\newblock


\bibitem[Papadakis and Le~Traon(2015)]%
        {papadakis2015metallaxis}
\bibfield{author}{\bibinfo{person}{Mike Papadakis} {and} \bibinfo{person}{Yves Le~Traon}.} \bibinfo{year}{2015}\natexlab{}.
\newblock \showarticletitle{Metallaxis-FL: mutation-based fault localization}.
\newblock \bibinfo{journal}{\emph{Software Testing, Verification and Reliability}} \bibinfo{volume}{25}, \bibinfo{number}{5-7} (\bibinfo{year}{2015}), \bibinfo{pages}{605--628}.
\newblock


\bibitem[Pendlebury et~al\mbox{.}(2019)]%
        {pendlebury2019tesseract}
\bibfield{author}{\bibinfo{person}{Feargus Pendlebury}, \bibinfo{person}{Fabio Pierazzi}, \bibinfo{person}{Roberto Jordaney}, \bibinfo{person}{Johannes Kinder}, {and} \bibinfo{person}{Lorenzo Cavallaro}.} \bibinfo{year}{2019}\natexlab{}.
\newblock \showarticletitle{$\{$TESSERACT$\}$: Eliminating experimental bias in malware classification across space and time}. In \bibinfo{booktitle}{\emph{28th USENIX Security}}. \bibinfo{pages}{729--746}.
\newblock


\bibitem[Pennington et~al\mbox{.}(2014)]%
        {pennington2014glove}
\bibfield{author}{\bibinfo{person}{Jeffrey Pennington}, \bibinfo{person}{Richard Socher}, {and} \bibinfo{person}{Christopher~D Manning}.} \bibinfo{year}{2014}\natexlab{}.
\newblock \showarticletitle{Glove: Global vectors for word representation}. In \bibinfo{booktitle}{\emph{Proceedings of the 2014 EMNLP}}. \bibinfo{pages}{1532--1543}.
\newblock


\bibitem[Polisetty et~al\mbox{.}(2019)]%
        {polisetty2019usefulness}
\bibfield{author}{\bibinfo{person}{Sravya Polisetty}, \bibinfo{person}{Andriy Miranskyy}, {and} \bibinfo{person}{Ay{\c{s}}e Ba{\c{s}}ar}.} \bibinfo{year}{2019}\natexlab{}.
\newblock \showarticletitle{On usefulness of the deep-learning-based bug localization models to practitioners}. In \bibinfo{booktitle}{\emph{Proceedings of the Fifteenth International Conference on Predictive Models and Data Analytics in Software Engineering}}. \bibinfo{pages}{16--25}.
\newblock


\bibitem[Qi et~al\mbox{.}(2021)]%
        {qi2021dreamloc}
\bibfield{author}{\bibinfo{person}{Binhang Qi}, \bibinfo{person}{Hailong Sun}, \bibinfo{person}{Wei Yuan}, \bibinfo{person}{Hongyu Zhang}, {and} \bibinfo{person}{Xiangxin Meng}.} \bibinfo{year}{2021}\natexlab{}.
\newblock \showarticletitle{Dreamloc: A deep relevance matching-based framework for bug localization}.
\newblock \bibinfo{journal}{\emph{IEEE Transactions on Reliability}} \bibinfo{volume}{71}, \bibinfo{number}{1} (\bibinfo{year}{2021}), \bibinfo{pages}{235--249}.
\newblock


\bibitem[Rahman et~al\mbox{.}(2011)]%
        {rahman2011bugcache}
\bibfield{author}{\bibinfo{person}{Foyzur Rahman}, \bibinfo{person}{Daryl Posnett}, \bibinfo{person}{Abram Hindle}, \bibinfo{person}{Earl Barr}, {and} \bibinfo{person}{Premkumar Devanbu}.} \bibinfo{year}{2011}\natexlab{}.
\newblock \showarticletitle{BugCache for inspections: hit or miss?}. In \bibinfo{booktitle}{\emph{FSE}}. \bibinfo{pages}{322--331}.
\newblock


\bibitem[Rao and Kak(2011)]%
        {rao2011retrieval}
\bibfield{author}{\bibinfo{person}{Shivani Rao} {and} \bibinfo{person}{Avinash Kak}.} \bibinfo{year}{2011}\natexlab{}.
\newblock \showarticletitle{Retrieval from software libraries for bug localization: a comparative study of generic and composite text models}. In \bibinfo{booktitle}{\emph{Proceedings of the 8th MSR}}. \bibinfo{pages}{43--52}.
\newblock


\bibitem[Rath et~al\mbox{.}(2018)]%
        {rath2018analyzing}
\bibfield{author}{\bibinfo{person}{Michael Rath}, \bibinfo{person}{David Lo}, {and} \bibinfo{person}{Patrick M{\"a}der}.} \bibinfo{year}{2018}\natexlab{}.
\newblock \showarticletitle{Analyzing requirements and traceability information to improve bug localization}. In \bibinfo{booktitle}{\emph{Proceedings of the 15th MSR}}. \bibinfo{pages}{442--453}.
\newblock


\bibitem[Reimers(2023)]%
        {sentencetransformer}
\bibfield{author}{\bibinfo{person}{Nils Reimers}.} \bibinfo{year}{2023}\natexlab{}.
\newblock \bibinfo{booktitle}{\emph{SentenceTransformer}}.
\newblock
\urldef\tempurl%
\url{https://huggingface.co/sentence-transformers}
\showURL{%
\tempurl}


\bibitem[Renieres and Reiss(2003)]%
        {renieres2003fault}
\bibfield{author}{\bibinfo{person}{Manos Renieres} {and} \bibinfo{person}{Steven~P Reiss}.} \bibinfo{year}{2003}\natexlab{}.
\newblock \showarticletitle{Fault localization with nearest neighbor queries}. In \bibinfo{booktitle}{\emph{18th IEEE ASE.}} IEEE, \bibinfo{pages}{30--39}.
\newblock


\bibitem[Rumelhart et~al\mbox{.}(1986)]%
        {rumelhart1986learning}
\bibfield{author}{\bibinfo{person}{David~E Rumelhart}, \bibinfo{person}{Geoffrey~E Hinton}, {and} \bibinfo{person}{Ronald~J Williams}.} \bibinfo{year}{1986}\natexlab{}.
\newblock \showarticletitle{Learning representations by back-propagating errors}.
\newblock \bibinfo{journal}{\emph{nature}} \bibinfo{volume}{323}, \bibinfo{number}{6088} (\bibinfo{year}{1986}), \bibinfo{pages}{533--536}.
\newblock


\bibitem[Saha et~al\mbox{.}(2014)]%
        {saha2014effectiveness}
\bibfield{author}{\bibinfo{person}{Ripon~K Saha}, \bibinfo{person}{Julia Lawall}, \bibinfo{person}{Sarfraz Khurshid}, {and} \bibinfo{person}{Dewayne~E Perry}.} \bibinfo{year}{2014}\natexlab{}.
\newblock \showarticletitle{On the effectiveness of information retrieval based bug localization for c programs}. In \bibinfo{booktitle}{\emph{2014 IEEE ICSME}}. IEEE, \bibinfo{pages}{161--170}.
\newblock


\bibitem[Saha et~al\mbox{.}(2013)]%
        {saha2013improving}
\bibfield{author}{\bibinfo{person}{Ripon~K Saha}, \bibinfo{person}{Matthew Lease}, \bibinfo{person}{Sarfraz Khurshid}, {and} \bibinfo{person}{Dewayne~E Perry}.} \bibinfo{year}{2013}\natexlab{}.
\newblock \showarticletitle{Improving bug localization using structured information retrieval}. In \bibinfo{booktitle}{\emph{28th IEEE/ACM ASE}}. IEEE, \bibinfo{pages}{345--355}.
\newblock


\bibitem[Salton(1989)]%
        {salton1989automatic}
\bibfield{author}{\bibinfo{person}{Gerard Salton}.} \bibinfo{year}{1989}\natexlab{}.
\newblock \showarticletitle{Automatic text processing: The transformation, analysis, and retrieval of}.
\newblock \bibinfo{journal}{\emph{Reading: Addison-Wesley}}  \bibinfo{volume}{169} (\bibinfo{year}{1989}).
\newblock


\bibitem[Sangle et~al\mbox{.}(2020)]%
        {sangle2020drast}
\bibfield{author}{\bibinfo{person}{Shubham Sangle}, \bibinfo{person}{Sandeep Muvva}, \bibinfo{person}{Sridhar Chimalakonda}, \bibinfo{person}{Karthikeyan Ponnalagu}, {and} \bibinfo{person}{Vijendran~Gopalan Venkoparao}.} \bibinfo{year}{2020}\natexlab{}.
\newblock \showarticletitle{DRAST--A Deep Learning and AST Based Approach for Bug Localization}.
\newblock \bibinfo{journal}{\emph{arXiv:2011.03449}} (\bibinfo{year}{2020}).
\newblock


\bibitem[Schroter et~al\mbox{.}(2010)]%
        {schroter2010stack}
\bibfield{author}{\bibinfo{person}{Adrian Schroter}, \bibinfo{person}{Adrian Schr{\"o}ter}, \bibinfo{person}{Nicolas Bettenburg}, {and} \bibinfo{person}{Rahul Premraj}.} \bibinfo{year}{2010}\natexlab{}.
\newblock \showarticletitle{Do stack traces help developers fix bugs?}. In \bibinfo{booktitle}{\emph{2010 7th IEEE MSR}}. IEEE, \bibinfo{pages}{118--121}.
\newblock


\bibitem[Sch{\"u}tze et~al\mbox{.}(2008)]%
        {schutze2008introduction}
\bibfield{author}{\bibinfo{person}{Hinrich Sch{\"u}tze}, \bibinfo{person}{Christopher~D Manning}, {and} \bibinfo{person}{Prabhakar Raghavan}.} \bibinfo{year}{2008}\natexlab{}.
\newblock \bibinfo{booktitle}{\emph{Introduction to information retrieval}}. Vol.~\bibinfo{volume}{39}.
\newblock \bibinfo{publisher}{Cambridge University Press Cambridge}.
\newblock


\bibitem[Selva~Birunda and Kanniga~Devi(2021)]%
        {selva2021review}
\bibfield{author}{\bibinfo{person}{S Selva~Birunda} {and} \bibinfo{person}{R Kanniga~Devi}.} \bibinfo{year}{2021}\natexlab{}.
\newblock \showarticletitle{A review on word embedding techniques for text classification}.
\newblock \bibinfo{journal}{\emph{Innovative Data Communication Technologies and Application: Proceedings of ICIDCA 2020}} (\bibinfo{year}{2021}), \bibinfo{pages}{267--281}.
\newblock


\bibitem[Shi et~al\mbox{.}(2022)]%
        {shi2022semirfl}
\bibfield{author}{\bibinfo{person}{Xiangyu Shi}, \bibinfo{person}{Xiaolin Ju}, \bibinfo{person}{Xiang Chen}, \bibinfo{person}{Guilong Lu}, {and} \bibinfo{person}{Mengqi Xu}.} \bibinfo{year}{2022}\natexlab{}.
\newblock \showarticletitle{SemirFL: Boosting Fault Localization via Combining Semantic Information and Information Retrieval}. In \bibinfo{booktitle}{\emph{2022 IEEE 22nd QRS-C}}. IEEE, \bibinfo{pages}{324--332}.
\newblock


\bibitem[Sisman and Kak(2012)]%
        {sisman2012incorporating}
\bibfield{author}{\bibinfo{person}{Bunyamin Sisman} {and} \bibinfo{person}{Avinash~C Kak}.} \bibinfo{year}{2012}\natexlab{}.
\newblock \showarticletitle{Incorporating version histories in information retrieval based bug localization}. In \bibinfo{booktitle}{\emph{2012 9th IEEE MSR}}. IEEE, \bibinfo{pages}{50--59}.
\newblock


\bibitem[{\'S}liwerski et~al\mbox{.}(2005)]%
        {sliwerski2005changes}
\bibfield{author}{\bibinfo{person}{Jacek {\'S}liwerski}, \bibinfo{person}{Thomas Zimmermann}, {and} \bibinfo{person}{Andreas Zeller}.} \bibinfo{year}{2005}\natexlab{}.
\newblock \showarticletitle{When do changes induce fixes?}
\newblock \bibinfo{journal}{\emph{ACM sigsoft software engineering notes}} \bibinfo{volume}{30}, \bibinfo{number}{4} (\bibinfo{year}{2005}), \bibinfo{pages}{1--5}.
\newblock


\bibitem[Sonbol et~al\mbox{.}(2022)]%
        {sonbol2022use}
\bibfield{author}{\bibinfo{person}{Riad Sonbol}, \bibinfo{person}{Ghaida Rebdawi}, {and} \bibinfo{person}{Nada Ghneim}.} \bibinfo{year}{2022}\natexlab{}.
\newblock \showarticletitle{The use of nlp-based text representation techniques to support requirement engineering tasks: A systematic mapping review}.
\newblock \bibinfo{journal}{\emph{IEEE Access}} (\bibinfo{year}{2022}).
\newblock


\bibitem[Stein et~al\mbox{.}(2019)]%
        {stein2019analysis}
\bibfield{author}{\bibinfo{person}{Roger~Alan Stein}, \bibinfo{person}{Patricia~A Jaques}, {and} \bibinfo{person}{Joao~Francisco Valiati}.} \bibinfo{year}{2019}\natexlab{}.
\newblock \showarticletitle{An analysis of hierarchical text classification using word embeddings}.
\newblock \bibinfo{journal}{\emph{Information Sciences}}  \bibinfo{volume}{471} (\bibinfo{year}{2019}), \bibinfo{pages}{216--232}.
\newblock


\bibitem[Tabassum et~al\mbox{.}(2020)]%
        {tabassum2020code}
\bibfield{author}{\bibinfo{person}{Jeniya Tabassum}, \bibinfo{person}{Mounica Maddela}, \bibinfo{person}{Wei Xu}, {and} \bibinfo{person}{Alan Ritter}.} \bibinfo{year}{2020}\natexlab{}.
\newblock \showarticletitle{Code and named entity recognition in stackoverflow}.
\newblock \bibinfo{journal}{\emph{arXiv:2005.01634}} (\bibinfo{year}{2020}).
\newblock


\bibitem[Tao et~al\mbox{.}(2024)]%
        {tao2024magis}
\bibfield{author}{\bibinfo{person}{Wei Tao}, \bibinfo{person}{Yucheng Zhou}, \bibinfo{person}{Yanlin Wang}, \bibinfo{person}{Wenqiang Zhang}, \bibinfo{person}{Hongyu Zhang}, {and} \bibinfo{person}{Yu Cheng}.} \bibinfo{year}{2024}\natexlab{}.
\newblock \showarticletitle{Magis: Llm-based multi-agent framework for github issue resolution}.
\newblock \bibinfo{journal}{\emph{arXiv:2403.17927}} (\bibinfo{year}{2024}).
\newblock


\bibitem[Tsumita et~al\mbox{.}(2023)]%
        {tsumita2023large}
\bibfield{author}{\bibinfo{person}{Shizuka Tsumita}, \bibinfo{person}{Shinpei Hayashi}, {and} \bibinfo{person}{Sousuke Amasaki}.} \bibinfo{year}{2023}\natexlab{}.
\newblock \showarticletitle{Large-Scale Evaluation of Method-Level Bug Localization with FinerBench4BL}. In \bibinfo{booktitle}{\emph{2023 IEEE SANER}}. IEEE, \bibinfo{pages}{815--824}.
\newblock


\bibitem[Vaswani et~al\mbox{.}(2017)]%
        {vaswani2017attention}
\bibfield{author}{\bibinfo{person}{Ashish Vaswani}, \bibinfo{person}{Noam Shazeer}, \bibinfo{person}{Niki Parmar}, \bibinfo{person}{Jakob Uszkoreit}, \bibinfo{person}{Llion Jones}, \bibinfo{person}{Aidan~N Gomez}, \bibinfo{person}{{\L}ukasz Kaiser}, {and} \bibinfo{person}{Illia Polosukhin}.} \bibinfo{year}{2017}\natexlab{}.
\newblock \showarticletitle{Attention is all you need}.
\newblock \bibinfo{journal}{\emph{Advances in neural information processing systems}}  \bibinfo{volume}{30} (\bibinfo{year}{2017}).
\newblock


\bibitem[Voorhees et~al\mbox{.}(1999)]%
        {voorhees1999trec}
\bibfield{author}{\bibinfo{person}{Ellen~M Voorhees} {et~al\mbox{.}}} \bibinfo{year}{1999}\natexlab{}.
\newblock \showarticletitle{The trec-8 question answering track report.}. In \bibinfo{booktitle}{\emph{Trec}}, Vol.~\bibinfo{volume}{99}. \bibinfo{pages}{77--82}.
\newblock


\bibitem[Wang et~al\mbox{.}(2020)]%
        {wang2020multimd-cnn}
\bibfield{author}{\bibinfo{person}{Bei Wang}, \bibinfo{person}{Ling Xu}, \bibinfo{person}{Meng Yan}, \bibinfo{person}{Chao Liu}, {and} \bibinfo{person}{Ling Liu}.} \bibinfo{year}{2020}\natexlab{}.
\newblock \showarticletitle{Multi-dimension convolutional neural network for bug localization}.
\newblock \bibinfo{journal}{\emph{IEEE TSC}} \bibinfo{volume}{15}, \bibinfo{number}{3} (\bibinfo{year}{2020}), \bibinfo{pages}{1649--1663}.
\newblock


\bibitem[Wang et~al\mbox{.}(2015)]%
        {wang2015evaluating}
\bibfield{author}{\bibinfo{person}{Qianqian Wang}, \bibinfo{person}{Chris Parnin}, {and} \bibinfo{person}{Alessandro Orso}.} \bibinfo{year}{2015}\natexlab{}.
\newblock \showarticletitle{Evaluating the usefulness of ir-based fault localization techniques}. In \bibinfo{booktitle}{\emph{Proceedings of the 2015 ISSTA}}. \bibinfo{pages}{1--11}.
\newblock


\bibitem[Wang and Lo(2014)]%
        {wang2014version}
\bibfield{author}{\bibinfo{person}{Shaowei Wang} {and} \bibinfo{person}{David Lo}.} \bibinfo{year}{2014}\natexlab{}.
\newblock \showarticletitle{Version history, similar report, and structure: Putting them together for improved bug localization}. In \bibinfo{booktitle}{\emph{Proceedings of the 22nd ICPC}}. \bibinfo{pages}{53--63}.
\newblock


\bibitem[Wang and Lo(2016)]%
        {wang2016amalgam+}
\bibfield{author}{\bibinfo{person}{Shaowei Wang} {and} \bibinfo{person}{David Lo}.} \bibinfo{year}{2016}\natexlab{}.
\newblock \showarticletitle{Amalgam+: Composing rich information sources for accurate bug localization}.
\newblock \bibinfo{journal}{\emph{Journal of Software: Evolution and Process}} \bibinfo{volume}{28}, \bibinfo{number}{10} (\bibinfo{year}{2016}), \bibinfo{pages}{921--942}.
\newblock


\bibitem[Wang et~al\mbox{.}(2008)]%
        {wang2008approach}
\bibfield{author}{\bibinfo{person}{Xiaoyin Wang}, \bibinfo{person}{Lu Zhang}, \bibinfo{person}{Tao Xie}, \bibinfo{person}{John Anvik}, {and} \bibinfo{person}{Jiasu Sun}.} \bibinfo{year}{2008}\natexlab{}.
\newblock \showarticletitle{An approach to detecting duplicate bug reports using natural language and execution information}. In \bibinfo{booktitle}{\emph{Proceedings of the 30th ICSE}}. \bibinfo{pages}{461--470}.
\newblock


\bibitem[Wang et~al\mbox{.}(2023)]%
        {wang2023codet5+}
\bibfield{author}{\bibinfo{person}{Yue Wang}, \bibinfo{person}{Hung Le}, \bibinfo{person}{Akhilesh~Deepak Gotmare}, \bibinfo{person}{Nghi~DQ Bui}, \bibinfo{person}{Junnan Li}, {and} \bibinfo{person}{Steven~CH Hoi}.} \bibinfo{year}{2023}\natexlab{}.
\newblock \showarticletitle{Codet5+: Open code large language models for code understanding and generation}.
\newblock \bibinfo{journal}{\emph{arXiv:2305.07922}} (\bibinfo{year}{2023}).
\newblock


\bibitem[Wen et~al\mbox{.}(2016)]%
        {wen2016locus}
\bibfield{author}{\bibinfo{person}{Ming Wen}, \bibinfo{person}{Rongxin Wu}, {and} \bibinfo{person}{Shing-Chi Cheung}.} \bibinfo{year}{2016}\natexlab{}.
\newblock \showarticletitle{Locus: Locating bugs from software changes}. In \bibinfo{booktitle}{\emph{Proceedings of the 31st IEEE/ACM ASE}}. \bibinfo{pages}{262--273}.
\newblock


\bibitem[Widyasari et~al\mbox{.}(2022)]%
        {widyasari2022influence}
\bibfield{author}{\bibinfo{person}{Ratnadira Widyasari}, \bibinfo{person}{Stefanus~Agus Haryono}, \bibinfo{person}{Ferdian Thung}, \bibinfo{person}{Jieke Shi}, \bibinfo{person}{Constance Tan}, \bibinfo{person}{Fiona Wee}, \bibinfo{person}{Jack Phan}, {and} \bibinfo{person}{David Lo}.} \bibinfo{year}{2022}\natexlab{}.
\newblock \showarticletitle{On the influence of biases in bug localization: Evaluation and benchmark}. In \bibinfo{booktitle}{\emph{IEEE SANER}}. \bibinfo{pages}{128--139}.
\newblock


\bibitem[Widyasari et~al\mbox{.}(2020)]%
        {widyasari2020bugsinpy}
\bibfield{author}{\bibinfo{person}{Ratnadira Widyasari}, \bibinfo{person}{Sheng~Qin Sim}, \bibinfo{person}{Camellia Lok}, \bibinfo{person}{Haodi Qi}, \bibinfo{person}{Jack Phan}, \bibinfo{person}{Qijin Tay}, \bibinfo{person}{Constance Tan}, \bibinfo{person}{Fiona Wee}, \bibinfo{person}{Jodie~Ethelda Tan}, \bibinfo{person}{Yuheng Yieh}, {et~al\mbox{.}}} \bibinfo{year}{2020}\natexlab{}.
\newblock \showarticletitle{Bugsinpy: a database of existing bugs in python programs to enable controlled testing and debugging studies}. In \bibinfo{booktitle}{\emph{Proceedings of the 28th FSE}}. \bibinfo{pages}{1556--1560}.
\newblock


\bibitem[Wohlin(2014)]%
        {wohlin2014guidelines}
\bibfield{author}{\bibinfo{person}{Claes Wohlin}.} \bibinfo{year}{2014}\natexlab{}.
\newblock \showarticletitle{Guidelines for snowballing in systematic literature studies and a replication in software engineering}. In \bibinfo{booktitle}{\emph{Proceedings of the 18th EASE}}. \bibinfo{pages}{1--10}.
\newblock


\bibitem[Wong et~al\mbox{.}(2014)]%
        {wong2014boosting}
\bibfield{author}{\bibinfo{person}{Chu-Pan Wong}, \bibinfo{person}{Yingfei Xiong}, \bibinfo{person}{Hongyu Zhang}, \bibinfo{person}{Dan Hao}, \bibinfo{person}{Lu Zhang}, {and} \bibinfo{person}{Hong Mei}.} \bibinfo{year}{2014}\natexlab{}.
\newblock \showarticletitle{Boosting bug-report-oriented fault localization with segmentation and stack-trace analysis}. In \bibinfo{booktitle}{\emph{2014 IEEE ICSME}}. IEEE, \bibinfo{pages}{181--190}.
\newblock


\bibitem[Wong et~al\mbox{.}(2016)]%
        {wong2016survey}
\bibfield{author}{\bibinfo{person}{W~Eric Wong}, \bibinfo{person}{Ruizhi Gao}, \bibinfo{person}{Yihao Li}, \bibinfo{person}{Rui Abreu}, {and} \bibinfo{person}{Franz Wotawa}.} \bibinfo{year}{2016}\natexlab{}.
\newblock \showarticletitle{A survey on software fault localization}.
\newblock \bibinfo{journal}{\emph{IEEE TSE}} \bibinfo{volume}{42}, \bibinfo{number}{8} (\bibinfo{year}{2016}), \bibinfo{pages}{707--740}.
\newblock


\bibitem[Wong et~al\mbox{.}(2023)]%
        {wong2023software}
\bibfield{author}{\bibinfo{person}{W~Eric Wong}, \bibinfo{person}{Ruizhi Gao}, \bibinfo{person}{Yihao Li}, \bibinfo{person}{Rui Abreu}, \bibinfo{person}{Franz Wotawa}, {and} \bibinfo{person}{Dongcheng Li}.} \bibinfo{year}{2023}\natexlab{}.
\newblock \showarticletitle{Software fault localization: An overview of research, techniques, and tools}.
\newblock \bibinfo{journal}{\emph{Handbook of Software Fault Localization: Foundations and Advances}} (\bibinfo{year}{2023}), \bibinfo{pages}{1--117}.
\newblock


\bibitem[Wu et~al\mbox{.}(2014)]%
        {wu2014crashlocator}
\bibfield{author}{\bibinfo{person}{Rongxin Wu}, \bibinfo{person}{Hongyu Zhang}, \bibinfo{person}{Shing-Chi Cheung}, {and} \bibinfo{person}{Sunghun Kim}.} \bibinfo{year}{2014}\natexlab{}.
\newblock \showarticletitle{Crashlocator: Locating crashing faults based on crash stacks}. In \bibinfo{booktitle}{\emph{Proceedings of the 2014 ISSTA}}. \bibinfo{pages}{204--214}.
\newblock


\bibitem[Wu et~al\mbox{.}(2018)]%
        {wu2018unsupervised}
\bibfield{author}{\bibinfo{person}{Zhirong Wu}, \bibinfo{person}{Yuanjun Xiong}, \bibinfo{person}{Stella~X Yu}, {and} \bibinfo{person}{Dahua Lin}.} \bibinfo{year}{2018}\natexlab{}.
\newblock \showarticletitle{Unsupervised feature learning via non-parametric instance discrimination}. In \bibinfo{booktitle}{\emph{Proceedings of the IEEE conference on computer vision and pattern recognition}}. \bibinfo{pages}{3733--3742}.
\newblock


\bibitem[Xia and Lo(2023)]%
        {xia2023information}
\bibfield{author}{\bibinfo{person}{Xin Xia} {and} \bibinfo{person}{David Lo}.} \bibinfo{year}{2023}\natexlab{}.
\newblock \showarticletitle{Information Retrieval-Based Techniques for Software Fault Localization}.
\newblock \bibinfo{journal}{\emph{Handbook of Software Fault Localization: Foundations and Advances}} (\bibinfo{year}{2023}), \bibinfo{pages}{365--391}.
\newblock


\bibitem[Xia et~al\mbox{.}(2014)]%
        {xia2014cross}
\bibfield{author}{\bibinfo{person}{Xin Xia}, \bibinfo{person}{David Lo}, \bibinfo{person}{Xingen Wang}, \bibinfo{person}{Chenyi Zhang}, {and} \bibinfo{person}{Xinyu Wang}.} \bibinfo{year}{2014}\natexlab{}.
\newblock \showarticletitle{Cross-language bug localization}. In \bibinfo{booktitle}{\emph{Proceedings of the 22nd ICPC}}. \bibinfo{pages}{275--278}.
\newblock


\bibitem[Xiao et~al\mbox{.}(2023)]%
        {xiao2023bugradar}
\bibfield{author}{\bibinfo{person}{Xi Xiao}, \bibinfo{person}{Renjie Xiao}, \bibinfo{person}{Qing Li}, \bibinfo{person}{Jianhui Lv}, \bibinfo{person}{Shunyan Cui}, {and} \bibinfo{person}{Qixu Liu}.} \bibinfo{year}{2023}\natexlab{}.
\newblock \showarticletitle{BugRadar: Bug localization by knowledge graph link prediction}.
\newblock \bibinfo{journal}{\emph{IST}} (\bibinfo{year}{2023}), \bibinfo{pages}{107274}.
\newblock


\bibitem[Xiao and Keung(2018)]%
        {xiao2018improving}
\bibfield{author}{\bibinfo{person}{Yan Xiao} {and} \bibinfo{person}{Jacky Keung}.} \bibinfo{year}{2018}\natexlab{}.
\newblock \showarticletitle{Improving bug localization with character-level convolutional neural network and recurrent neural network}. In \bibinfo{booktitle}{\emph{2018 25th APSEC}}. IEEE, \bibinfo{pages}{703--704}.
\newblock


\bibitem[Xiao et~al\mbox{.}(2018a)]%
        {xiao2018machinebugtranslator}
\bibfield{author}{\bibinfo{person}{Yan Xiao}, \bibinfo{person}{Jacky Keung}, \bibinfo{person}{Kwabena~E Bennin}, {and} \bibinfo{person}{Qing Mi}.} \bibinfo{year}{2018}\natexlab{a}.
\newblock \showarticletitle{Machine translation-based bug localization technique for bridging lexical gap}.
\newblock \bibinfo{journal}{\emph{IST}}  \bibinfo{volume}{99} (\bibinfo{year}{2018}), \bibinfo{pages}{58--61}.
\newblock


\bibitem[Xiao et~al\mbox{.}(2019)]%
        {xiao2019improvingdeeploc}
\bibfield{author}{\bibinfo{person}{Yan Xiao}, \bibinfo{person}{Jacky Keung}, \bibinfo{person}{Kwabena~E Bennin}, {and} \bibinfo{person}{Qing Mi}.} \bibinfo{year}{2019}\natexlab{}.
\newblock \showarticletitle{Improving bug localization with word embedding and enhanced convolutional neural networks}.
\newblock \bibinfo{journal}{\emph{IST}}  \bibinfo{volume}{105} (\bibinfo{year}{2019}), \bibinfo{pages}{17--29}.
\newblock


\bibitem[Xiao et~al\mbox{.}(2017)]%
        {xiao2017improvingdeeplocator}
\bibfield{author}{\bibinfo{person}{Yan Xiao}, \bibinfo{person}{Jacky Keung}, \bibinfo{person}{Qing Mi}, {and} \bibinfo{person}{Kwabena~E Bennin}.} \bibinfo{year}{2017}\natexlab{}.
\newblock \showarticletitle{Improving bug localization with an enhanced convolutional neural network}. In \bibinfo{booktitle}{\emph{2017 24th APSEC}}. IEEE, \bibinfo{pages}{338--347}.
\newblock


\bibitem[Xiao et~al\mbox{.}(2018b)]%
        {xiao2018bugcnn_forest}
\bibfield{author}{\bibinfo{person}{Yan Xiao}, \bibinfo{person}{Jacky Keung}, \bibinfo{person}{Qing Mi}, {and} \bibinfo{person}{Kwabena~E Bennin}.} \bibinfo{year}{2018}\natexlab{b}.
\newblock \showarticletitle{Bug localization with semantic and structural features using convolutional neural network and cascade forest}. In \bibinfo{booktitle}{\emph{Proceedings of the 22nd EASE}}. \bibinfo{pages}{101--111}.
\newblock


\bibitem[Xie et~al\mbox{.}(2013)]%
        {xie2013theoretical}
\bibfield{author}{\bibinfo{person}{Xiaoyuan Xie}, \bibinfo{person}{Tsong~Yueh Chen}, \bibinfo{person}{Fei-Ching Kuo}, {and} \bibinfo{person}{Baowen Xu}.} \bibinfo{year}{2013}\natexlab{}.
\newblock \showarticletitle{A theoretical analysis of the risk evaluation formulas for spectrum-based fault localization}.
\newblock \bibinfo{journal}{\emph{ACM TOSEM}} \bibinfo{volume}{22}, \bibinfo{number}{4} (\bibinfo{year}{2013}), \bibinfo{pages}{1--40}.
\newblock


\bibitem[Xu et~al\mbox{.}(2023)]%
        {xu2023buglocfront}
\bibfield{author}{\bibinfo{person}{Guoqing Xu}, \bibinfo{person}{Xingqi Wang}, \bibinfo{person}{Dan Wei}, \bibinfo{person}{Yanli Shao}, {and} \bibinfo{person}{Bin Chen}.} \bibinfo{year}{2023}\natexlab{}.
\newblock \showarticletitle{Bug Localization with Features Crossing and Structured Semantic Information Matching}.
\newblock \bibinfo{journal}{\emph{SEKE}} (\bibinfo{year}{2023}).
\newblock


\bibitem[Yamaguchi et~al\mbox{.}(2014)]%
        {yamaguchi2014modeling}
\bibfield{author}{\bibinfo{person}{Fabian Yamaguchi}, \bibinfo{person}{Nico Golde}, \bibinfo{person}{Daniel Arp}, {and} \bibinfo{person}{Konrad Rieck}.} \bibinfo{year}{2014}\natexlab{}.
\newblock \showarticletitle{Modeling and discovering vulnerabilities with code property graphs}. In \bibinfo{booktitle}{\emph{2014 IEEE Symposium on Security and Privacy}}. IEEE, \bibinfo{pages}{590--604}.
\newblock


\bibitem[Yan et~al\mbox{.}(2023)]%
        {yan2023bug}
\bibfield{author}{\bibinfo{person}{Xuefeng Yan}, \bibinfo{person}{Shasha Cheng}, {and} \bibinfo{person}{Liqin Guo}.} \bibinfo{year}{2023}\natexlab{}.
\newblock \showarticletitle{Bug localization based on syntactical and semantic information of source code}.
\newblock \bibinfo{journal}{\emph{Journal of Systems Engineering and Electronics}} \bibinfo{volume}{34}, \bibinfo{number}{1} (\bibinfo{year}{2023}), \bibinfo{pages}{236--246}.
\newblock


\bibitem[Yang and Lee(2021)]%
        {yang2021utilizing}
\bibfield{author}{\bibinfo{person}{Geunseok Yang} {and} \bibinfo{person}{Byungjeong Lee}.} \bibinfo{year}{2021}\natexlab{}.
\newblock \showarticletitle{Utilizing topic-based similar commit information and CNN-LSTM algorithm for bug localization}.
\newblock \bibinfo{journal}{\emph{Symmetry}} \bibinfo{volume}{13}, \bibinfo{number}{3} (\bibinfo{year}{2021}), \bibinfo{pages}{406}.
\newblock


\bibitem[Yang et~al\mbox{.}(2020)]%
        {yang2020applying}
\bibfield{author}{\bibinfo{person}{Geunseok Yang}, \bibinfo{person}{Kyeongsic Min}, {and} \bibinfo{person}{Byungjeong Lee}.} \bibinfo{year}{2020}\natexlab{}.
\newblock \showarticletitle{Applying deep learning algorithm to automatic bug localization and repair}. In \bibinfo{booktitle}{\emph{Proceedings of the 35th Annual ACM symposium on applied computing}}. \bibinfo{pages}{1634--1641}.
\newblock


\bibitem[Yang et~al\mbox{.}(2021)]%
        {yang2021locatingmram}
\bibfield{author}{\bibinfo{person}{Shouliang Yang}, \bibinfo{person}{Junming Cao}, \bibinfo{person}{Hushuang Zeng}, \bibinfo{person}{Beijun Shen}, {and} \bibinfo{person}{Hao Zhong}.} \bibinfo{year}{2021}\natexlab{}.
\newblock \showarticletitle{Locating faulty methods with a mixed RNN and attention model}. In \bibinfo{booktitle}{\emph{2021 IEEE/ACM 29th ICPC}}. IEEE, \bibinfo{pages}{207--218}.
\newblock


\bibitem[Ye et~al\mbox{.}(2014)]%
        {ye2014learning}
\bibfield{author}{\bibinfo{person}{Xin Ye}, \bibinfo{person}{Razvan Bunescu}, {and} \bibinfo{person}{Chang Liu}.} \bibinfo{year}{2014}\natexlab{}.
\newblock \showarticletitle{Learning to rank relevant files for bug reports using domain knowledge}. In \bibinfo{booktitle}{\emph{Proceedings of the 22nd ACM SIGSOFT FSE}}. \bibinfo{pages}{689--699}.
\newblock


\bibitem[Ye et~al\mbox{.}(2015)]%
        {ye2015mapping}
\bibfield{author}{\bibinfo{person}{Xin Ye}, \bibinfo{person}{Razvan Bunescu}, {and} \bibinfo{person}{Chang Liu}.} \bibinfo{year}{2015}\natexlab{}.
\newblock \showarticletitle{Mapping bug reports to relevant files: A ranking model, a fine-grained benchmark, and feature evaluation}.
\newblock \bibinfo{journal}{\emph{IEEE TSE}} \bibinfo{volume}{42}, \bibinfo{number}{4} (\bibinfo{year}{2015}), \bibinfo{pages}{379--402}.
\newblock


\bibitem[Yong et~al\mbox{.}(2023)]%
        {yong2023decomposings-buglocator}
\bibfield{author}{\bibinfo{person}{Jian Yong}, \bibinfo{person}{Ziye Zhu}, {and} \bibinfo{person}{Yun Li}.} \bibinfo{year}{2023}\natexlab{}.
\newblock \showarticletitle{Decomposing Source Codes by Program Slicing for Bug Localization}. In \bibinfo{booktitle}{\emph{2023 IJCNN}}. IEEE, \bibinfo{pages}{1--8}.
\newblock


\bibitem[Youm et~al\mbox{.}(2015)]%
        {youm2015bug}
\bibfield{author}{\bibinfo{person}{Klaus~Changsun Youm}, \bibinfo{person}{June Ahn}, \bibinfo{person}{Jeongho Kim}, {and} \bibinfo{person}{Eunseok Lee}.} \bibinfo{year}{2015}\natexlab{}.
\newblock \showarticletitle{Bug localization based on code change histories and bug reports}. In \bibinfo{booktitle}{\emph{2015 APSEC}}. IEEE, \bibinfo{pages}{190--197}.
\newblock


\bibitem[Yu et~al\mbox{.}(2017)]%
        {yu2017refining}
\bibfield{author}{\bibinfo{person}{Liang-Chih Yu}, \bibinfo{person}{Jin Wang}, \bibinfo{person}{K~Robert Lai}, {and} \bibinfo{person}{Xuejie Zhang}.} \bibinfo{year}{2017}\natexlab{}.
\newblock \showarticletitle{Refining word embeddings for sentiment analysis}. In \bibinfo{booktitle}{\emph{Proceedings of the 2017 EMNLP}}. \bibinfo{pages}{534--539}.
\newblock


\bibitem[Yu et~al\mbox{.}(2024)]%
        {yu2024dataset}
\bibfield{author}{\bibinfo{person}{Xiao Yu}, \bibinfo{person}{Zexian Zhang}, \bibinfo{person}{Feifei Niu}, \bibinfo{person}{Xing Hu}, \bibinfo{person}{Xin Xia}, {and} \bibinfo{person}{John Grundy}.} \bibinfo{year}{2024}\natexlab{}.
\newblock \showarticletitle{What Makes a High-Quality Training Dataset for Large Language Models: A Practitioners' Perspective}. In \bibinfo{booktitle}{\emph{Proceedings of the 39th IEEE/ACM ASE}}. \bibinfo{pages}{656--668}.
\newblock


\bibitem[Yuan et~al\mbox{.}(2020)]%
        {yuan2020dependloc}
\bibfield{author}{\bibinfo{person}{Wei Yuan}, \bibinfo{person}{Binhang Qi}, \bibinfo{person}{Hailong Sun}, {and} \bibinfo{person}{Xudong Liu}.} \bibinfo{year}{2020}\natexlab{}.
\newblock \showarticletitle{Dependloc: A dependency-based framework for bug localization}. In \bibinfo{booktitle}{\emph{2020 27th APSEC}}. IEEE, \bibinfo{pages}{61--70}.
\newblock


\bibitem[Zakari et~al\mbox{.}(2020)]%
        {zakari2020multiple}
\bibfield{author}{\bibinfo{person}{Abubakar Zakari}, \bibinfo{person}{Sai~Peck Lee}, \bibinfo{person}{Rui Abreu}, \bibinfo{person}{Babiker~Hussien Ahmed}, {and} \bibinfo{person}{Rasheed~Abubakar Rasheed}.} \bibinfo{year}{2020}\natexlab{}.
\newblock \showarticletitle{Multiple fault localization of software programs: A systematic literature review}.
\newblock \bibinfo{journal}{\emph{IST}}  \bibinfo{volume}{124} (\bibinfo{year}{2020}), \bibinfo{pages}{106312}.
\newblock


\bibitem[Zakari et~al\mbox{.}(2019)]%
        {zakari2019software}
\bibfield{author}{\bibinfo{person}{Abubakar Zakari}, \bibinfo{person}{Sai~Peck Lee}, \bibinfo{person}{Khubaib~Amjad Alam}, {and} \bibinfo{person}{Rodina Ahmad}.} \bibinfo{year}{2019}\natexlab{}.
\newblock \showarticletitle{Software fault localisation: a systematic mapping study}.
\newblock \bibinfo{journal}{\emph{IET Software}} \bibinfo{volume}{13}, \bibinfo{number}{1} (\bibinfo{year}{2019}), \bibinfo{pages}{60--74}.
\newblock


\bibitem[Zamfirov(2022)]%
        {zamfirov2022literature}
\bibfield{author}{\bibinfo{person}{Filip Zamfirov}.} \bibinfo{year}{2022}\natexlab{}.
\newblock \showarticletitle{A literature review on different types of empirically evaluated bug localization approaches}.
\newblock \bibinfo{journal}{\emph{arXiv:2212.11774}} (\bibinfo{year}{2022}).
\newblock


\bibitem[Zeng et~al\mbox{.}(2021)]%
        {zeng2021deep}
\bibfield{author}{\bibinfo{person}{Zhengran Zeng}, \bibinfo{person}{Yuqun Zhang}, \bibinfo{person}{Haotian Zhang}, {and} \bibinfo{person}{Lingming Zhang}.} \bibinfo{year}{2021}\natexlab{}.
\newblock \showarticletitle{Deep just-in-time defect prediction: how far are we?}. In \bibinfo{booktitle}{\emph{30th ISSTA}}. \bibinfo{pages}{427--438}.
\newblock


\bibitem[Zhang et~al\mbox{.}(2024)]%
        {zhang2024code}
\bibfield{author}{\bibinfo{person}{Dejiao Zhang}, \bibinfo{person}{Wasi~Uddin Ahmad}, \bibinfo{person}{Ming Tan}, \bibinfo{person}{Hantian Ding}, \bibinfo{person}{Ramesh Nallapati}, \bibinfo{person}{Dan Roth}, \bibinfo{person}{Xiaofei Ma}, {and} \bibinfo{person}{Bing Xiang}.} \bibinfo{year}{2024}\natexlab{}.
\newblock \showarticletitle{{CODE} {REPRESENTATION} {LEARNING} {AT} {SCALE}}. In \bibinfo{booktitle}{\emph{The Twelfth ICLR}}.
\newblock


\bibitem[Zhang et~al\mbox{.}(2011)]%
        {zhang2011identifying}
\bibfield{author}{\bibinfo{person}{He Zhang}, \bibinfo{person}{Muhammad~Ali Babar}, {and} \bibinfo{person}{Paolo Tell}.} \bibinfo{year}{2011}\natexlab{}.
\newblock \showarticletitle{Identifying relevant studies in software engineering}.
\newblock \bibinfo{journal}{\emph{IST}} \bibinfo{volume}{53}, \bibinfo{number}{6} (\bibinfo{year}{2011}), \bibinfo{pages}{625--637}.
\newblock


\bibitem[Zhang et~al\mbox{.}(2020)]%
        {zhang2020exploitingkgbuglocator}
\bibfield{author}{\bibinfo{person}{Jinglei Zhang}, \bibinfo{person}{Rui Xie}, \bibinfo{person}{Wei Ye}, \bibinfo{person}{Yuhan Zhang}, {and} \bibinfo{person}{Shikun Zhang}.} \bibinfo{year}{2020}\natexlab{}.
\newblock \showarticletitle{Exploiting code knowledge graph for bug localization via bi-directional attention}. In \bibinfo{booktitle}{\emph{Proceedings of the 28th ICPC}}. \bibinfo{pages}{219--229}.
\newblock


\bibitem[Zhang et~al\mbox{.}(2021)]%
        {zhang2021kcrec}
\bibfield{author}{\bibinfo{person}{Lisa Zhang}, \bibinfo{person}{Zhe Kang}, \bibinfo{person}{Xiaoxin Sun}, \bibinfo{person}{Hong Sun}, \bibinfo{person}{Bangzuo Zhang}, {and} \bibinfo{person}{Dongbing Pu}.} \bibinfo{year}{2021}\natexlab{}.
\newblock \showarticletitle{KCRec: Knowledge-aware representation graph convolutional network for recommendation}.
\newblock \bibinfo{journal}{\emph{Knowledge-Based Systems}}  \bibinfo{volume}{230} (\bibinfo{year}{2021}), \bibinfo{pages}{107399}.
\newblock


\bibitem[Zhang et~al\mbox{.}(2006)]%
        {zhang2006locating}
\bibfield{author}{\bibinfo{person}{Xiangyu Zhang}, \bibinfo{person}{Neelam Gupta}, {and} \bibinfo{person}{Rajiv Gupta}.} \bibinfo{year}{2006}\natexlab{}.
\newblock \showarticletitle{Locating faults through automated predicate switching}. In \bibinfo{booktitle}{\emph{Proceedings of the 28th ICSE}}. \bibinfo{pages}{272--281}.
\newblock


\bibitem[Zhang et~al\mbox{.}(2023)]%
        {zhang2023enhancing}
\bibfield{author}{\bibinfo{person}{Xia Zhang}, \bibinfo{person}{Ziye Zhu}, {and} \bibinfo{person}{Yun Li}.} \bibinfo{year}{2023}\natexlab{}.
\newblock \showarticletitle{Enhancing Bug Localization through Bug Report Summarization}. In \bibinfo{booktitle}{\emph{2023 IEEE ICDM}}. IEEE, \bibinfo{pages}{1541--1546}.
\newblock


\bibitem[Zhao et~al\mbox{.}(2024)]%
        {zhao2024fineibl}
\bibfield{author}{\bibinfo{person}{Yaqiang Zhao}, \bibinfo{person}{Xiaozhuo Li}, \bibinfo{person}{Wei Deng}, \bibinfo{person}{Ying Li}, \bibinfo{person}{Xiaobo Guo}, \bibinfo{person}{Qing Tian}, {and} \bibinfo{person}{Ying Fan}.} \bibinfo{year}{2024}\natexlab{}.
\newblock \showarticletitle{Fine-Grained Bug Localization Based on Rich Context using Attention Tree-GRU}. In \bibinfo{booktitle}{\emph{2024 5th ICCEA}}. IEEE, \bibinfo{pages}{640--646}.
\newblock


\bibitem[Zhou et~al\mbox{.}(2024)]%
        {zhou2024multimacl-irfl}
\bibfield{author}{\bibinfo{person}{Chunying Zhou}, \bibinfo{person}{Xiaoyuan Xie}, \bibinfo{person}{Gong Chen}, \bibinfo{person}{Peng He}, {and} \bibinfo{person}{Bing Li}.} \bibinfo{year}{2024}\natexlab{}.
\newblock \showarticletitle{Multi-View Adaptive Contrastive Learning for Information Retrieval Based Fault Localization}.
\newblock \bibinfo{journal}{\emph{arXiv:2409.12519}} (\bibinfo{year}{2024}).
\newblock


\bibitem[Zhou et~al\mbox{.}(2012)]%
        {zhou2012should}
\bibfield{author}{\bibinfo{person}{Jian Zhou}, \bibinfo{person}{Hongyu Zhang}, {and} \bibinfo{person}{David Lo}.} \bibinfo{year}{2012}\natexlab{}.
\newblock \showarticletitle{Where should the bugs be fixed? more accurate information retrieval-based bug localization based on bug reports}. In \bibinfo{booktitle}{\emph{2012 34th ICSE}}. IEEE, \bibinfo{pages}{14--24}.
\newblock


\bibitem[Zhu et~al\mbox{.}(2020)]%
        {zhu2020cooba}
\bibfield{author}{\bibinfo{person}{Ziye Zhu}, \bibinfo{person}{Yun Li}, \bibinfo{person}{Hanghang Tong}, {and} \bibinfo{person}{Yu Wang}.} \bibinfo{year}{2020}\natexlab{}.
\newblock \showarticletitle{Cooba: Cross-project bug localization via adversarial transfer learning}. In \bibinfo{booktitle}{\emph{IJCAI}}.
\newblock


\bibitem[Zhu et~al\mbox{.}(2021a)]%
        {zhu2021deepdemob}
\bibfield{author}{\bibinfo{person}{Ziye Zhu}, \bibinfo{person}{Yun Li}, \bibinfo{person}{Yu Wang}, \bibinfo{person}{Yaojing Wang}, {and} \bibinfo{person}{Hanghang Tong}.} \bibinfo{year}{2021}\natexlab{a}.
\newblock \showarticletitle{A deep multimodal model for bug localization}.
\newblock \bibinfo{journal}{\emph{Data Mining and Knowledge Discovery}} \bibinfo{volume}{35}, \bibinfo{number}{4} (\bibinfo{year}{2021}), \bibinfo{pages}{1369--1392}.
\newblock


\bibitem[Zhu et~al\mbox{.}(2022a)]%
        {zhu2022bl-gan}
\bibfield{author}{\bibinfo{person}{Ziye Zhu}, \bibinfo{person}{Hanghang Tong}, \bibinfo{person}{Yu Wang}, {and} \bibinfo{person}{Yun Li}.} \bibinfo{year}{2022}\natexlab{a}.
\newblock \showarticletitle{BL-GAN: Semi-Supervised Bug Localization Via Generative Adversarial Network}.
\newblock \bibinfo{journal}{\emph{IEEE TKDE}} (\bibinfo{year}{2022}).
\newblock


\bibitem[Zhu et~al\mbox{.}(2022b)]%
        {zhu2022enhancingbloco}
\bibfield{author}{\bibinfo{person}{Ziye Zhu}, \bibinfo{person}{Hanghang Tong}, \bibinfo{person}{Yu Wang}, {and} \bibinfo{person}{Yun Li}.} \bibinfo{year}{2022}\natexlab{b}.
\newblock \showarticletitle{Enhancing bug localization with bug report decomposition and code hierarchical network}.
\newblock \bibinfo{journal}{\emph{Knowledge-Based Systems}}  \bibinfo{volume}{248} (\bibinfo{year}{2022}), \bibinfo{pages}{108741}.
\newblock


\bibitem[Zhu et~al\mbox{.}(2021b)]%
        {zhu2021trobo}
\bibfield{author}{\bibinfo{person}{Ziye Zhu}, \bibinfo{person}{Yu Wang}, {and} \bibinfo{person}{Yun Li}.} \bibinfo{year}{2021}\natexlab{b}.
\newblock \showarticletitle{TroBo: A Novel Deep Transfer Model for Enhancing Cross-Project Bug Localization}. In \bibinfo{booktitle}{\emph{International Conference on Knowledge Science, Engineering and Management}}. Springer, \bibinfo{pages}{529--541}.
\newblock


\bibitem[Zimmermann et~al\mbox{.}(2009)]%
        {zimmermann2009cross}
\bibfield{author}{\bibinfo{person}{Thomas Zimmermann}, \bibinfo{person}{Nachiappan Nagappan}, \bibinfo{person}{Harald Gall}, \bibinfo{person}{Emanuel Giger}, {and} \bibinfo{person}{Brendan Murphy}.} \bibinfo{year}{2009}\natexlab{}.
\newblock \showarticletitle{Cross-project defect prediction: a large scale experiment on data vs. domain vs. process}. In \bibinfo{booktitle}{\emph{FSE}}. \bibinfo{pages}{91--100}.
\newblock


\bibitem[Zou et~al\mbox{.}(2019)]%
        {zou2019empirical}
\bibfield{author}{\bibinfo{person}{Daming Zou}, \bibinfo{person}{Jingjing Liang}, \bibinfo{person}{Yingfei Xiong}, \bibinfo{person}{Michael~D Ernst}, {and} \bibinfo{person}{Lu Zhang}.} \bibinfo{year}{2019}\natexlab{}.
\newblock \showarticletitle{An empirical study of fault localization families and their combinations}.
\newblock \bibinfo{journal}{\emph{IEEE TSE}} \bibinfo{volume}{47}, \bibinfo{number}{2} (\bibinfo{year}{2019}), \bibinfo{pages}{332--347}.
\newblock


\bibitem[Zou et~al\mbox{.}(2021)]%
        {zou2021bleser}
\bibfield{author}{\bibinfo{person}{Weiqin Zou}, \bibinfo{person}{Enming Li}, {and} \bibinfo{person}{Chunrong Fang}.} \bibinfo{year}{2021}\natexlab{}.
\newblock \showarticletitle{BLESER: Bug localization based on enhanced semantic retrieval}.
\newblock \bibinfo{journal}{\emph{arXiv:2109.03555}} (\bibinfo{year}{2021}).
\newblock


\bibitem[Zou et~al\mbox{.}(2018)]%
        {zou2018practitioners}
\bibfield{author}{\bibinfo{person}{Weiqin Zou}, \bibinfo{person}{David Lo}, \bibinfo{person}{Zhenyu Chen}, \bibinfo{person}{Xin Xia}, \bibinfo{person}{Yang Feng}, {and} \bibinfo{person}{Baowen Xu}.} \bibinfo{year}{2018}\natexlab{}.
\newblock \showarticletitle{How practitioners perceive automated bug report management techniques}.
\newblock \bibinfo{journal}{\emph{IEEE TSE}} \bibinfo{volume}{46}, \bibinfo{number}{8} (\bibinfo{year}{2018}), \bibinfo{pages}{836--862}.
\newblock


\end{thebibliography}
\end{document}